\documentclass[a4paper,fleqn]{cas-sc}

\usepackage[authoryear,round]{natbib}
\usepackage{units} 
\def\tsc#1{\csdef{#1}{\textsc{\lowercase{#1}}\xspace}}
\tsc{WGM}
\tsc{QE}

\begin{document}

\let\WriteBookmarks\relax
\def\floatpagepagefraction{1}
\def\textpagefraction{.001}

\shorttitle{}    

\shortauthors{}

\title [mode = title]{Beyond Lanes: Traffic Flow Dynamics in Disordered Conditions Based on High-Resolution Trajectory Data}



%

\author[1]{Shrey Agrawal}



\ead{shrey20@iitk.ac.in}

\ead[url]{}

\credit{}

\author[2]{Gowri Asaithambi}


\ead{gowri@iittp.ac.in}
\cormark[1]

\ead[url]{}

\credit{}

\affiliation[1]{organization={Indian Institute of Technology Kanpur},
            city={Kanpur},
            postcode={208016}, 
            state={Uttar Pradesh},
            country={India}}

\author[1]{Venkatesan Kanagaraj}


\ead{venkatk@iitk.ac.in}

\ead[url]{}

\credit{}

\affiliation[2]{organization={Indian Institute of Technology Tirupati},
            city={Tirupati},
            postcode={517619}, 
            state={Andhra Pradesh},
            country={India}}

\author[3]{Martin Treiber}


\ead{martin@mtreiber.de}

\ead[url]{}

\credit{}

\author[3]{Ostap Okhrin}


\ead{ostap.okhrin@tu-dresden.de}

\ead[url]{}

\credit{}
\author[1]{Harish Babu Kumara}


\ead{harishbk24@iitk.ac.in}

\ead[url]{}

\credit{}

\affiliation[3]{organization={Technical University of Dresden},
            city={Dresden},
            postcode={01062}, 
            country={Germany}}

\cortext[1]{Corresponding author}

\fntext[1]{}


\begin{abstract}
 Disordered traffic flow is characterized by weak or non-existent lane discipline in the presence of strong vehicle heterogeneity and continuous lateral interactions, challenging traditional lane-based modeling assumptions. This study presents an empirical study of macroscopic and microscopic aspects of disordered traffic using high-resolution UAV trajectory data collected on an urban arterial. A two-dimensional extension of Edie’s framework is applied to quantify aggregate traffic variables and produce a two-dimensional fundamental diagram, revealing that traffic states cannot be adequately represented using one-dimensional formulations and highlighting the persistent role of lateral redistribution. The propagation of congestion is estimated directly from the spatiotemporal speed fields, demonstrating the emergence of coherent stop-and-go waves and showing a similar dynamics as conventional lane-based flow, in spite of the heterogeneous vehicle interactions. At the microscopic level, steady-state follower-leader identification is used to examine desired time gaps and minimum lateral spacing, vehicle dimension distributions, and kinematic characteristics, revealing pronounced inter-class heterogeneity that explains disordered traffic behavior. The study provides an empirical framework linking vehicle-level interactions and aggregate traffic dynamics and establishes a data-driven basis for the calibration and validation of traffic models for disordered mixed traffic systems.

\end{abstract}




\maketitle

\section{Introduction}
\noindent
Vehicle traffic is a complex system in which individual road users continuously interact with surrounding vehicles and the roadway environment while pursuing their own travel objectives. Although these interactions are inherently stochastic at the individual level, they give rise to recognizable patterns and collective behavior at larger scales. Such emergent structure makes it possible to analyze traffic flow, develop mathematical descriptions, and design strategies for prediction, management, and control. Therefore, detailed understanding of traffic flow dynamics remains central to improving road safety, operational efficiency, and the performance of urban transport systems. Traffic flow dynamics describes how vehicle movement evolves over space and time as a result of interactions among drivers, vehicles, and the road environment. These dynamics are commonly examined through a set of macroscopic variables, such as flow, density, and average speed, that describe aggregate system behavior, as well as microscopic variables, such as spacing, time gaps, speed, acceleration, and lateral positioning, that capture vehicle-level interactions and driver responses. Together, these measures provide complementary views of how traffic states form, evolve, and transition. In traffic systems with strong lane discipline, vehicle interactions are largely constrained in the longitudinal direction, leading to relatively structured and predictable dynamics. Whereas, under disordered conditions marked by weak lane adherence and heterogeneous vehicle composition, interactions extend laterally and become inherently two-dimensional. This added complexity affects spacing behavior, traffic characteristics, and congestion formation, making simplified representations insufficient. Capturing such dynamics therefore requires detailed trajectory data that provides individual vehicle motion with high spatial and temporal fidelity, enabling consistent analysis across both microscopic and macroscopic scales.\\

\noindent
Extensive efforts have been made to collect high-resolution vehicle trajectory data to support the study of traffic flow dynamics. Early large-scale initiatives, such as the Next Generation SIMulation (NGSIM) program \citep{USDOT2016NGSIM}, provided detailed continuous vehicle trajectories on freeways and urban segments, enabling foundational work on car-following and lane-changing behavior. Since then, a wide range of public trajectory datasets have been developed using stationary and aerial sensing platforms. These include highway-focused datasets such as highD \citep{highD2018}, HIGH-SIM \citep{HighSIM2021}, AUTOMATUM \citep{AUTOMATUM2021}, and I-24 MOTION \citep{I242023}, which capture detailed longitudinal interactions over extended road sections, as well as urban datasets such as pNEUMA \citep{pNEUMA2020}, CitySim \citep{CitySIM2024}, City Scale \citep{City2023}, TUMDOT-MUC \citep{tumdot2024}, MiTra \citep{Mitra2025}, MAGIC \citep{MAGIC2022}, and Zen Traffic Data \citep{Zen2020}, which incorporate intersections, ramps, and weaving areas. Collectively, these datasets span diverse traffic states and road environments and have enabled substantial advances in traffic modeling, model calibration, and data-driven analysis. Despite differences in sensing platforms and road settings, most of these datasets represent traffic systems with clear lane structure, relatively stable longitudinal interactions, and limited lateral freedom. Conversely, trajectory datasets for disordered or lane-free traffic remain limited.\\

\noindent
Early efforts to collect trajectory data under disordered  traffic conditions have primarily relied on fixed video cameras installed along urban road sections. One of the first publicly available midblock datasets was reported by \citet{Venkatesan2015Traj}, who extracted trajectories over a $245$~m stretch in Chennai using stationary cameras, followed by later studies covering similar environments over both shorter and longer road segments \citep{Amrutsamanvar2021, Raju2022}. While these datasets offered valuable early insights, oblique camera views often introduced occlusions and perspective distortions, particularly under dense traffic, making it difficult to resolve closely spaced vehicles. The increasing availability of unmanned aerial vehicles (UAVs) has enabled more reliable trajectory extraction through near-orthographic views, motivating several recent studies that captured long disordered-traffic trajectories over stretches ranging from about $560$ to $605$~m using single or multiple UAVs \citep{Chouhan2023, Kashyap2023Traj, RAJPUT2026Traj}. Despite these advances, most existing datasets predominantly represent free-flow to moderately congested states, with limited coverage of sustained congestion and stop-and-go dynamics. Such regimes are critical for understanding non-steady state conditions, queue formation, and congestion propagation, as they involve strong vehicle interactions and frequent speed adjustments. The dataset reported by \citet{Kumar2025StopNGo} addresses this limitation by capturing extended stop-and-go traffic alongside other traffic states over a long midblock section, providing a richer basis for analyzing traffic flow dynamics and for calibrating models tailored to disordered traffic conditions.\\

\noindent The present study investigates traffic flow dynamics in disordered, lane-free conditions using high-resolution stop-and-go trajectory data reported by \citet{Kumar2025StopNGo}. The aim is to examine how heterogeneous vehicle interactions shape traffic behavior at both the macroscopic and microscopic scales under sustained congestion. The contributions of this study are threefold. First, at the macroscopic level, a two-dimensional fundamental diagram of the flow-density (flux-density) vector as a function of the 2D density is developed directly from trajectory data, enabling an aggregate representation of both longitudinal and lateral interactions without imposing lane-based assumptions. In addition, congestion propagation speed is estimated from spatiotemporal speed fields using a cross-correlation approach, providing a direct measure of collective traffic dynamics, in contrast to earlier studies that primarily inferred wave speed indirectly from aggregate fundamental relationships. Such direct estimation is important for understanding traffic instability and calibrating macroscopic traffic flow models. At the microscopic level, vehicle interactions are analyzed through desired time gaps, minimum standstill gaps, geometric dimensions, and kinematic trends, with an emphasis on steady-state follower–leader conditions to reduce the influence of transient effects and enable a more reliable characterization of driver behavior across heterogeneous vehicle classes. Together, these contributions address key limitations in existing empirical studies, where macroscopic and microscopic analyses are often treated separately or are derived under restrictive assumptions. The resulting framework links vehicle-level interactions with aggregate traffic behavior, supporting improved modeling, calibration, and management of disordered urban traffic systems.\\

\section{Literature Review}
\noindent
Weak lane discipline is a defining feature of traffic streams where vehicles of different sizes and performance characteristics share the same roadway without strict lateral segregation even if lane markers are visible on the road surface. In such settings, drivers continuously adapt their longitudinal and lateral velocity components and their position based not only on the vehicle ahead but also on the local vehicle neighborhood in any direction.These adaptations introduce strong spatial variability and interaction complexity that cannot be adequately described using assumptions developed for homogeneous lane-based traffic \citep{asaithambi2012characteristics, Kiran2016Review}. As a result, traffic behavior under weak lane discipline exhibits greater dispersion, higher susceptibility to traffic instabilities, and strong dependence on vehicle composition, motivating sustained research into models and metrics suited to disordered conditions.\\

\noindent
Much of the early work on disordered traffic on macroscopic modeling sought extensions of classical flow theory that account for heterogeneity and lateral freedom. Multiclass formulations were initially explored to represent different vehicle types, but these approaches alone were insufficient to capture key features such as gap filling and uneven road usage \citep{Bhavathrathan2012EvolutionPerspective}. Subsequent studies introduced concepts such as porous flow, where smaller vehicles exploit fine gaps, and regime-dependent behavior, where vehicle interactions differ between free-flow and congested states \citep{Nair2011ASystems, Mayakuntla2019CTM}. More recent efforts have emphasized area-based and two-dimensional representations, recognizing that lateral motion and spatial distribution play a direct role in traffic evolution \citep{Mohan2021Multi-classSurface, Chakroborty2019Disorderly, Vikram2022StabilizedFlow, Agrawal2023TwoD}. These developments, supported by empirical trajectory observations \citep{Venkatesan2015Traj, RAJPUT2026Traj} and microscopic models \citep{TreiberChaudhari2022IAM, KASHYAPNR2024HSSFM, Kanagaraj2018SelfDriven}, collectively indicate that one-dimensional descriptions are inadequate for disordered traffic, especially under congestion where lateral interactions become pronounced.\\

\noindent
The fundamental diagram (FD) is a core concept in traffic flow theory that relates macroscopic variables such as flow, density, and speed to describe aggregate traffic behavior \citep{Lighthill1955OnRoads}. First introduced through empirical observations of highway traffic \citep{Greenshields1935ACapacity}, the FD framework has since been extended to urban arterials, signalized corridors, and network-level representations, including the macroscopic fundamental diagram (MFD) \citep{pedersen2011trb, WU2011AFD, Zhang2020MFD, Carrillo2024MFD, Mitra2025}. To better represent variability and mixed traffic conditions, stochastic and multimodal FDs have been proposed, offering probabilistic descriptions of speed and flow under similar density levels \citep{SUMALEE2011SCTM, GEROLIMINIS2014FD, Zhang2025SFD}. These formulations underpin widely used macroscopic models such as the LWR and cell transmission models and are routinely employed to estimate key parameters, such as, capacity, critical density, and jam density, that inform infrastructure design and traffic control strategies \citep{Richards1956ShockHighway, SAADULLAH2025NFD}. Whereas, FD development for heterogeneous and disordered traffic remains comparatively limited. Existing studies have proposed multimodal and stochastic FDs by scaling variables using passenger car units or by incorporating vehicle composition through area occupancy measures and data-driven clustering \citep{AHMED202FD, Nandan2023FD, Nandan2024FD}. While these approaches improve representation of heterogeneity, most still rely on implicit assumptions of one-dimensional, lane-based motion and estimate macroscopic variables using space–time averaging methods such as Edie’s formulation \citep{Edie1963DiscussionDefinitions}. Such assumptions are less suitable under weak lane discipline, where lateral movements play a direct role in traffic evolution. Recent work therefore supports for extending FD concepts to two dimensions by treating density as an areal quantity and flow as a vector, allowing both longitudinal and lateral interactions to be captured explicitly \citep{Chakroborty2019Disorderly, DELPIANO20202DCF, Vikram2022StabilizedFlow, Agrawal2023TwoD, KASHYAPNR2024HSSFM}. Similar extensions of space–time averaging have been explored in planar movement contexts such as pedestrian flows and traffic networks \citep{Saberi2014EstimatingTrajectories, VanWageningen2014ExtensionDynamics}. However, a consistent empirical formulation of a two-dimensional fundamental diagram based on trajectory data remains largely unexplored. This limits the ability to directly relate areal density and directional flow in disordered traffic, motivating the need for a two-dimensional FD framework that captures coupled longitudinal and lateral interactions while remaining consistent with trajectory-based measurements.\\

\noindent
Congestion wave propagation speed is a fundamental property of traffic flow, derived from kinematic wave theory where the speed of a shock equals the slope of the fundamental diagram \citep{Lighthill1955OnRoads, Richards1956ShockHighway}. This principle supports macroscopic models such as the Cell Transmission Model, in which queue growth and spillback are governed by backward-moving waves \citep{DAGANZO1994CTM}. Empirical studies across a wide range of traffic environments consistently report upstream stop-and-go wave speeds in the range of approximately $-3.33$ to $-5.56$~m/s, suggesting that wave speed is a stable characteristic of congested traffic rather than a site-specific artifact \citep{TreitererMyers1974hysteresis, Treiber2000IDM,Zielke2008StopandGo, Sugiyama2008Congestion, Laval2010propagation, Treiber2012Cong, CHEN2014Oscillations, I242023}. For disordered and mixed traffic, recent studies have estimated congestion wave speeds using area-occupancy–based fundamental diagrams and trajectory-based transformations, reporting class- and composition-dependent propagation characteristics \citep{Nandan2023FD, Nandan2024FD}. While these efforts extend wave speed analysis to heterogeneous traffic, estimates are typically obtained indirectly from aggregated relationships or optimized transformations rather than from direct observation of wave fronts. Further, vehicle-class-specific estimates often rely on implicit car-following assumptions. Therefore, the present study, estimates congestion wave speed from the spatiotemporal evolution of trajectory-derived speed fields, enabling direct observation of wave propagation without relying on fundamental diagram slopes or transformation assumptions. This provides a more consistent basis for analyzing congestion dynamics and for calibrating traffic flow models under weak lane discipline.\\

\noindent
Microscopic modeling of disordered (lane-free) traffic has evolved by relaxing the single-leader, lane-based assumptions of classical car-following theory. Early adaptations of stimulus–response, Gipps, Optimal Velocity, and IDM models \citep{Chandler1958GHR, Gipps1981, Bando1995OVM, Treiber2000IDM} incorporated vehicle-type-specific parameters and lateral offset terms to reflect heterogeneous interactions under weak lane discipline \citep{Gunay2007CFM}. However, these retained one-dimensional following logic and did not adequately capture dynamic leader switching or non-following regimes. Recent models have adopted interaction-based, geometry-aware formulations in which longitudinal acceleration depends on multiple surrounding vehicles, lateral separation, and vehicle dimensions \citep{Asaithambi2016Review, Azam2022Disordered, JIN2010CFM, Kanagaraj2018SelfDriven, Sangram2024CFM, KASHYAPNR2024HSSFM}. With improved trajectory data availability, calibration has gradually shifted toward position-based, global optimization methods rather than macroscopic fitting \citep{TREIBER2013Calibration, Li2016Calibration, Chaudhari2022Calibration}. Across these approaches, principal microscopic inputs remain consistent, which includes longitudinal distance gap, relative speed, desired time gap, minimum spacing, acceleration and deceleration limits, vehicle length and width, and lateral overlap measures \citep{Mahapatra2018ParametricStudy}.\\

\noindent
Empirical microscopic studies provide strong evidence for the behavioral assumptions embedded in disordered traffic car-following models. Trajectory-based analyses show that weak lane discipline leads to shorter and more dispersed time headways and longitudinal gaps, particularly under staggered following where leader identification depends on lateral overlap rather than strict lane alignment \citep{Venkatesan2015Traj,Das2019TimeHeadway}. Spacing and gap distributions vary systematically by vehicle class, with two-wheelers and auto-rickshaws maintaining smaller and more variable gaps than cars and heavy vehicles due to size and maneuverability differences \citep{Amrutsamanvar2021}. Empirical observations also document heterogeneity in free speeds, acceleration–deceleration behavior, vehicle dimensions, and transverse clearance, along with flow-dependent lateral positioning that constrains movement as density increases \citep{Venkatesan2008Microscopic,Raju2022}. At signalized intersections, vehicle-class-wise speed profiles further confirm state-dependent longitudinal responses across stopped, saturated, and unaffected regimes \citep{Chauhan2021SignalizedIntersection}. However, a recurring limitation is that many reported time-gap and spacing distributions are derived from transient or disturbance-affected regimes without ensuring steady-state flow or persistent follower–leader pairing. Since time gap and distance gap are meaningful descriptors only under sustained following conditions, estimates from non-steady states may bias calibration and validation assessment. In addition, intra-class variability in vehicle dimensions is often simplified despite its direct impact on packing density, gap acceptance, and the realism of microscopic simulation models for heterogeneous traffic systems.\\

\noindent Existing research on disordered and mixed traffic has advanced significantly with the development of trajectory datasets and lane-free traffic models; however, several key limitations remain. Macroscopic fundamental diagram studies for disordered traffic have largely relied on one-dimensional representations that implicitly assume predominantly longitudinal motion, making them inadequate for capturing the strong lateral interactions and spatial redistribution observed in lane-free traffic. Similarly, congestion wave propagation in disordered traffic has often been estimated indirectly from aggregated relationships rather than directly from spatiotemporal trajectory data. At the microscopic level, many reported spacing and time-gap distributions are derived from transient or disturbance-affected regimes without ensuring steady-state follower-leader interactions, potentially biasing model calibration. In addition, intra- and inter-class variability in vehicle dimensions and kinematics is frequently simplified despite its importance for space utilization and flow dynamics. These limitations highlight the need for approaches that directly link detailed trajectory data with both macroscopic and microscopic traffic behavior in disordered environments. Thus, providing an empirical foundation for the calibration and validation of traffic models for disordered mixed traffic.\\


\noindent The remainder of this paper is organized as follows. Section~3 describes the study site, UAV data collection, and trajectory extraction process. Section~4 presents the trajectory processing, coordinate transformation, and validation procedures. Section~5 provides the macroscopic and microscopic analyses of disordered traffic flow, including the two-dimensional fundamental relationships and congestion wave propagation. Finally, Section~6 summarizes the key findings and discusses implications for traffic modeling.

\section{Study Site and Empirical Data}\label{Sec: Data Collection}
\begin{figure}
    \centering
    \includegraphics[width=\textwidth]{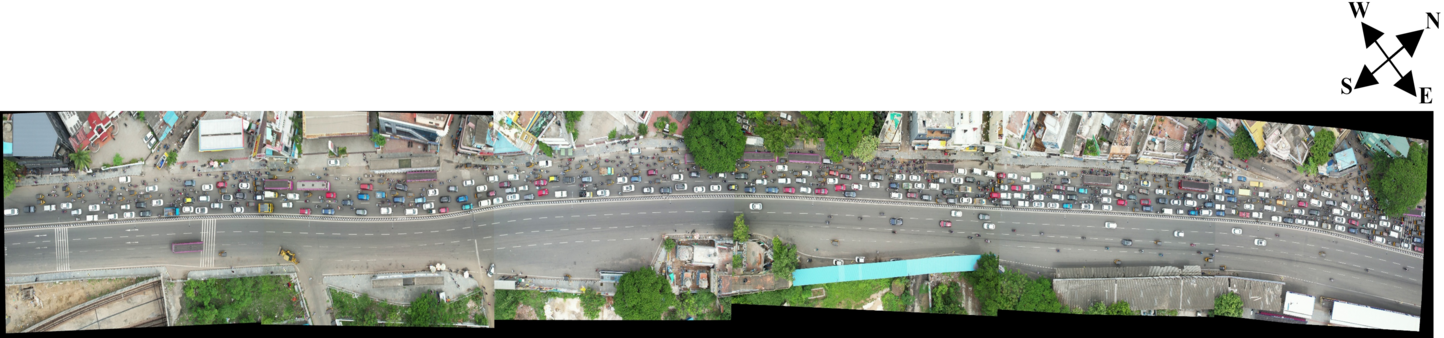}
    \caption{Study section on Anna Salai, Chennai, showing the six-lane
    divided urban arterial. The north-eastbound traffic stream forms
    the basis of the empirical analysis.}
    \label{fig:Data_site}
\end{figure}

\noindent
The video data used in this study were collected at a nominal six-lane divided mid-block section of an urban arterial located in Saidapet, Chennai, India ($13^{\circ}01'15.17''\,\text{N},\;80^{\circ}13'31.55''\,\text{E}$), during peak traffic periods (Figure~\ref{fig:Data_site}). Traffic data were acquired using a coordinated swarm of unmanned aerial vehicles (UAVs), collectively monitoring an approximately $550\,\text{m}$ long road segment. The multi-UAV deployment enabled near-orthographic observation of the traffic stream over the entire study stretch, thereby reducing perspective-related distortions and limiting trajectory loss due to vehicle occlusions. The recorded videos were first stabilized and synchronized, followed by image-based stitching to generate a continuous view of the study section. Deep learning–based methods were then employed to detect and track individual vehicles across frames, after which geometric correction was applied to obtain continuous trajectories over the entire road segment. The resulting vehicle trajectories were extracted as time-stamped position sequences and form the basis for all subsequent analyses. A detailed description of the UAV deployment strategy and the trajectory extraction framework is provided in \citet{Kumar2025StopNGo}.\\

\noindent The extracted trajectories were sampled at a temporal resolution of $\Delta t = 0.04\, \text{s}$ and represented in a global Cartesian coordinate system (CCS) approximately aligned with the roadway. Vehicle positions at each tim
e instant were defined by the geometric centers of the detected bounding boxes, assuming that the bounding boxes tightly circumscribe the vehicle. Under this assumption, the bounding-box centers provide a consistent and centrally aligned representation of vehicle position across successive frames. Each UAV flight covered a duration of approximately $18\,\text{min}$, and the present analysis is based on two such flight segments recorded between 09:30 and 10:30 hours on 26 June 2023, yielding a total of $36\,\text{min}$ of trajectory data. The dataset spans traffic regimes ranging from free-flow to heavily congested conditions and captures complex spatiotemporal phenomena such as the formation and propagation of stop-and-go waves. All data were collected under dry weather and consistent daylight conditions. 
Vehicles were categorized into five classes based on their physical and dynamic characteristics: two-wheelers (TW), cars (CAR), auto-rickshaws (AR), light commercial vehicles (LCV), and heavy commercial vehicles (HCV). In total, approximately $5{,}891$ vehicle trajectories were extracted across all flight segments, of which about $70.84\%$ correspond to two-wheelers, $20.37\%$ to cars, $7.32\%$ to auto-rickshaws, $0.54\%$ to light commercial vehicles, and $0.93\%$ to heavy commercial vehicles. This distribution reflects the heterogeneous traffic composition typically observed on urban arterial roads in Indian cities.

\section{Trajectory Processing and Validation}
\noindent While the extracted vehicle trajectories offer comprehensive spatiotemporal information on individual vehicle movements, their direct representation in a global Cartesian coordinate system is often inadequate for analyzing traffic dynamics on geometrically complex road segments. Furthermore, these trajectories are inherently vulnerable to errors resulting from video capture and automated extraction methods, such as detection noise, temporary occlusions, and tracking discrepancies. Thus, the raw trajectories necessitate geometric refinement and quantitative validation before empirical analysis. This section presents the trajectory processing framework utilized in this study, which includes coordinate system transformation and validation against manually generated ground-truth trajectories. These measures guarantee that the processed data are geometrically consistent and adequately precise for subsequent analyses of traffic flow dynamics and vehicle-level behavior.

\subsection{Coordinate Transformation}
When vehicle trajectories are expressed in a global Cartesian coordinate system, the recorded positions and derived kinematic quantities (such as velocity and acceleration) inherently embed the combined effects of road curvature and traffic interactions. As a result, motion components attributable to roadway geometry cannot be directly separated from those arising from vehicle interactions. To analyze vehicle dynamics relative to the traffic stream, it is therefore necessary to transform trajectories from the global reference frame to a curvilinear driver-centric, road-aligned reference frame. This transformation removes the geometric influence of horizontal curvature and enables vehicle motion to be interpreted primarily in terms of longitudinal and lateral responses to surrounding traffic and roadway constraints along the curved corridor.\\

\begin{figure}
    \centering
    \includegraphics[width=0.8\linewidth]{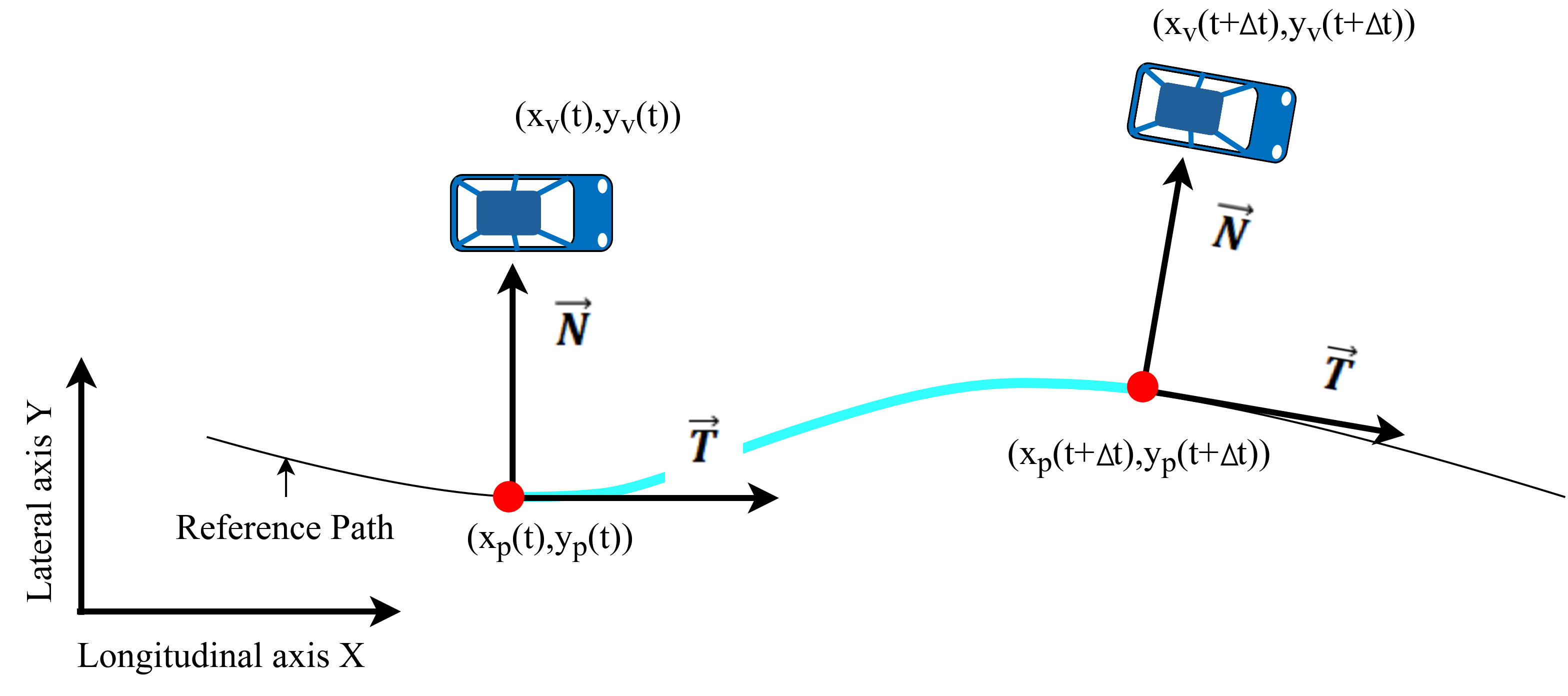}
    \caption{Application of the Frenet coordinate system to a vehicle trajectory}
    \label{fig:fig_2}
\end{figure}

\noindent The Frenet coordinate system (FCS) provides a moving reference frame attached to a point traveling along a continuous reference path, describing vehicle motion in terms of longitudinal distance along the path and lateral displacement normal to it (Figure~\ref{fig:fig_2}). The transformation is implemented by defining the roadway centerline as a continuous parametric curve \citep{Chen2021Fernet}. The position vector $\vec{r}(t)$ of a point on the reference path, parameterized by time $t$, is given by
\begin{equation}
\vec{r}(t) = x_p(t)\hat{i} + y_p(t)\hat{j}.
\end{equation}

\noindent Here, $x_p(t)$ and $y_p(t)$ represents longitudinal and lateral positions in the global coordinate system, repectively. The longitudinal position of a vehicle is measured by the arc length traced by the normal projection of the vehicle onto the reference path. Over a time interval $\Delta t$, this arc length is computed as
\begin{equation}
\text{arc}(\Delta t) = \int_t^{t+\Delta t} \sqrt{\qty(\dv{x_p}{t})^2 + \qty(\dv{y_p}{t})^2}\,\dd t.
\end{equation}
\noindent Accordingly, the longitudinal coordinate of the vehicle at time $t+\Delta t$ is obtained as
\begin{equation}\label{eqn:longitudinal position}
s(t+\Delta t) = s(t) + \text{arc}(\Delta t).
\end{equation}

\begin{figure}
     \centering
     \begin{subfigure}{\textwidth}
         \centering
         \includegraphics[width=\textwidth]{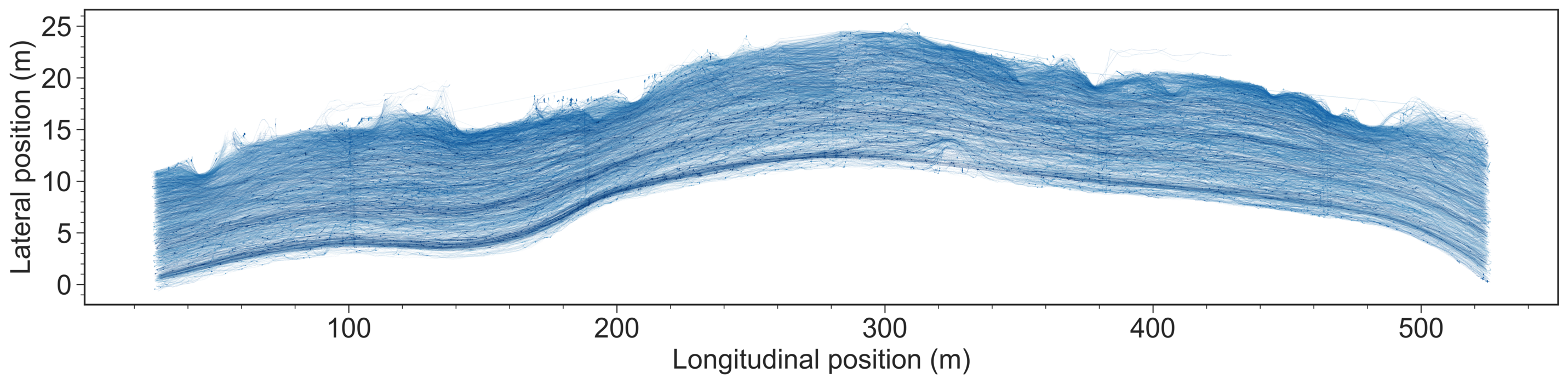}
         \caption{Cartesian representation of vehicle trajectories}
         \label{fig:fig_3a}
     \end{subfigure}
     \vspace{0.3cm}
     \begin{subfigure}{\textwidth}
         \centering
         \includegraphics[width=\textwidth]{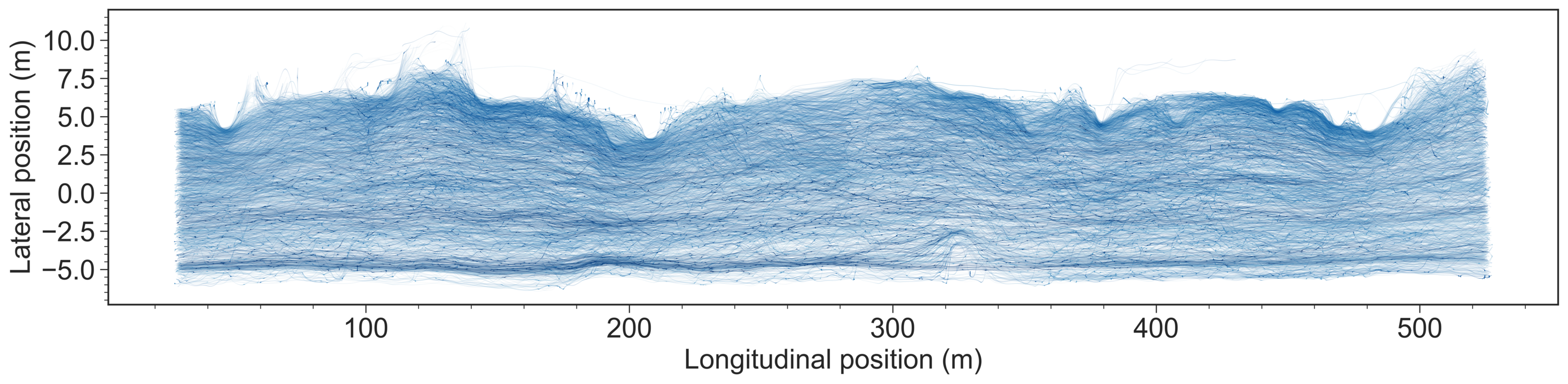}
         \caption{Frenet representation of vehicle trajectories}
         \label{fig:fig_3b}
     \end{subfigure}
        \caption{Coordinate transformation of vehicle trajectories.}
        \label{fig:fig_3}
\end{figure}

\noindent Since the distance traveled along the reference path $s(t)$ increases monotonically with time, the path can equivalently be parameterized as $\vec{r}\qty(s(t))$. Let $\vec{\mathbf{T}}\qty(s(t))$ and $\vec{\mathbf{N}}\qty(s(t))$ denote the unit tangent and normal vectors to the reference path at $\vec{r}\qty(s(t))$, respectively. Taking $\vec{r}\qty(s(t))$ as the local origin of the Frenet frame, the lateral position of a vehicle $\vec{r}_N$ relative to the reference path is expressed as

\begin{equation}\label{eqn:lateral position}
\vec{r}_N\qty(s(t), d(t)) = d(t)\vec{\mathbf{N}}\qty(s(t)),
\end{equation}
%
%


where $d(t)$ represents the perpendicular lateral offset of the vehicle from the reference path (positive if to the right). For the study section, a clothoid spline was fitted to the identified lane marking to obtain a continuous and smooth reference curve. Using this reference path, the longitudinal and lateral positions of vehicles were computed using Eq.~\ref{eqn:longitudinal position} and Eq.~\ref{eqn:lateral position}, respectively. Figure~\ref{fig:fig_3}(a) shows the raw vehicle trajectories in the global coordinate system, while Figure~\ref{fig:fig_3}(b) shows the corresponding trajectories after transformation into the Frenet coordinate system.\\



This separation is essential for subsequent analysis of microscopic interactions, as it ensures that measured longitudinal and lateral velocities, accelerations, and spacing variables represent traffic-induced behavior rather than artifacts of roadway geometry. Notice, however, that this decomposition only makes sense if (i) the local curvature radius $r=\left|\text{d}\vec{\mathbf{N}}/\text{d}s\right|^{-1} $ satisfies $v_f^2/r\ll a_{\rm comf}$ where $v_f$ is the maximum free-flow speed and $a_{\rm comf}$ is the comfortable lateral acceleration otherwise, the inertial forces caused by the lateral curvature would influence the lateral motion, and (ii) $d_{\rm max}/r \ll 1$, i.e., the maximum lateral displacement $(d_{\rm max})$ is much smaller than the curvature radius such that variations of the longitudinal velocity component for constant $\text{d}s/\text{d}t$ and ensuing Coriolis acceleration components are negligible. In our case, this is clearly satisfied.

\subsection{Validation of Extracted Trajectories} 
The inherent variability in object detection and tracking across consecutive frames leads to measurement inaccuracies, manifested as positional variations in the estimated centers of vehicles. The robustness of the detection and tracking methodology has been quantitatively evaluated using standard practices in the computer vision and artificial intelligence community, as reported in \citet{Kumar2025StopNGo}, demonstrating a vehicle detection precision rate of 97\%, a recall rate of 93\%, and a mean average precision (mAP) of 91\% at an intersection-over-union threshold of 0.5. The evaluation of tracking accuracy, compared to manually annotated ground truth trajectories, yielded a multi-object tracking accuracy (MOTA) of 94\% and a mean positional error (MOTP) of 0.15. Standard evaluation practices in the traffic research community assess trajectory accuracy in real-world coordinates, expressed in meters in both the longitudinal and lateral directions. This representation facilitates straightforward interpretation of positional errors and enables quantitative assessment of the accuracy of the processed trajectories at the trajectory level before their use in subsequent empirical analyses.\\

\noindent To this end, a benchmark dataset was constructed by manually annotating trajectories of 35 vehicles randomly sampled from the study dataset. These manually annotated trajectories are first subjected to the Frenet coordinate transformation and subsequently serve as ground truth for evaluating the accuracy of the automatically extracted and transformed trajectories (Figure~\ref{fig:fig_3b}). Trajectory accuracy is assessed using standard error metrics that quantify both overall positional deviations and directional consistency with respect to the ground truth. Let $i = 1, 2, \ldots, n$ denote the vehicle index, and let $(X_i^t, Y_i^t)$ and $(x_i^t, y_i^t)$ represent the blue{automatically} extracted and ground-truth vehicle positions, respectively, at time $t$ for vehicle $i$. The instantaneous positional deviation between the extracted and ground-truth trajectories is defined as the Euclidean distance between the corresponding vehicle centers, denoted by $d_i^t$. Using these deviations, trajectory accuracy is quantified using the root mean squared error (RMSE) and mean absolute error (MAE), computed over all vehicles and time instances.
\begin{equation}
\mathrm{RMSE} = \sqrt{\frac{1}{N}\sum_{i=1}^{n}\sum_{t=t_0^i}^{T^i}(d_i^t)^2},
\qquad
\mathrm{MAE} = \frac{1}{N}\sum_{i=1}^{n}\sum_{t=t_0^i}^{T^i} d_i^t,
\end{equation}
where $N=\sum_{i=1}^{n}(T^i-t_0^i+1)$ is the total number of trajectory points. While RMSE penalizes larger deviations more strongly and is sensitive to outliers, MAE provides a robust measure of the average positional error. To examine directional consistency, longitudinal and lateral deviations are evaluated separately by defining the absolute error vector
\begin{equation}
\boldsymbol{\varepsilon}_i^t=
\begin{bmatrix}
|\varepsilon_{x,i}^t| \\
|\varepsilon_{y,i}^t|
\end{bmatrix}
=
\begin{bmatrix}
|X_i^t - x_i^t| \\
|Y_i^t - y_i^t|
\end{bmatrix}.
\end{equation}

The mean absolute error vector is computed as:
\begin{equation}
\boldsymbol{\mathrm{MAE}}
=
\frac{1}{N}
\sum_{i=1}^{n}\sum_{t=t_0^i}^{T^i}
\boldsymbol{\varepsilon}_i^t
=
\begin{bmatrix}
\mathrm{MAE}_X \\
\mathrm{MAE}_Y
\end{bmatrix}.
\end{equation}
Here, $\mathrm{MAE}_X$ and $\mathrm{MAE}_Y$ quantify the average longitudinal and lateral positional accuracy, respectively. Together, these metrics provide a comprehensive assessment of both the magnitude and directional consistency of trajectory deviations relative to the ground truth. The trajectory validation demonstrates a high level of agreement between automatically extracted trajectories and manually annotated ground truth. As summarized in Table~\ref{tbl:trajectory_validation}, the overall root mean squared error is $0.316\,\mathrm{m}$, while the mean absolute error is $0.238\,\mathrm{m}$, indicating that positional deviations remain well below one-third of a meter for the majority of trajectory points. Given that vehicle positions are represented by bounding-box centers and sampled at a temporal resolution of $\Delta t = 0.04\,\mathrm{s}$, these error magnitudes are small relative to typical vehicle dimensions and inter-vehicle spacings observed in urban traffic streams.\\

\begin{figure}
    \centering
    \includegraphics[width=0.6\linewidth]{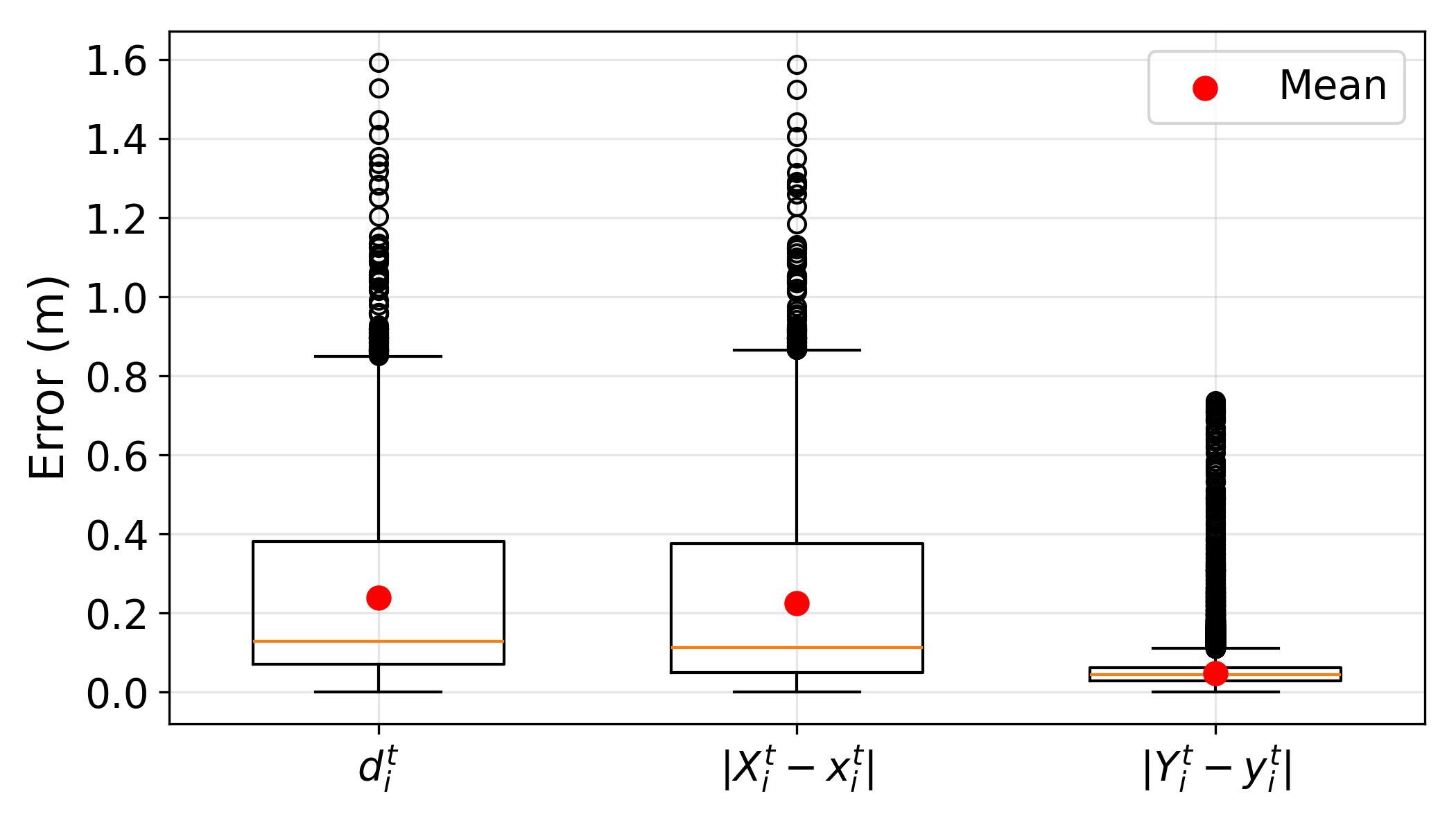}
    \caption{Trajectory Error distribution}
    \label{fig:fig_4}
\end{figure}

\begin{table}
\caption{Accuracy metrics for extracted vehicle trajectories evaluated against manually annotated ground-truth data.}
\label{tbl:trajectory_validation}
\begin{tabular*}{\tblwidth}{@{}lcccccc@{}}
\toprule
Metric & RMSE & MAE & MAE$_X$ & MAE$_Y$ \\
\midrule
Value (m) & 0.316 & 0.238 & 0.225 & 0.048\\
\bottomrule
\end{tabular*}
\end{table}

\begin{figure}
     \centering
     \begin{subfigure}{\textwidth}
         \centering
         \includegraphics[width=\textwidth]{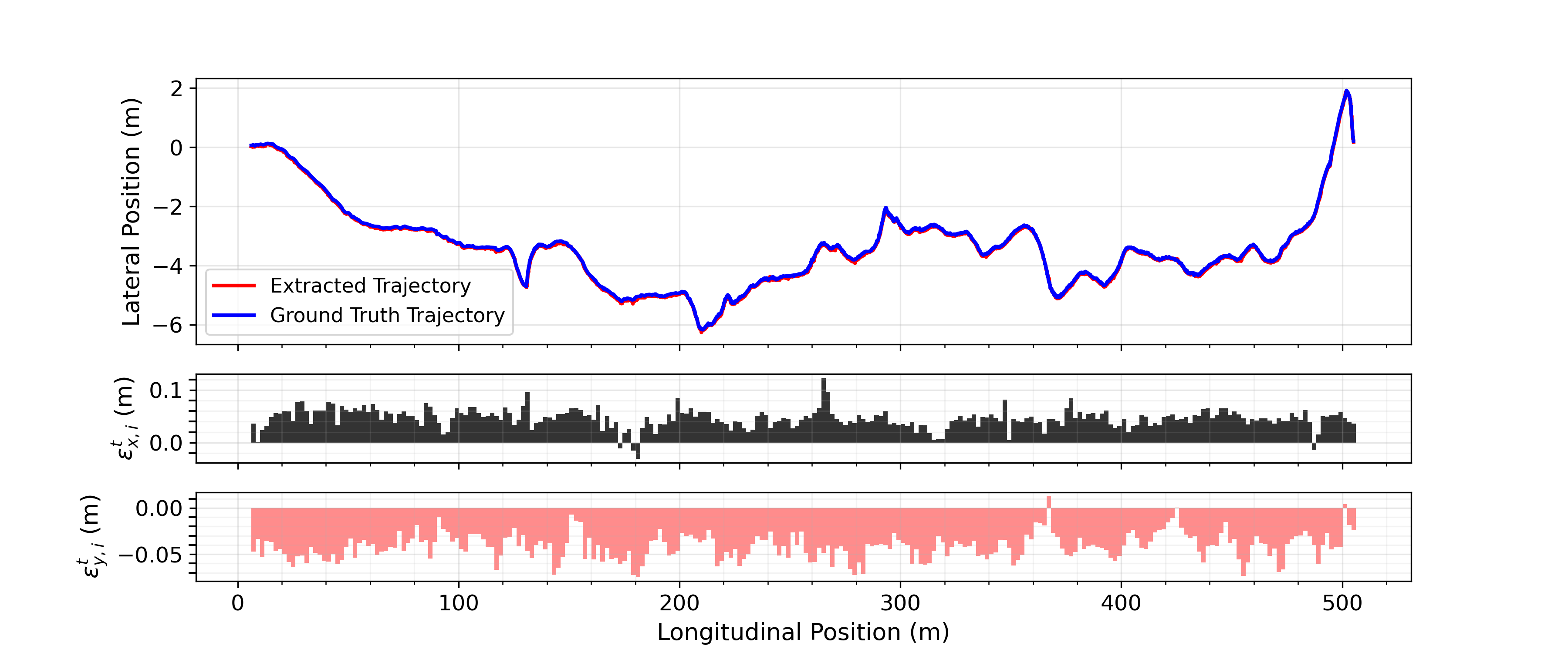}
         \caption{}
         \label{fig:fig_traja}
     \end{subfigure}
     \vspace{0.3cm}
     \begin{subfigure}{\textwidth}
         \centering
         \includegraphics[width=\textwidth]{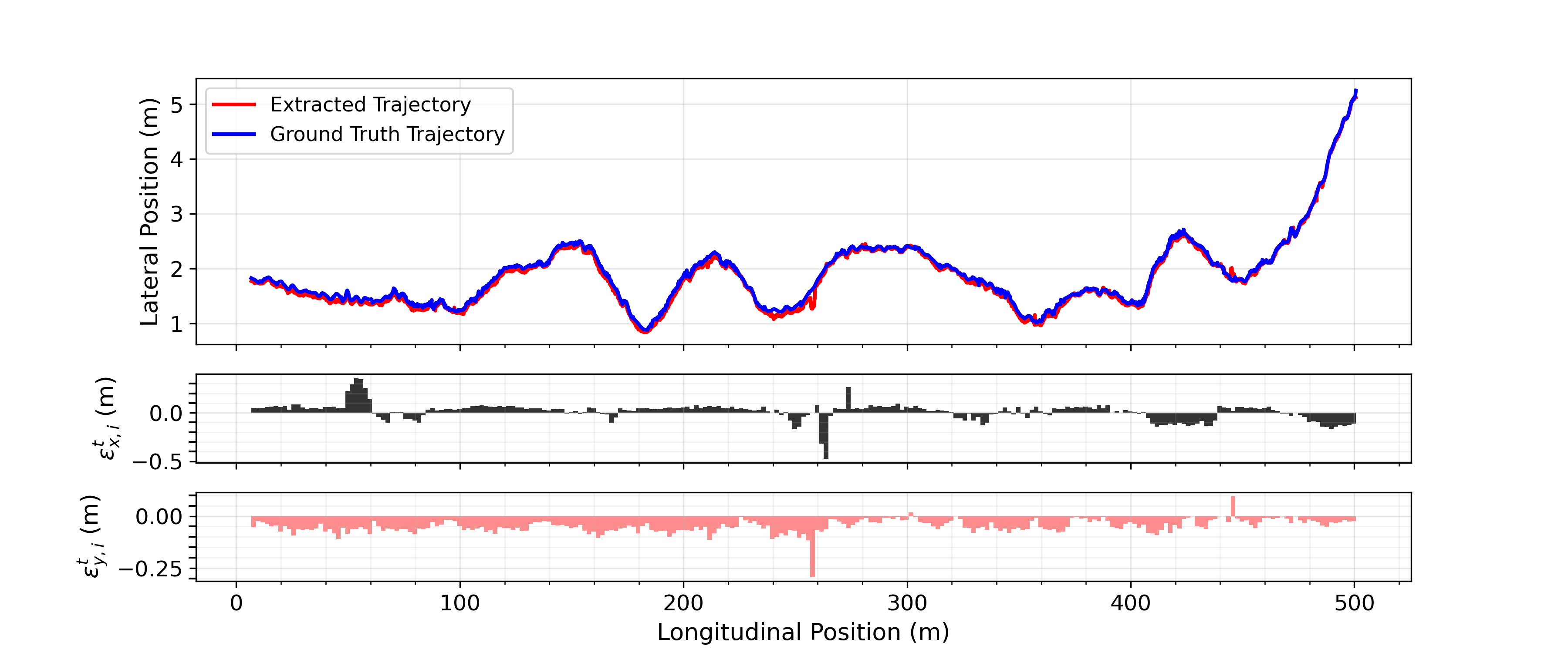}
         \caption{}
         \label{fig:fig_trajb}
     \end{subfigure}
        \caption{Comparison between automatically extracted vehicle trajectories \citep{Kumar2025StopNGo} and manually annotated ground-truth trajectories after coordinate transformation. Panel (a) illustrates a representative trajectory with close agreement between the extracted and ground-truth positions, while panel (b) shows a trajectory corresponding to a higher-error case, contributing to the tail of the error distribution shown in Figure~\ref{fig:fig_4}.}

        \label{fig:fig_traj}
\end{figure}

The error distribution shown in Figure~\ref{fig:fig_4} is strongly concentrated at low values, with only a limited number of larger deviations contributing to the upper tail. Directional decomposition reveals a clear anisotropy in positional accuracy: longitudinal errors dominate, with $\mathrm{MAE}_X = 0.225\,\mathrm{m}$ compared to $\mathrm{MAE}_Y = 0.048\,\mathrm{m}$. This disparity reflects the greater sensitivity of longitudinal position estimates to variations in the detected vehicle extent along the direction of motion, whereas lateral position estimates remain tightly constrained by the roadway geometry. The comparison plots in Figure~\ref{fig:fig_traj} illustrate these observations using two representative vehicle trajectories. Figure~\ref{fig:fig_traja} shows a two-wheeler trajectory for which the longitudinal deviation remains predominantly below $0.080\,\mathrm{m}$, resulting in a mean absolute longitudinal error of $\mathrm{MAE}_X = 0.050\,\mathrm{m}$. The corresponding lateral error is largely confined within $-0.060\,\mathrm{m}$, with a mean absolute lateral error of $\mathrm{MAE}_Y = 0.042\,\mathrm{m}$. These results indicate that, for small and spatially compact vehicles, positional deviations remain consistently bounded in both directions, yielding trajectories that closely follow the manually annotated ground truth.\\

Figure~\ref{fig:fig_trajb} presents a contrasting trajectory for a light commercial vehicle, which contributes to the upper tail of the error distribution shown in Figure~\ref{fig:fig_4}. In this case, intermittent longitudinal deviations exceeding $0.250\,\mathrm{m}$ are observed, leading to a higher mean absolute longitudinal error of $\mathrm{MAE}_X = 0.072\,\mathrm{m}$. In contrast, lateral deviations remain limited, with a mean absolute lateral error of $\mathrm{MAE}_Y = 0.050\,\mathrm{m}$. The observed increase in longitudinal variability for larger vehicles is consistent with their greater physical extent along the direction of motion, which amplifies small geometric offsets in center-position estimation. Importantly, these deviations remain bounded and localized, and lateral accuracy is preserved across vehicle classes. Overall, the magnitude and directional structure of the observed errors are consistent with vehicle geometry and the spatial scales relevant to traffic flow analysis. The positional deviations remain small relative to typical vehicle dimensions and inter-vehicle spacings observed in urban traffic streams. Consequently, the extracted trajectories provide sufficient spatial fidelity for macroscopic analyses, including the estimation of density, fundamental relationships, and congestion propagation, as well as for microscopic investigations of vehicle kinematics and interaction behavior. The validation results therefore establish that the dataset constitutes a robust empirical basis for the analyses presented in the subsequent sections.

\subsection{Trajectory Smoothing}
\begin{figure}
    \centering
    \begin{subfigure}{0.49\linewidth}
        \centering
        \includegraphics[width=\linewidth]{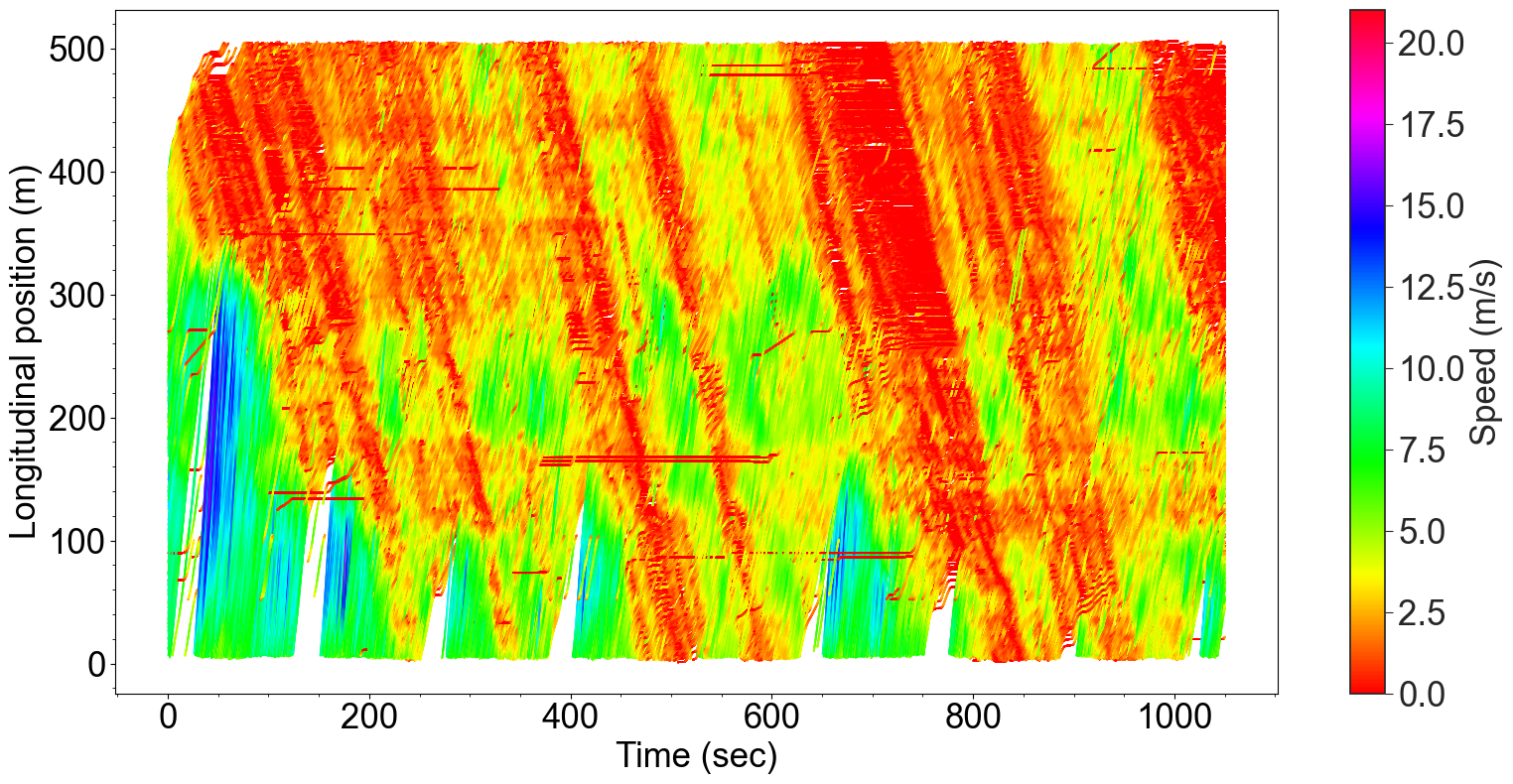}
        \caption{Flight 1, from 09:31–09:49.}
        \label{fig:fig_Traj_dataa}
    \end{subfigure}\hfill
    \begin{subfigure}{0.49\linewidth}
        \centering
        \includegraphics[width=\linewidth]{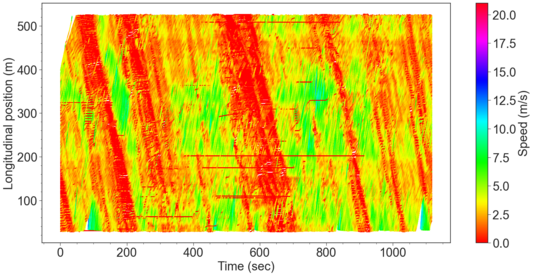}
        \caption{Flight 2, from 10:01–10:20.}
        \label{fig:fig_Traj_datab}
    \end{subfigure}
    \caption{Time–space representation of vehicle longitudinal trajectories in disordered traffic, illustrating the formation and dissipation of stop-and-go congestion waves.}
    \label{fig:Traj_data}
\end{figure}

\noindent Vehicle trajectory data obtained from video-based sensing systems are subject to measurement uncertainties arising from noise. Although these uncertainties may be small at the positional level, their influence becomes pronounced when temporal differentiation is used to compute kinematic quantities such as speed and acceleration, often resulting in values that are inconsistent with physically realistic vehicle motion. Several empirical investigations have reported that accelerations derived from unsmoothed trajectory data frequently exceed plausible bounds under normal driving conditions, highlighting the necessity of preprocessing \citep{Venkatesan2015Traj, Montanino2015Traj, Kashyap2023Traj, RAJPUT2026Traj}. To address this issue, the present study applies a symmetric exponential moving average (sEMA) filter to the extracted vehicle trajectories \citep{Thiemann2008Traj, Kashyap2023Traj, RAJPUT2026Traj}. The smoothing procedure is applied uniformly across all trajectories. Vehicle speeds and accelerations in both longitudinal and lateral directions are subsequently estimated using central difference approximations applied to the smoothed position data. Figure~\ref{fig:Traj_data} presents representative time-space trajectories of vehicle motion along the study section, revealing sustained congested traffic conditions punctuated by brief episodes of higher longitudinal speeds prior to congestion onset. The continuous coverage over a road length of approximately $550\,\mathrm{m}$ enables direct observation of congestion formation and upstream propagation within a lane-free traffic stream. The trajectories exhibit repeated stop-and-go patterns characterized by localized speed reductions that evolve into coherent congestion waves, highlighting the spatiotemporal structure of traffic instabilities under disordered conditions. In addition, several long horizontal trajectory segments indicate some vehicles remaining stationary for extended periods. These trajectories most likely correspond to vehicles that temporarily stopped along the roadside, such as for parking or passenger pick-up/drop-off, before re-entering the traffic stream. Such continuous trajectory coverage over an extended spatial domain allows vehicle motion to be examined directly in the time-space plane, forming a consistent empirical basis for subsequent macroscopic and microscopic analyses without reliance on point-based measurements or spatial aggregation at fixed detector locations.

\section{Disordered Traffic Flow Dynamics}
\noindent
Disordered traffic flow is characterized by weak lane discipline and a heterogeneous mix of vehicle types, resulting in interaction patterns that extend beyond predominantly longitudinal movement. In such environments, traffic dynamics emerge from the coupled longitudinal and lateral movements of vehicles rendering classical lane-based traffic flow theories insufficient \citep{Azam2022Disordered}. Addressing traffic flow under these conditions, therefore, requires methodologies that explicitly account for two-dimensional interactions and the spatiotemporal variability introduced by traffic heterogeneity \citep{Chakroborty2019Disorderly, Agrawal2023TwoD}. This section outlines a systematic approach for examining the dynamics of disordered traffic flow at both macroscopic and microscopic levels. The macroscopic analysis provides a two-dimensional representation of fundamental traffic relationships and an investigation into the propagation velocity of congestion in lane-free scenarios. The following microscopic analysis attempts to explain these macroscopic descriptions by examining vehicle-level attributes and dynamic behavior within various traffic flow states. Collectively, these components form a coherent methodological framework for studying traffic flow dynamics in disordered environments.\par

\subsection{Macroscopic Dynamics}
\noindent Macroscopic traffic dynamics are examined by representing the traffic stream as a two-dimensional continuum, where aggregate behavior arises from collective vehicle motion unconstrained by lane boundaries. In disordered traffic, lateral movements contribute directly to traffic evolution, motivating macroscopic descriptions that extend beyond one-dimensional formulations. Based on this perspective, a two-dimensional macroscopic representation is adopted and used to examine fundamental relationships among traffic variables. In addition, congestion propagation characteristics are analyzed, as they are central to understanding queue formation, improving model calibration, and supporting traffic management and control. In the present study, congestion propagation speed is estimated using one-dimensional macroscopic variables through a cross-correlation-based approach.

\subsubsection{Two-Dimensional Macroscopic Representation and Fundamental Diagram}
\begin{figure}
    \centering
    \includegraphics[width=0.7\linewidth]{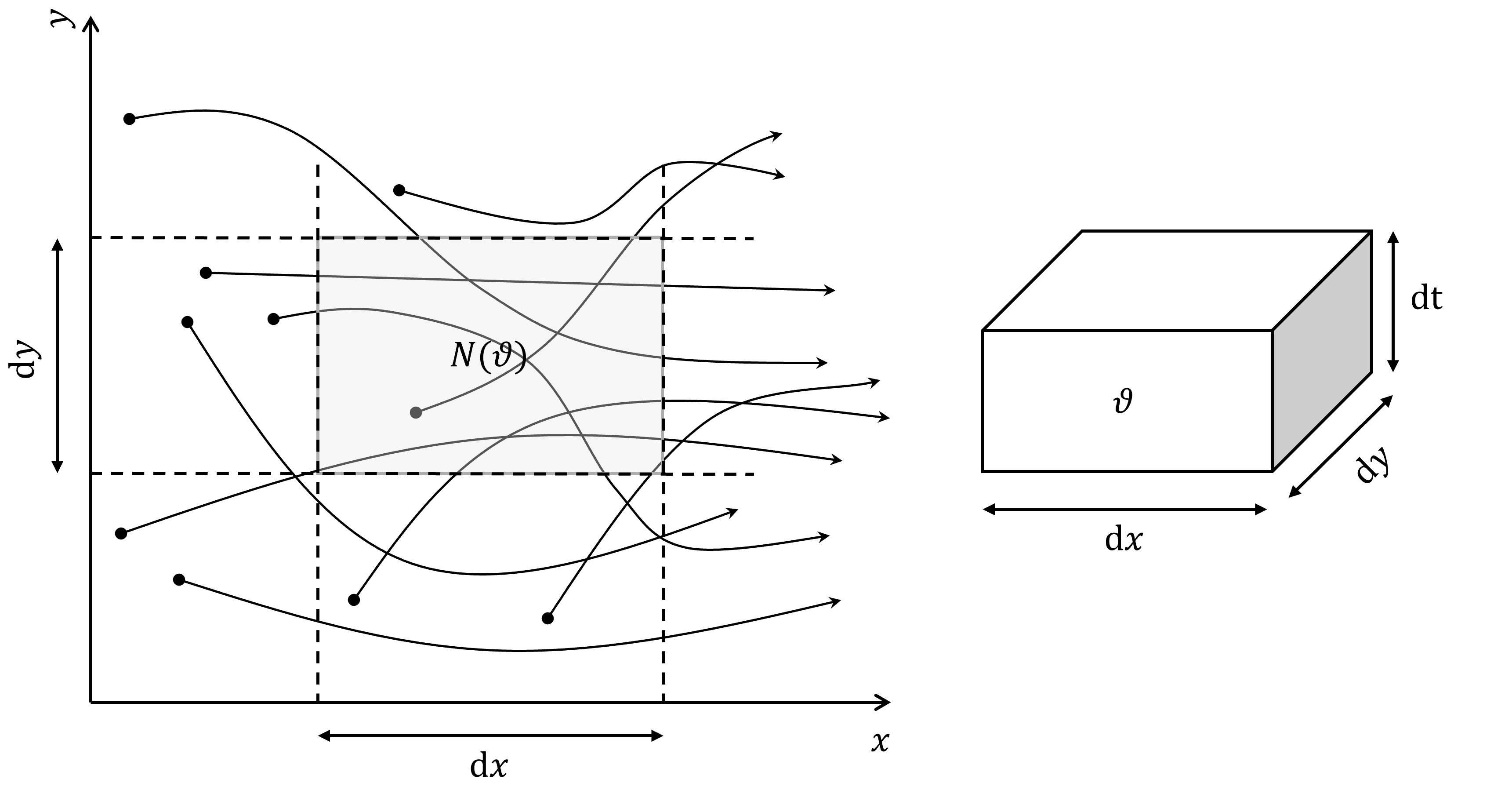}
    \caption{Extended Edie's method for estimating two-dimensional macroscopic traffic variables}
    \label{fig:extended_edie}
\end{figure}

\noindent
To quantify aggregate traffic behavior under lane-free conditions, we adopt a two-dimensional extension of Edie’s space-time averaging framework, which provides a consistent link between microscopic vehicle trajectories and macroscopic traffic variables. Unlike lane-based traffic, vehicle trajectories in disordered traffic are not constrained to align with the road centerline and frequently exhibit lateral drift, overtaking, and local reorganization. A macroscopic formulation must therefore account for vehicle motion in both longitudinal and lateral directions. Macroscopic variables are defined over a finite space-time control volume
\[
\vartheta = \dd x \times \dd y \times \dd t,
\]
where $\dd x$ and $\dd y$ denote the longitudinal and lateral extents of the spatial region, respectively, and $\dd t$ denotes the observation interval (Figure~\ref{fig:extended_edie}). Let $n = 1,2,\ldots,N$ 
index the vehicles whose trajectories intersect this control volume during $\dd t$. Each vehicle contributes (i) a residence time $T_n$, defined as the total time spent by the vehicle $n$ within the spatial region $\dd x \times \dd y$ during the observation time interval $\dd t$, and (ii) a displacement vector $\vec{d}_n=\langle \bar{X}_n, \bar{Y}_n \rangle$, where $\bar{X}_n$ and $\bar{Y}_n$ represent the net longitudinal and lateral displacements of vehicle $n$ within $\vartheta$, respectively.\\

\noindent\textbf{Density Definition:} Traffic density $\rho$ is defined as the ratio of the total time spent by all vehicles within a finite space-time domain to the volume of that space-time domain. It is given by
\begin{equation}
\rho = \frac{\sum_{n=1}^{N} T_n}{\dd x\,\dd y\,\dd t},
\end{equation}
This definition represents the vehicle concentration over the spatial region and is expressed in units of veh./m\textsuperscript{2}.\\

\noindent\textbf{flow density Definition:}
flow density is defined as a vector quantity consisting of longitudinal and lateral components. The longitudinal and lateral flow-densities are defined as the ratios of the total longitudinal and total lateral displacements of all vehicles within a finite space–time domain to the volume of that space–time domain, respectively.$\langle Q_x, Q_y \rangle$, are expressed as
\begin{equation}
\vec{Q} = \langle Q_x, Q_y \rangle
=
\left\langle
\frac{\sum_{n=1}^{N} \bar{X}_n}{\dd x\,\dd y\,\dd t},
\frac{\sum_{n=1}^{N} \bar{Y}_n}{\dd x\,\dd y\,\dd t}
\right\rangle,
\end{equation}
The resulting components $Q_x$ and $Q_y$ are expressed in units of veh./(m·s).\\

\noindent\textbf{Velocity Definition:}
The velocity field is defined as a vector quantity that represents the total displacement of all vehicles within a finite space-time volume divided by the total residence time spent by all vehicles in that domain. The longitudinal and lateral components of velocity are obtained by taking the ratio of the total longitudinal and lateral displacements of vehicles to the total time they spend within the spatial region during the observation interval. The components $V_x$ and $V_y$ are obtained as
\begin{equation}
\vec{V} = \langle V_x, V_y \rangle
=
\left\langle
\frac{\sum_{n=1}^{N} \bar{X}_n}{\sum_{n=1}^{N} T_n},
\frac{\sum_{n=1}^{N} \bar{Y}_n}{\sum_{n=1}^{N} T_n}
\right\rangle,
\end{equation}
which is equivalent to the conventional macroscopic kinematic relation $\vec{V}=\vec{Q}/\rho$. The resulting components represent the longitudinal and lateral local speeds and are expressed in units of m/s. Together, the density $\rho$, flow density $\vec{Q}$, and velocity field $\vec{V}$ form a consistent two-dimensional macroscopic representation of traffic flow under disordered conditions. Fundamental relationships are evaluated directly between these variables without imposing lane-based constraints, allowing lateral vehicle redistribution to be explicitly reflected in aggregate traffic states. The resulting two-dimensional fundamental diagrams provide a macroscopic characterization of traffic behavior consistent with lane-free operation.\\

\begin{figure}
    \centering

    \begin{subfigure}[t]{0.48\linewidth}
        \centering
        \includegraphics[width=\linewidth]{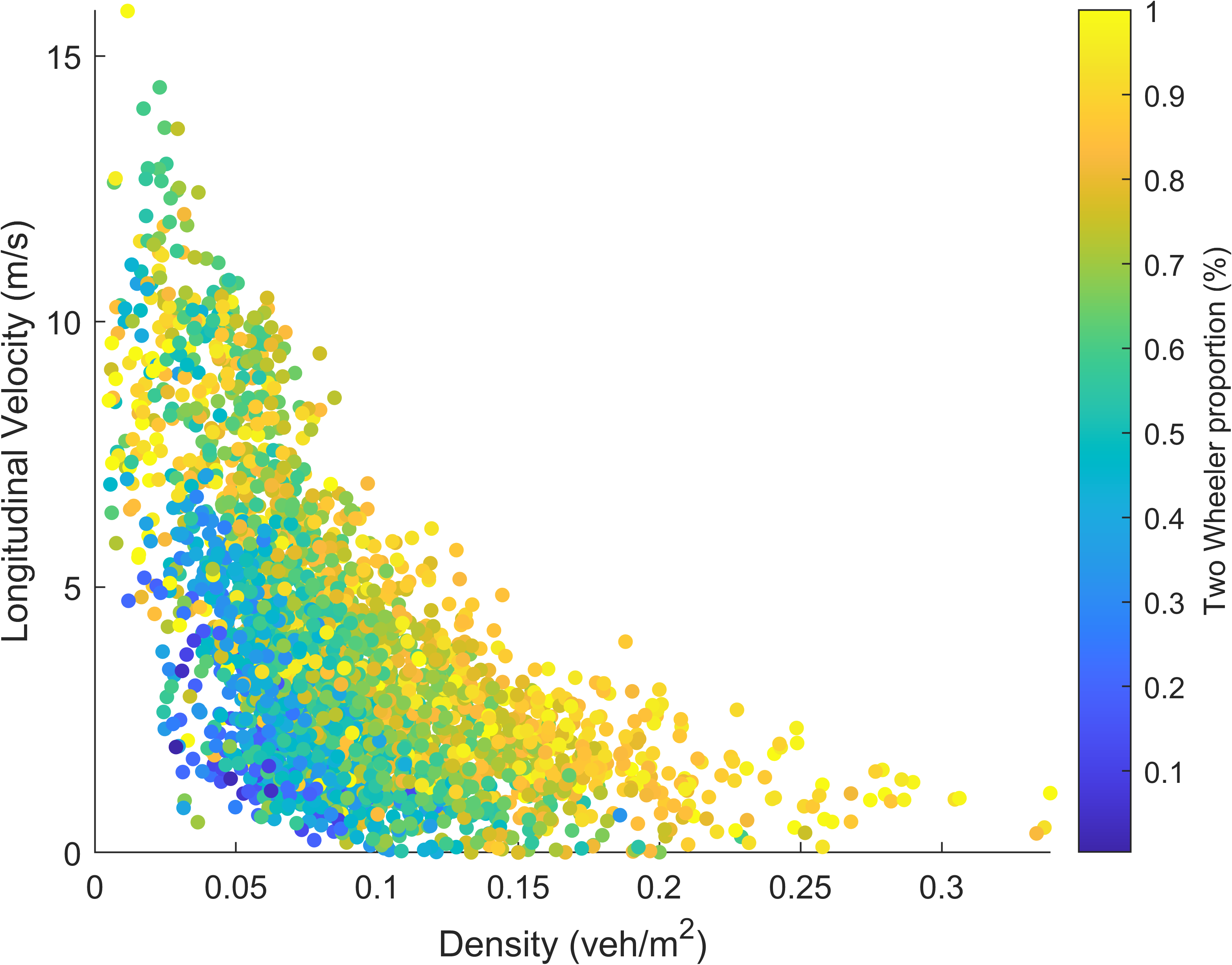}
        \caption{Longitudinal velocity vs density relationship.}
        \label{fig:fig_7a}
    \end{subfigure}\hfill
    \begin{subfigure}[t]{0.48\linewidth}
        \centering
        \includegraphics[width=\linewidth]{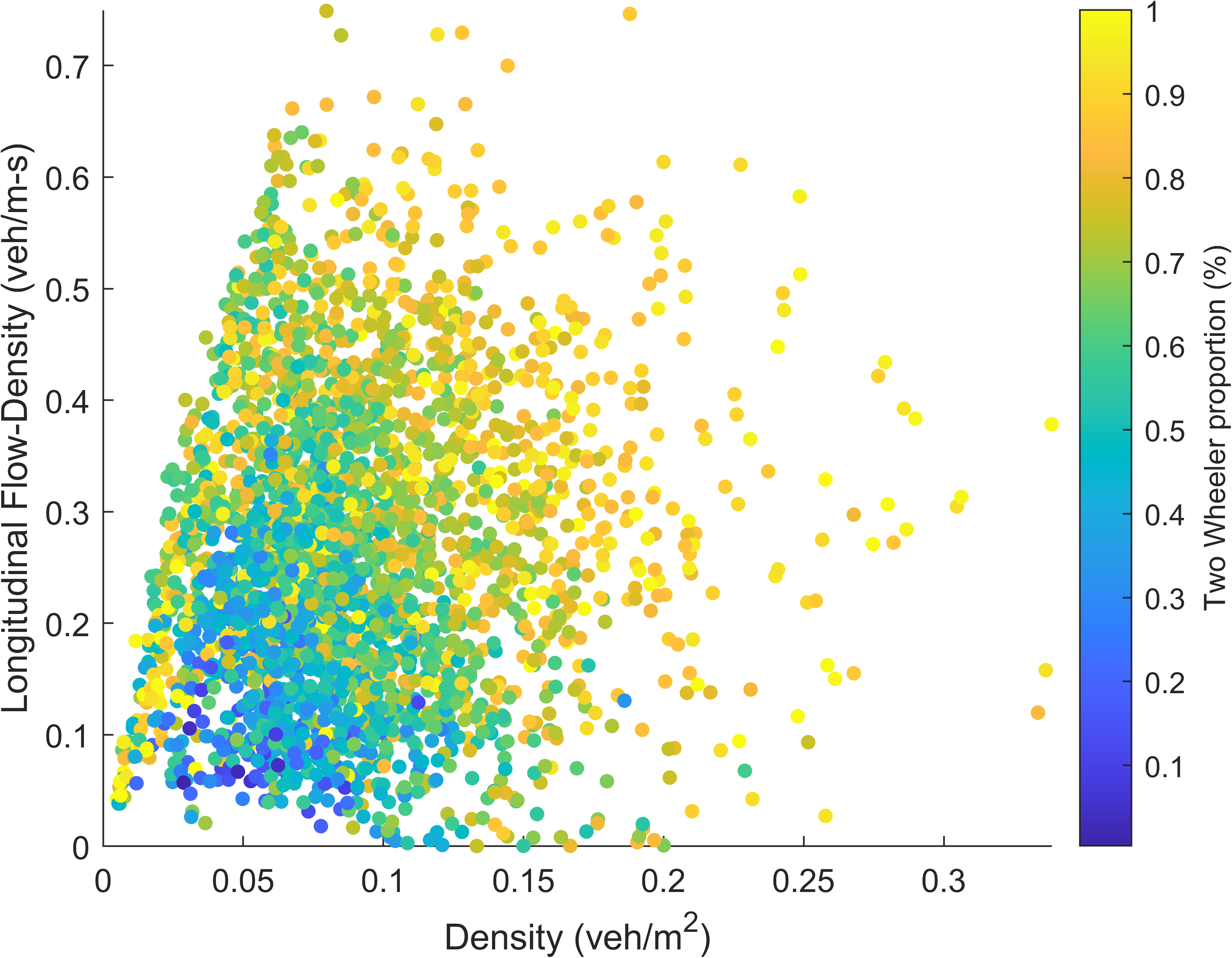}
        \caption{Longitudinal flow density vs density relationship.}
        \label{fig:fig_7b}
    \end{subfigure}

    \vspace{3mm}

    \begin{subfigure}[t]{0.48\linewidth}
        \centering
        \includegraphics[width=\linewidth]{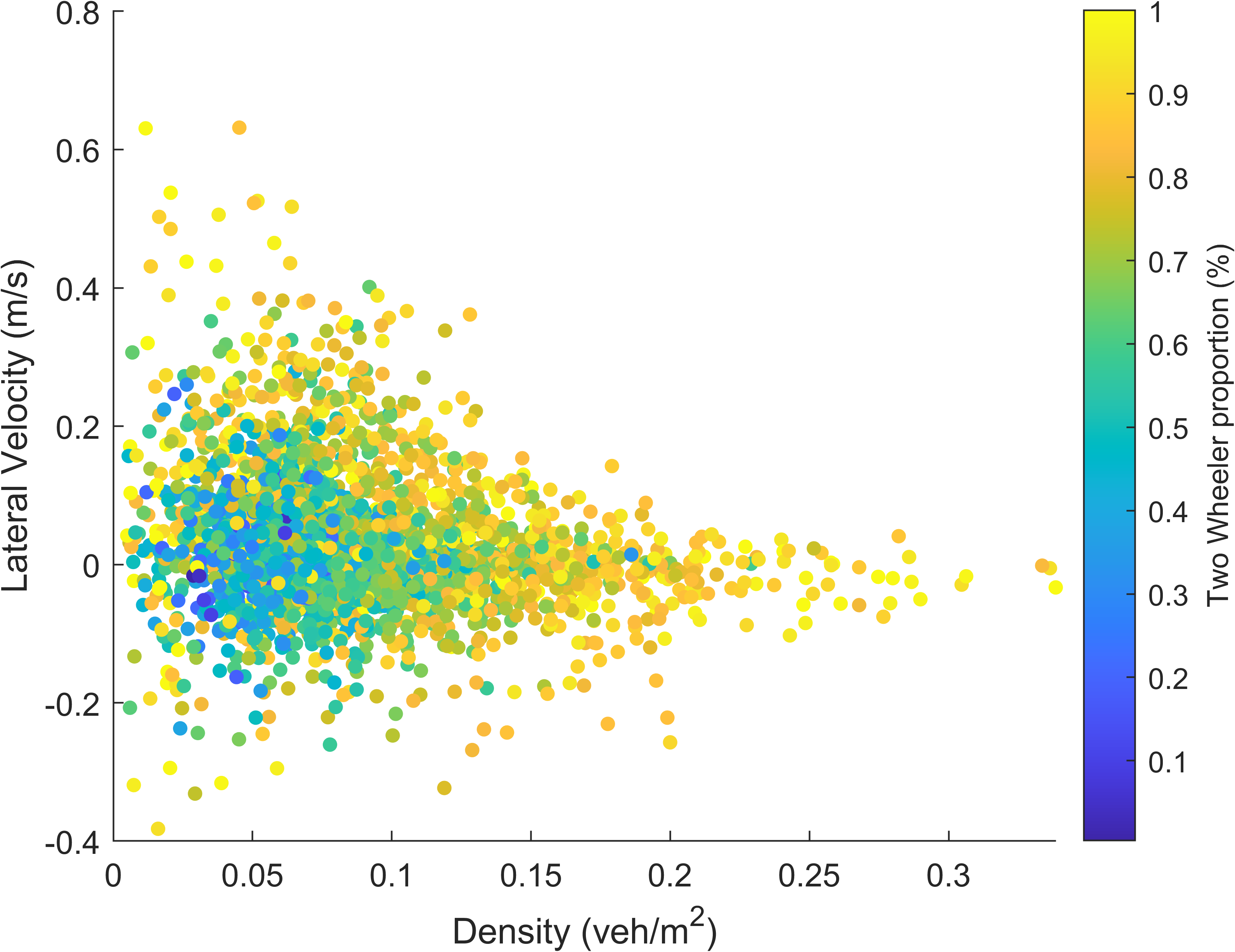}
        \caption{Lateral velocity vs density relationship.}
        \label{fig:fig_7c}
    \end{subfigure}\hfill
    \begin{subfigure}[t]{0.48\linewidth}
        \centering
        \includegraphics[width=\linewidth]{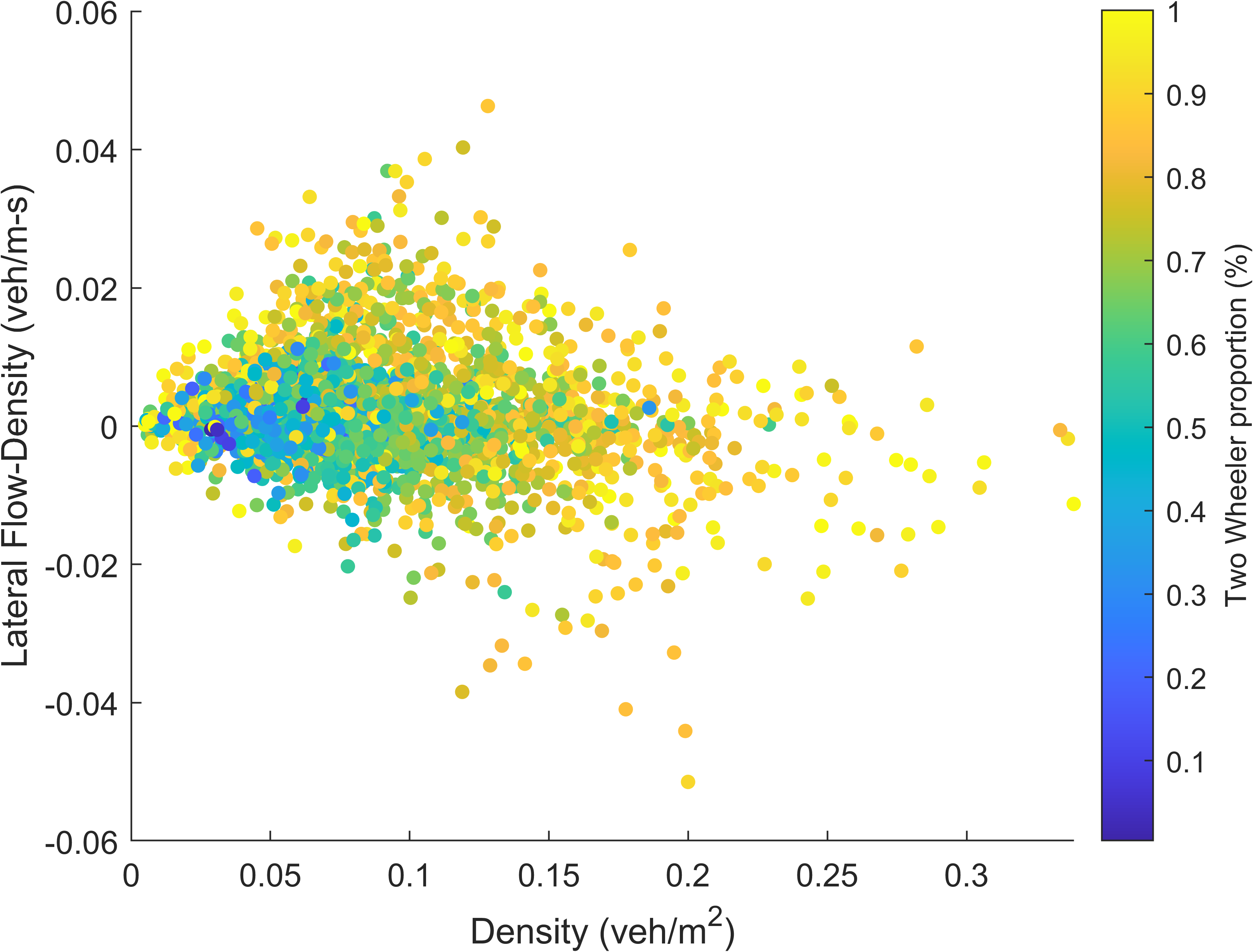}
        \caption{Lateral flow density vs density relationship.}
        \label{fig:fig_7d}
    \end{subfigure}

    \caption{Two-dimensional fundamental relationships obtained using space--time aggregation from Flight 1 dataset.}
    \label{fig:fig_7}
\end{figure}

\begin{figure}
    \centering

    \begin{subfigure}[t]{0.48\linewidth}
        \centering
        \includegraphics[width=\linewidth]{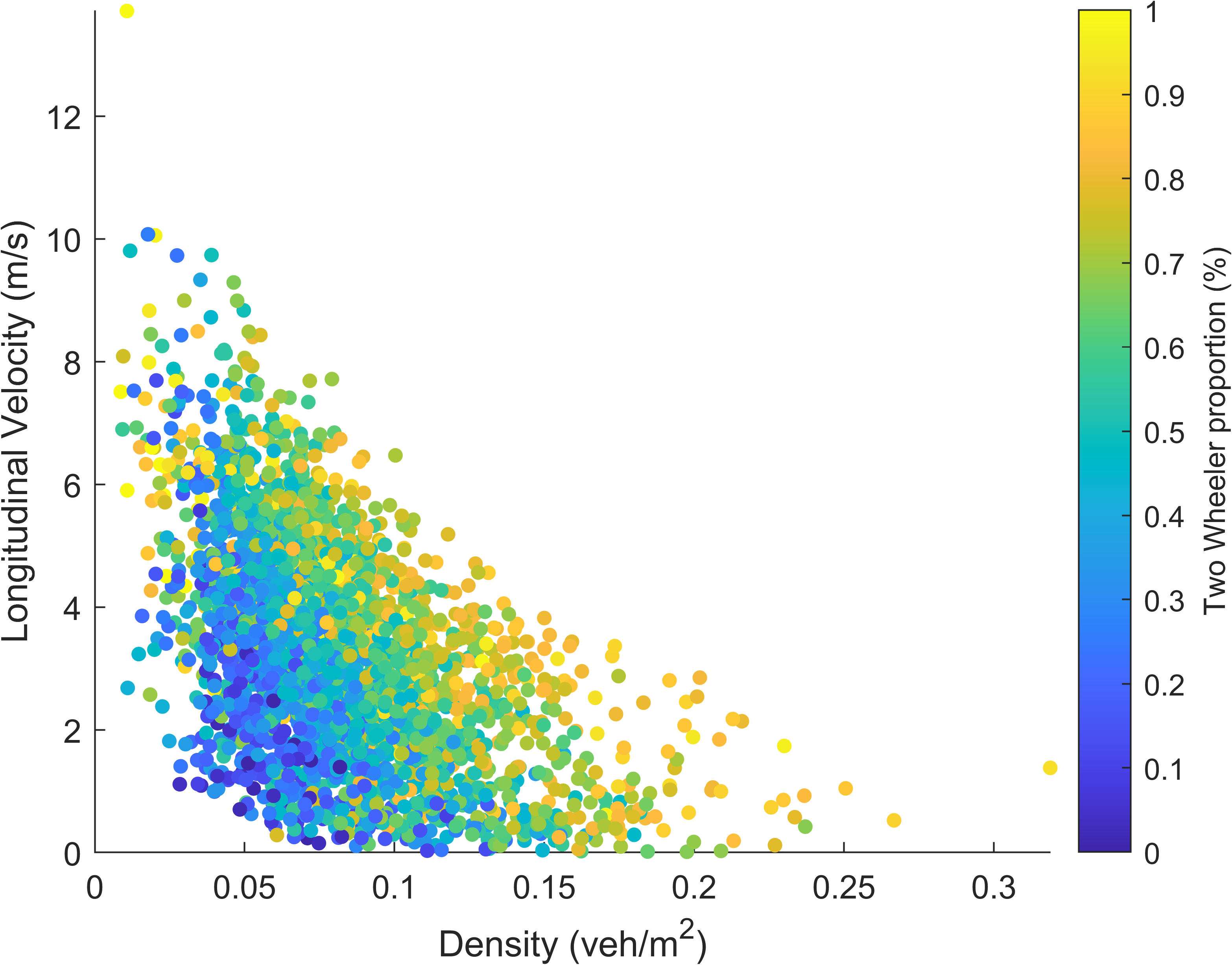}
        \caption{Longitudinal velocity vs density relationship.}
        \label{fig:fig_8a}
    \end{subfigure}\hfill
    \begin{subfigure}[t]{0.48\linewidth}
        \centering
        \includegraphics[width=\linewidth]{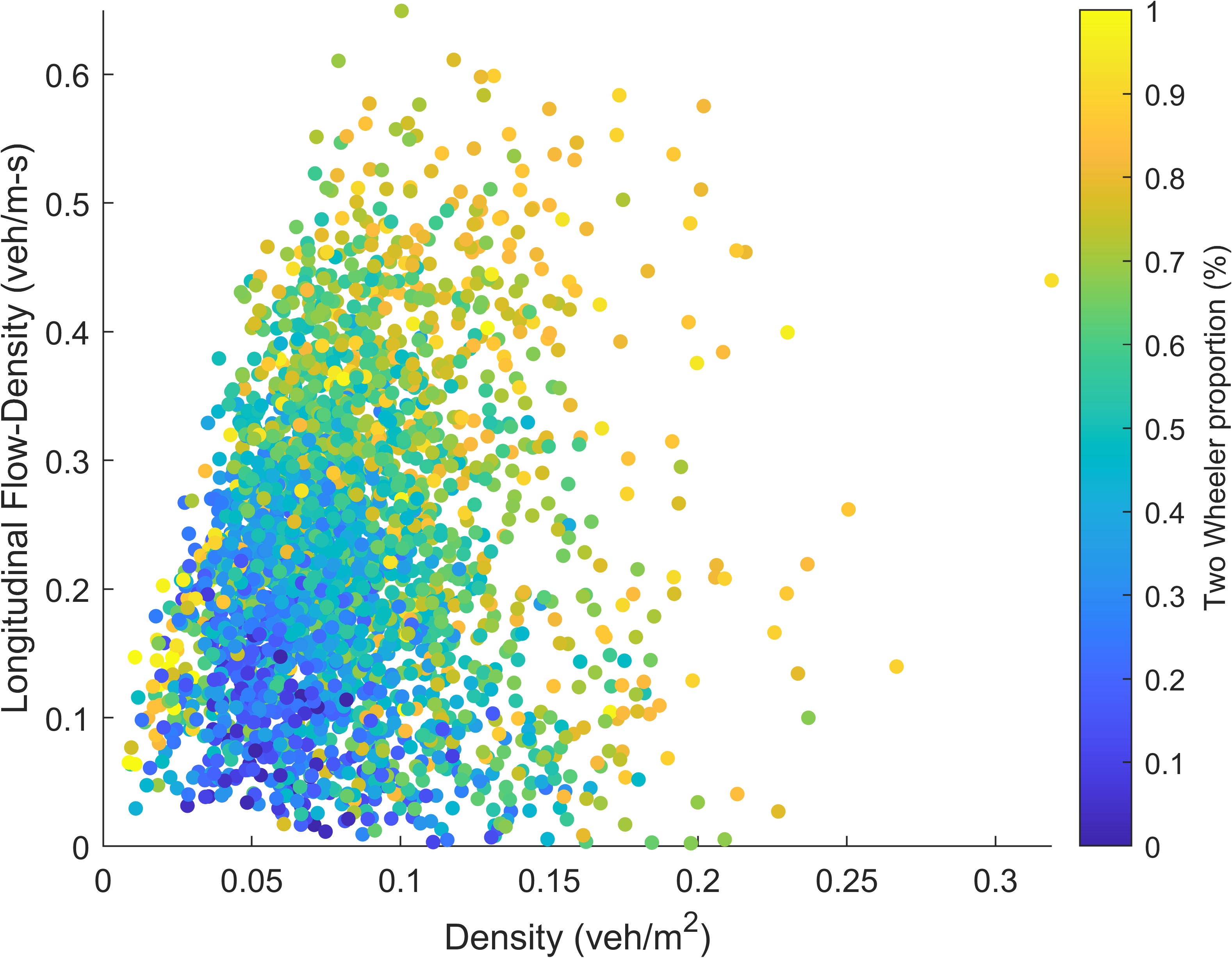}
        \caption{Longitudinal flow density vs density relationship.}
        \label{fig:fig_8b}
    \end{subfigure}

    \vspace{3mm}

    \begin{subfigure}[t]{0.48\linewidth}
        \centering
        \includegraphics[width=\linewidth]{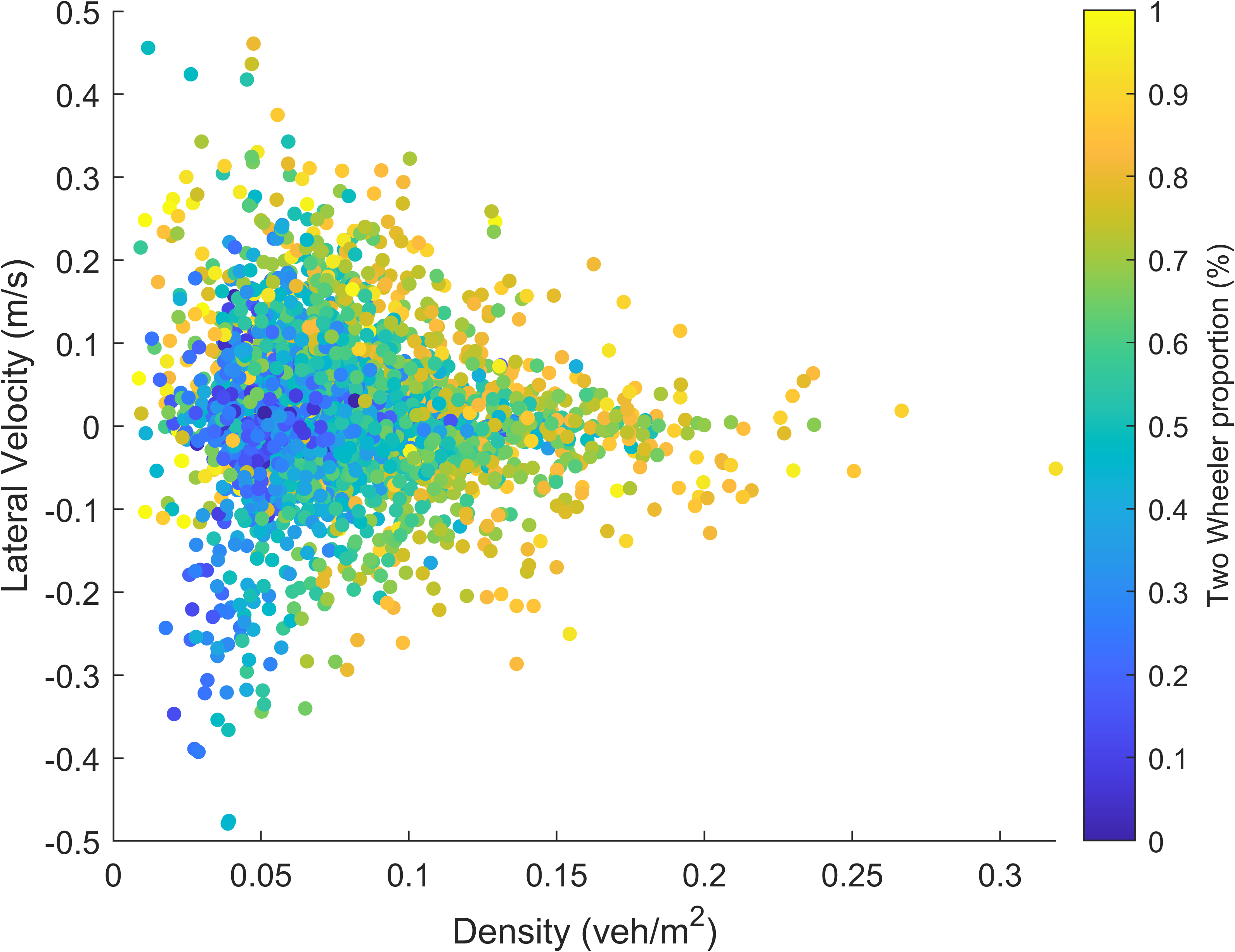}
        \caption{Lateral velocity vs density relationship.}
        \label{fig:fig_8c}
    \end{subfigure}\hfill
    \begin{subfigure}[t]{0.48\linewidth}
        \centering
        \includegraphics[width=\linewidth]{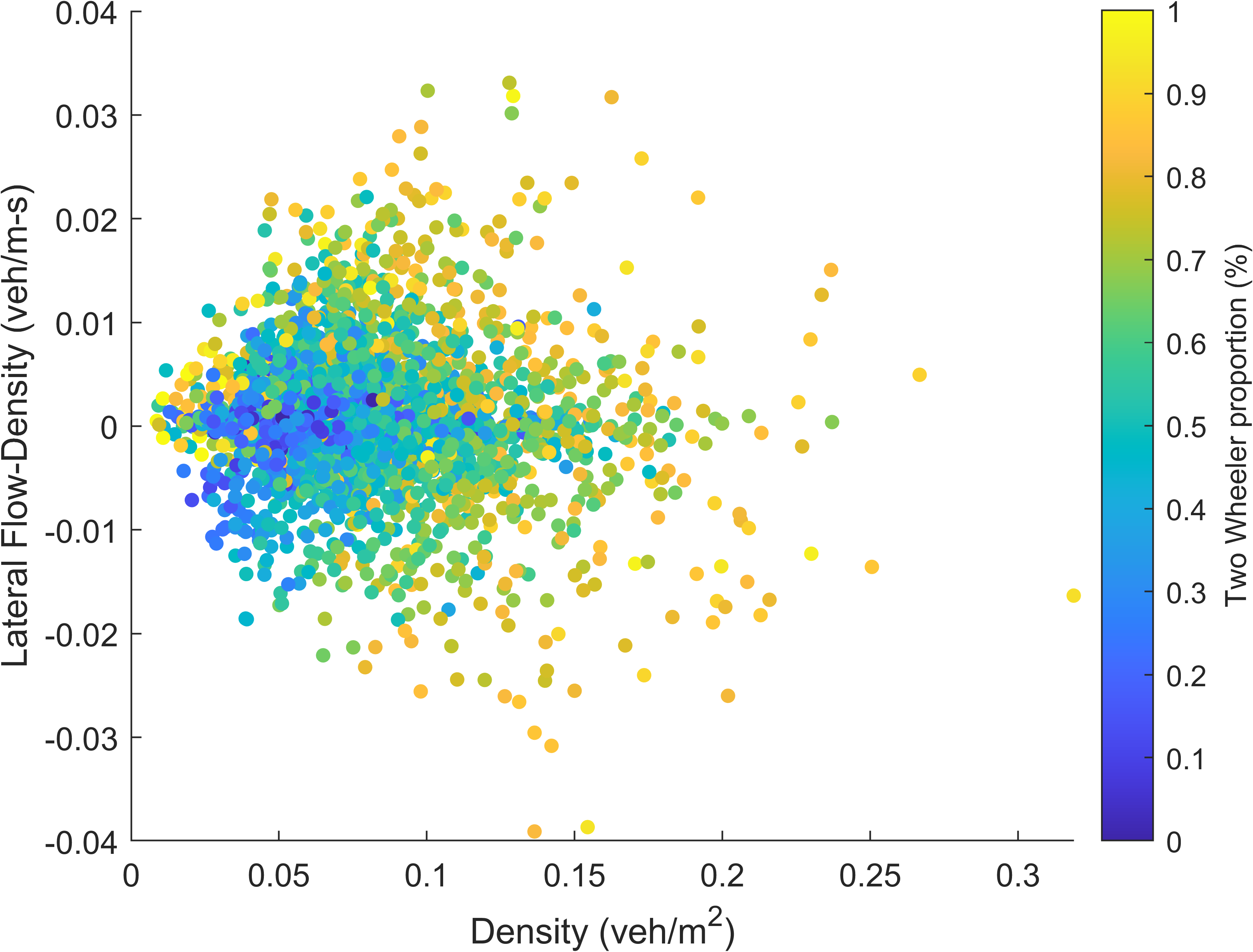}
        \caption{Lateral flow density vs density relationship.}
        \label{fig:fig_8d}
    \end{subfigure}

    \caption{Two-dimensional fundamental relationships obtained using space-time aggregation from Flight 2 dataset.}
    \label{fig:fig_8}
\end{figure}

\noindent A space-time discretization with spatial resolutions $\dd x = 20\,\mathrm{m}$ and $\dd y = 3\,\mathrm{m}$, and a temporal resolution of $\dd t = 30\,\mathrm{s}$, is adopted for the estimation of two-dimensional traffic variables. This resolution is sufficiently large to support a continuum approximation of vehicle motion, while remaining small enough to preserve the essential dynamics of traffic evolution. The study corridor is partitioned into uniform spatial cells of size $\dd x \times \dd y$, thereby establishing a grid of virtual detectors across the entire road section. Two-dimensional macroscopic variables are estimated at each virtual detector by aggregating vehicle trajectories over successive time intervals of duration $\dd t$. The resulting relationships among velocity, flow density, and density are illustrated in Figures~\ref{fig:fig_7} and~\ref{fig:fig_8}.\\

\noindent Figure~\ref{fig:fig_7} presents the two-dimensional fundamental relationships obtained from the Flight~1 dataset,
with density ranging approximately from 0 to 0.32~veh./m$^2$. Panels (a) and (b) show the longitudinal velocity-density and longitudinal flow density-density relationships, while panels (c) and (d) show the corresponding lateral relationships. The color scale denotes the proportion of two-wheelers within each macroscopic traffic state. In Figure~\ref{fig:fig_7}(a), the longitudinal velocity reaches values as high as approximately 14--16~m/s at very low densities (below about 0.03~veh./m$^2$), indicating near free-flow conditions. As density increases beyond roughly 0.05~veh./m$^2$, longitudinal velocity decreases rapidly, with most observations falling below 5~m/s for densities exceeding 0.10~veh./m$^2$. At higher densities (above about 0.20~veh./m$^2$), longitudinal velocities are predominantly below 2~m/s, reflecting strongly constrained traffic states. The substantial vertical scatter observed at low and intermediate densities indicates the coexistence of multiple traffic states under similar density levels, consistent with non-equilibrium mixed traffic conditions.\\

\noindent The longitudinal flow density-density relationship in Figure~\ref{fig:fig_7}(b) shows flow density values reaching up to approximately 0.7~veh./(m$\cdot$s), with the highest sustained values of $\approx  \unit[0.5]{veh./(m\cdot s)}$  occurring at intermediate densities in the range of 0.05--0.12~veh./m$^2$. Interestingly, the observed maximum flow density values (which can be interpreted as a "capacity density") are about half the values observed in pure pedestrian crowds \cite{Treiber_Pedestrian2015}. Unlike classical one-dimensional fundamental diagrams, no sharply defined capacity point is observed; instead, flow density exhibits significant dispersion across a wide density range. Besides depending on the size of the spatio-temporal aggregation element, this dispersion reflects the influence of vehicle heterogeneity and maneuvering\\

\noindent Figure~\ref{fig:fig_7}(c) shows that lateral velocity remains non-negligible over a broad density range, particularly if the percentage of the agile two wheelers is high. At low densities (below about 0.05~veh./m$^2$), lateral velocities span approximately $-0.3$~m/s to 0.6~m/s, indicating active lateral repositioning and overtaking behavior. As density increases beyond roughly 0.15~veh./m$^2$, the spread in lateral velocity narrows and values cluster closer to zero, suggesting progressive suppression of lateral motion under space-constrained conditions. The lateral flow density-density relationship in Figure~\ref{fig:fig_7}(d) is centered near zero but exhibits measurable dispersion, with lateral flow density values extending up to approximately $\pm 0.05$~veh./(m$\cdot$s). Although the lateral flow density is small relative to the longitudinal flow density, lateral vehicle redistribution remains observable at moderate densities and is progressively suppressed at higher densities due to spatial constraints. Such behavior is a defining feature of lane-free traffic and cannot be captured by one-dimensional macroscopic descriptions. Notice, however, that, unlike the longitudinal flow density, the lateral flow density depends strongly on the spatio-temporal aggregation volume. Still, the observed values indicate that there are local regions where flow moves systematically to the left or right.\\

\noindent Across all panels, traffic states with a higher proportion of two-wheelers are associated with higher observed densities and broader dispersion in both longitudinal and lateral macroscopic quantities. This reflects the compact physical dimensions and high maneuverability of two-wheelers, which enable denser space utilization and sustained motion even under congested conditions. Overall, Figure~\ref{fig:fig_7} demonstrates that macroscopic traffic behavior in mixed, lane-free conditions emerges from the coupled evolution of longitudinal progression and lateral redistribution, reinforcing the need for a two-dimensional macroscopic representation. Figure~\ref{fig:fig_8} exhibits trends consistent with those observed in Figure~\ref{fig:fig_7}, indicating that the qualitative structure of the two-dimensional fundamental relationships is preserved.\\

\begin{figure}
    \centering

    \begin{subfigure}[t]{0.48\linewidth}
        \centering
        \includegraphics[width=\linewidth]{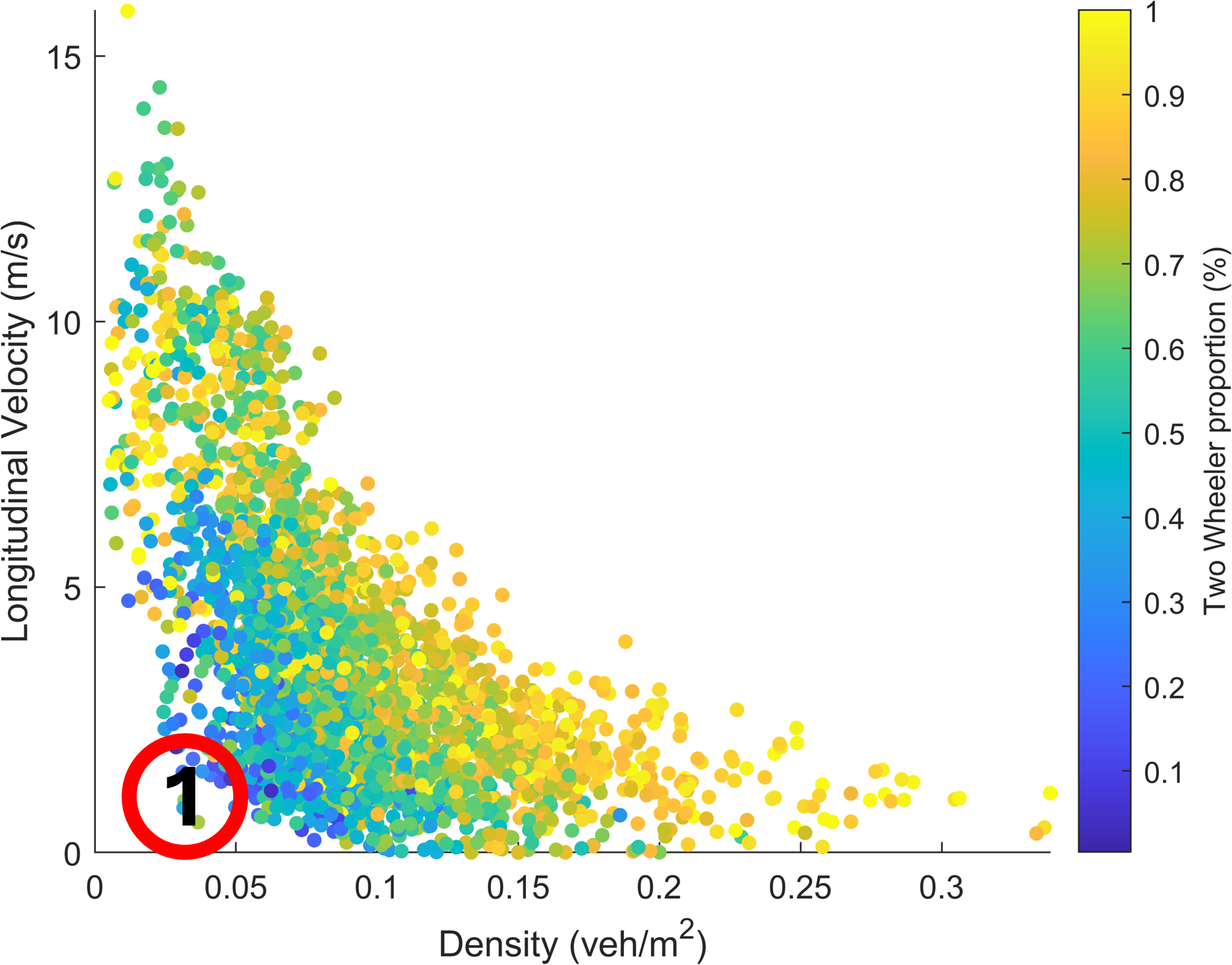}
        \caption{Longitudinal velocity--density diagram highlighting selected regions of interest.}
        \label{fig:fig_9a}
    \end{subfigure}\hfill
    \begin{subfigure}[t]{0.48\linewidth}
        \centering
        \includegraphics[width=\linewidth]{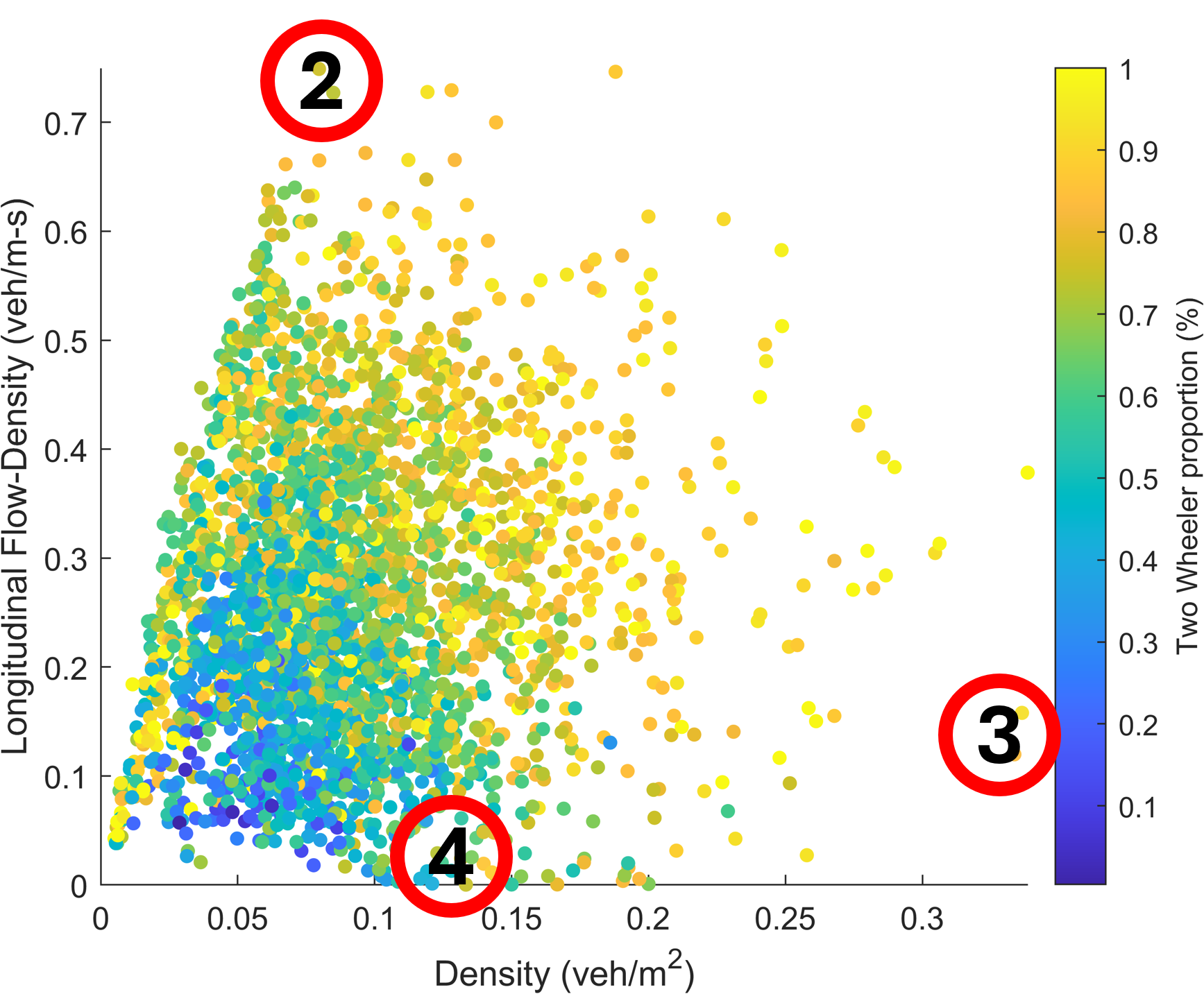}
        \caption{Longitudinal flow density--density diagram highlighting selected regions of interest.}
        \label{fig:fig_9b}
    \end{subfigure}

    \vspace{3mm}

    \begin{subfigure}[t]{0.48\linewidth}
        \centering
        \includegraphics[width=\linewidth]{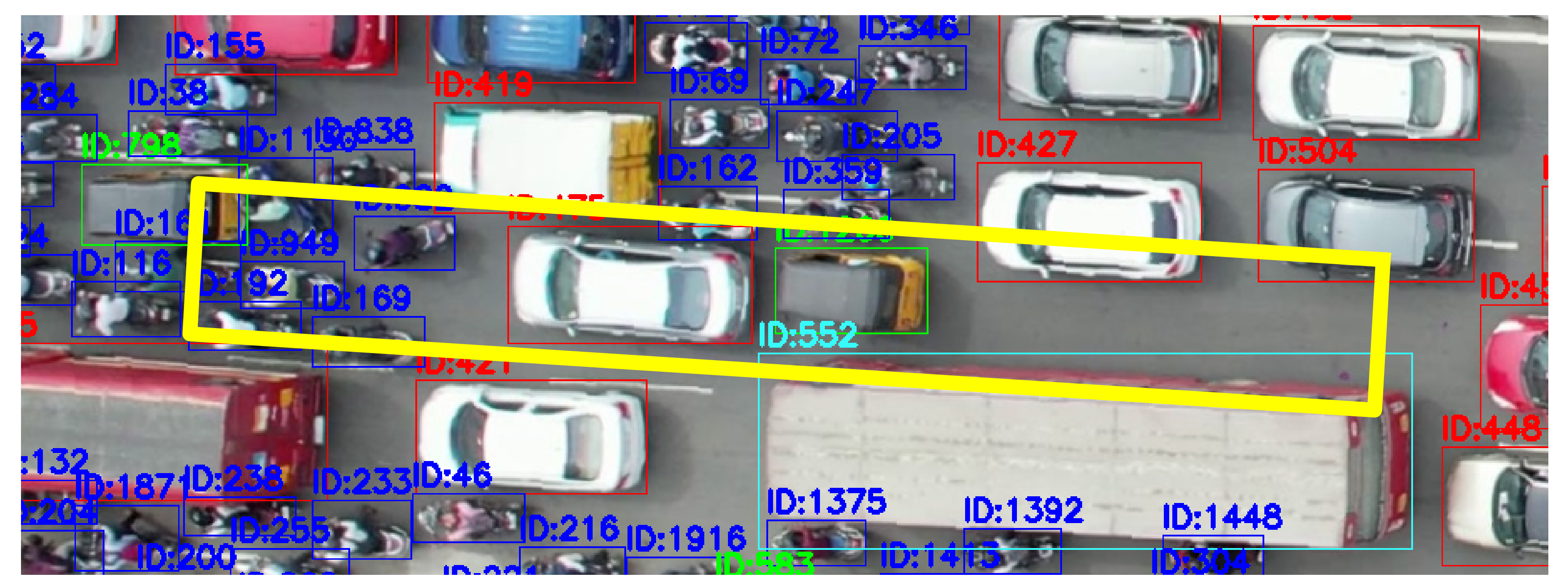}
        \caption{Region~1: Low-speed traffic state observed at low density.}
        \label{fig:fig_9c}
    \end{subfigure}\hfill
    \begin{subfigure}[t]{0.48\linewidth}
        \centering
        \includegraphics[width=\linewidth]{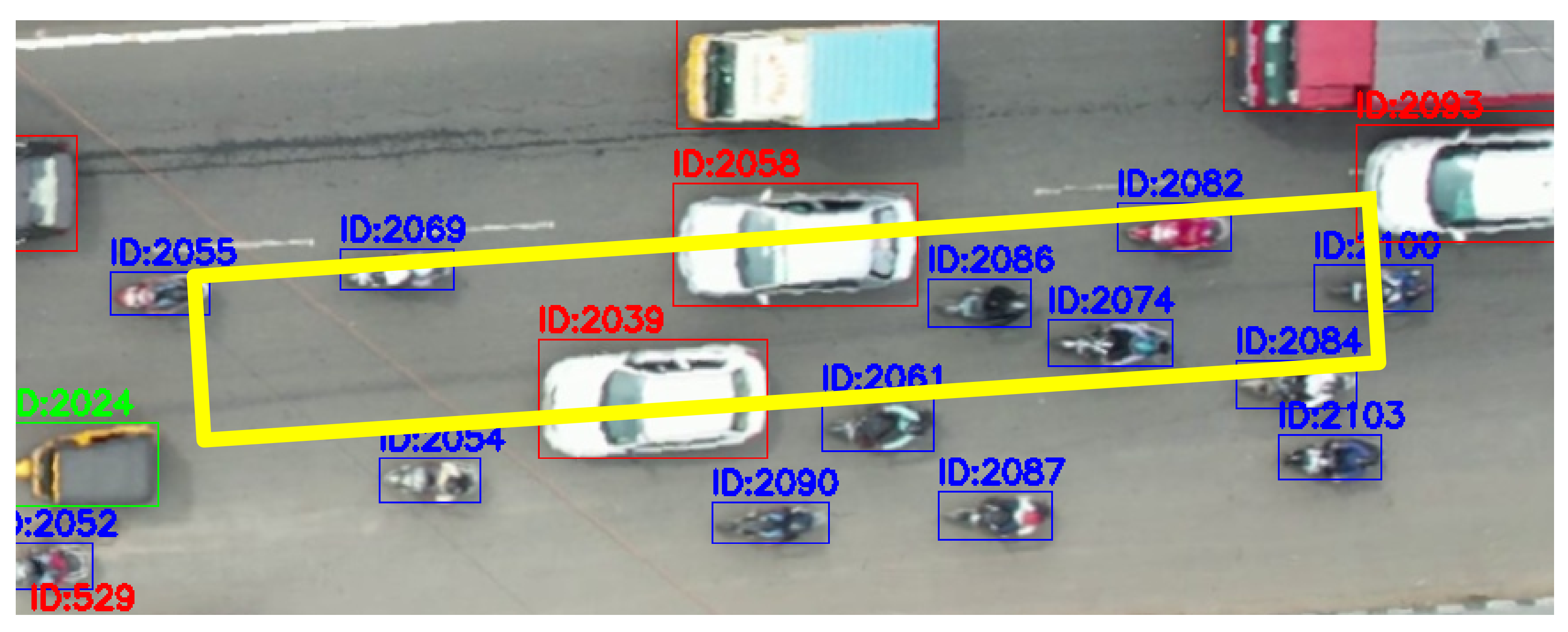}
        \caption{Region~2: Near-capacity traffic state.}
        \label{fig:fig_9d}
    \end{subfigure}

    \vspace{3mm}

    \begin{subfigure}[t]{0.48\linewidth}
        \centering
        \includegraphics[width=\linewidth]{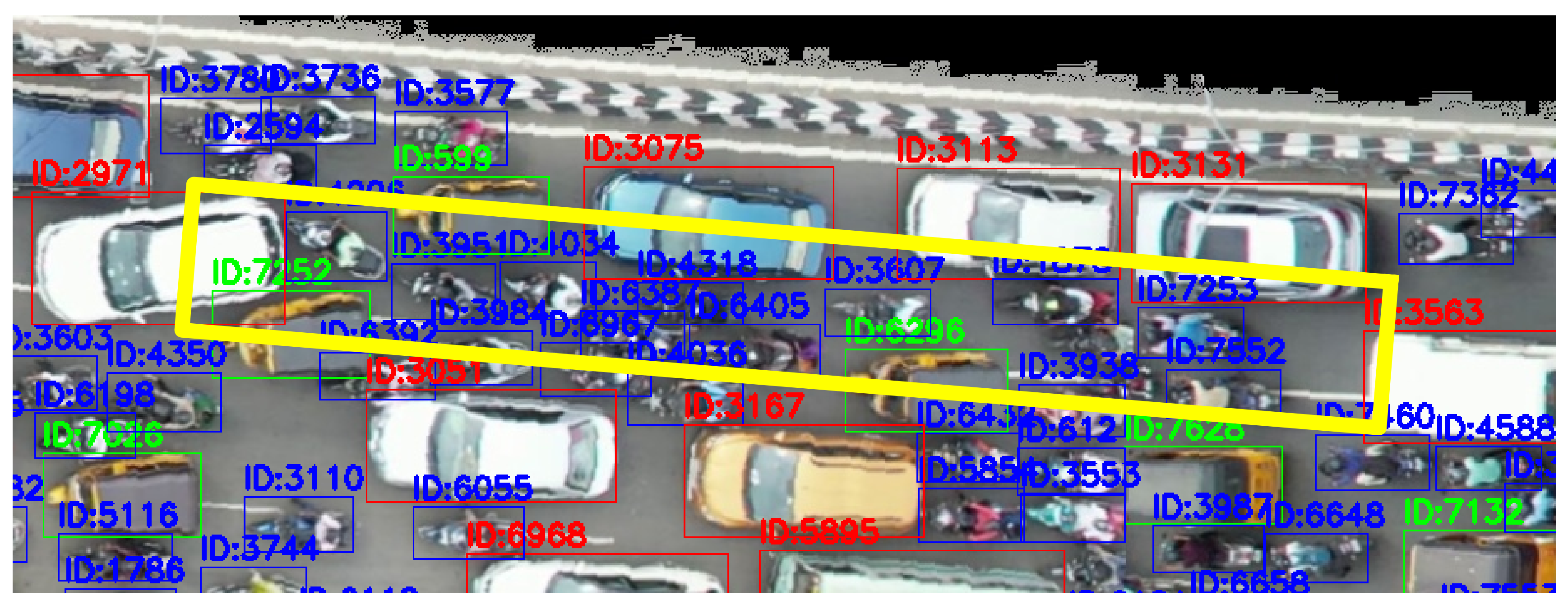}
        \caption{Region~3: High-density traffic state with sustained longitudinal flow density.}
        \label{fig:fig_9e}
    \end{subfigure}\hfill
    \begin{subfigure}[t]{0.48\linewidth}
        \centering
        \includegraphics[width=\linewidth]{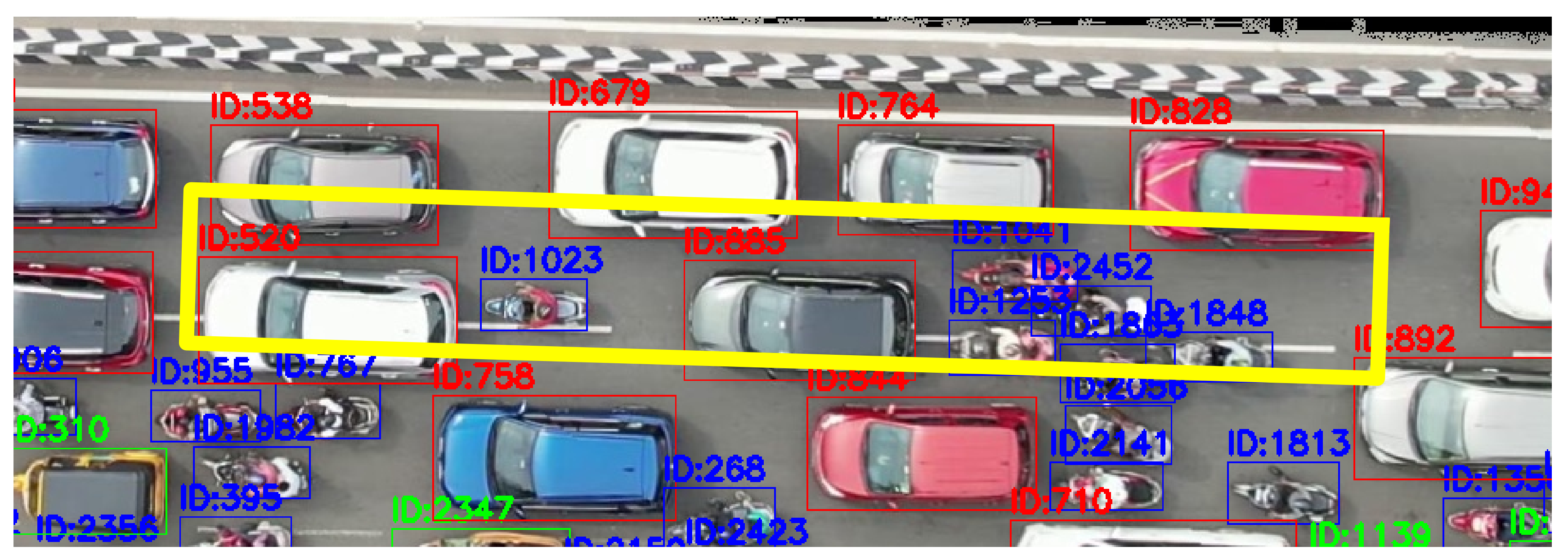}
        \caption{Region~4: Moderate-density traffic state with negligible longitudinal flow density.}
        \label{fig:fig_9f}
    \end{subfigure}

    \caption{Selected regions from the two-dimensional fundamental diagrams and corresponding on-road snapshots, illustrating distinct traffic states used to interpret dispersion in macroscopic relationships under lane-free conditions.}
    \label{fig:fig_9}
\end{figure}

\noindent
Figure~\ref{fig:fig_9} highlights selected regions from the two-dimensional fundamental diagrams, together with corresponding on-road snapshots, to associate distinct macroscopic traffic states with their underlying spatial vehicle configurations. The highlighted yellow box in the snapshots indicates the location of the virtual two-dimensional detector aligned with the lane marking, referenced to the Frenet-transformed trajectory data.
The identified regions correspond to four representative traffic states, summarized as follows:

\begin{itemize}
    \item \textbf{Region 1: Low-density state with reduced longitudinal velocity (Figures~\ref{fig:fig_9a} and~\ref{fig:fig_9c}).}  
    Region~1 corresponds to a traffic state with a low density of 0.036~veh./m$^2$ but a markedly reduced longitudinal velocity of 0.517~m/s. The associated on-road snapshot (Figure~\ref{fig:fig_9c}) shows that large vehicles, including cars, light commercial vehicles, and heavy commercial vehicles, occupy a substantial fraction of the roadway width. This spatial configuration generates voids that are inaccessible to smaller vehicles, leading to inefficient space utilization and suppressed longitudinal motion despite the low overall density.

    \item \textbf{Region 2: Near-capacity state with high longitudinal flow density (Figures~\ref{fig:fig_9b} and~\ref{fig:fig_9d}).}  
    Region~2 represents a near-capacity traffic state characterized by a density of 0.085~veh./m$^2$ and a high longitudinal flow density of 0.727~veh./(m$\cdot$s), while sustaining a mean longitudinal velocity of 8.564~m/s. In this state, two-wheelers constitute approximately 73.2\% of the traffic stream. The high proportion of space-efficient vehicles enables dense packing and sustained longitudinal motion, resulting in elevated flow density values under lane-free conditions.

    \item \textbf{Region 3: High-density state with sustained flow density (Figure~\ref{fig:fig_9e}).}  
    Region~3 corresponds to a high-density traffic state with a density of 0.333~veh./m$^2$, dominated by two-wheelers and auto-rickshaws. Despite a substantial reduction in mean longitudinal velocity, a non-negligible longitudinal flow density of 0.120~veh./(m$\cdot$s) is maintained. The proportion of two-wheelers in this region reaches 82.7\%, indicating that maneuverability and compact vehicle dimensions play a critical role in sustaining traffic flow under severely constrained spatial conditions.

    \item \textbf{Region 4: Moderate-density state with negligible longitudinal flow density (Figure~\ref{fig:fig_9f}).}  
    Region~4 illustrates a traffic state with a moderate density of 0.102~veh./m$^2$ but minimal longitudinal flow density. The on-road snapshot reveals that vehicles with larger physical dimensions and limited maneuverability occupy a significant portion of the available roadway space. This configuration restricts both longitudinal progression and lateral redistribution, resulting in low longitudinal speeds and suppressed flow density despite the density not being particularly high.
\end{itemize}

\noindent Together, these representative regions demonstrate that, under lane-free mixed traffic conditions, macroscopic traffic behavior is governed not solely by density but also by the combined effects of vehicle composition, spatial occupancy, and maneuverability. The two-dimensional fundamental diagrams reveal a continuous evolution of traffic states rather than distinct traffic regimes, with multiple flow states coexisting at similar density levels. The observed spread of traffic states reflects the influence of vehicle heterogeneity, lateral redistribution across the road width, and variations in local space utilization and speed adaptation. Such features cannot be captured by conventional one-dimensional macroscopic representations. This analysis represents one of the early empirical efforts to characterize traffic behavior in lane-free mixed traffic using a two-dimensional fundamental diagram derived from high-resolution trajectory data. The results provide additional empirical evidence supporting the need for two-dimensional macroscopic descriptions under weak lane discipline.

\subsubsection{Congestion Propagation Speed Estimation}

\begin{figure}
    \centering
    \begin{subfigure}{0.33\linewidth}
        \centering
        \includegraphics[width=\linewidth]{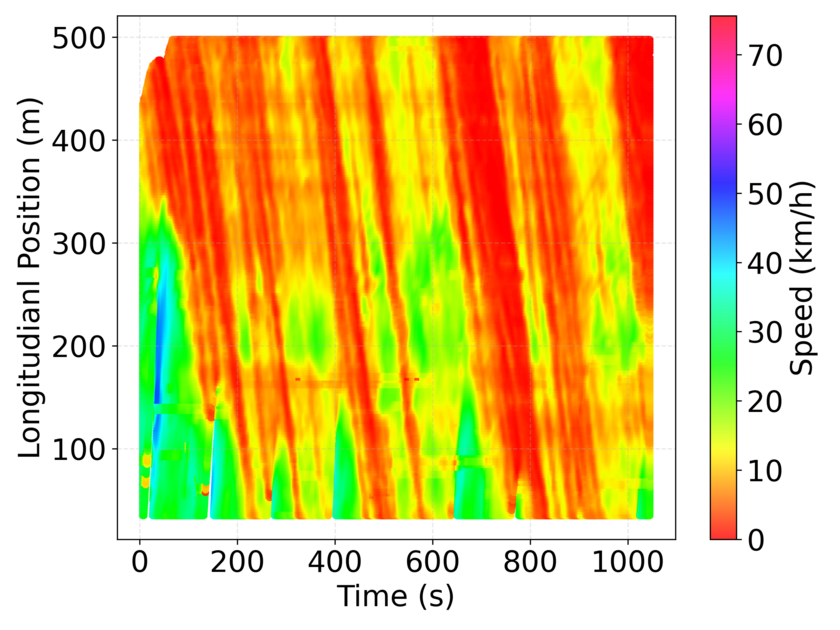}
        \caption{}
        \label{fig:fig_10a}
    \end{subfigure}\hfill
    \begin{subfigure}{0.33\linewidth}
        \centering
        \includegraphics[width=\linewidth]{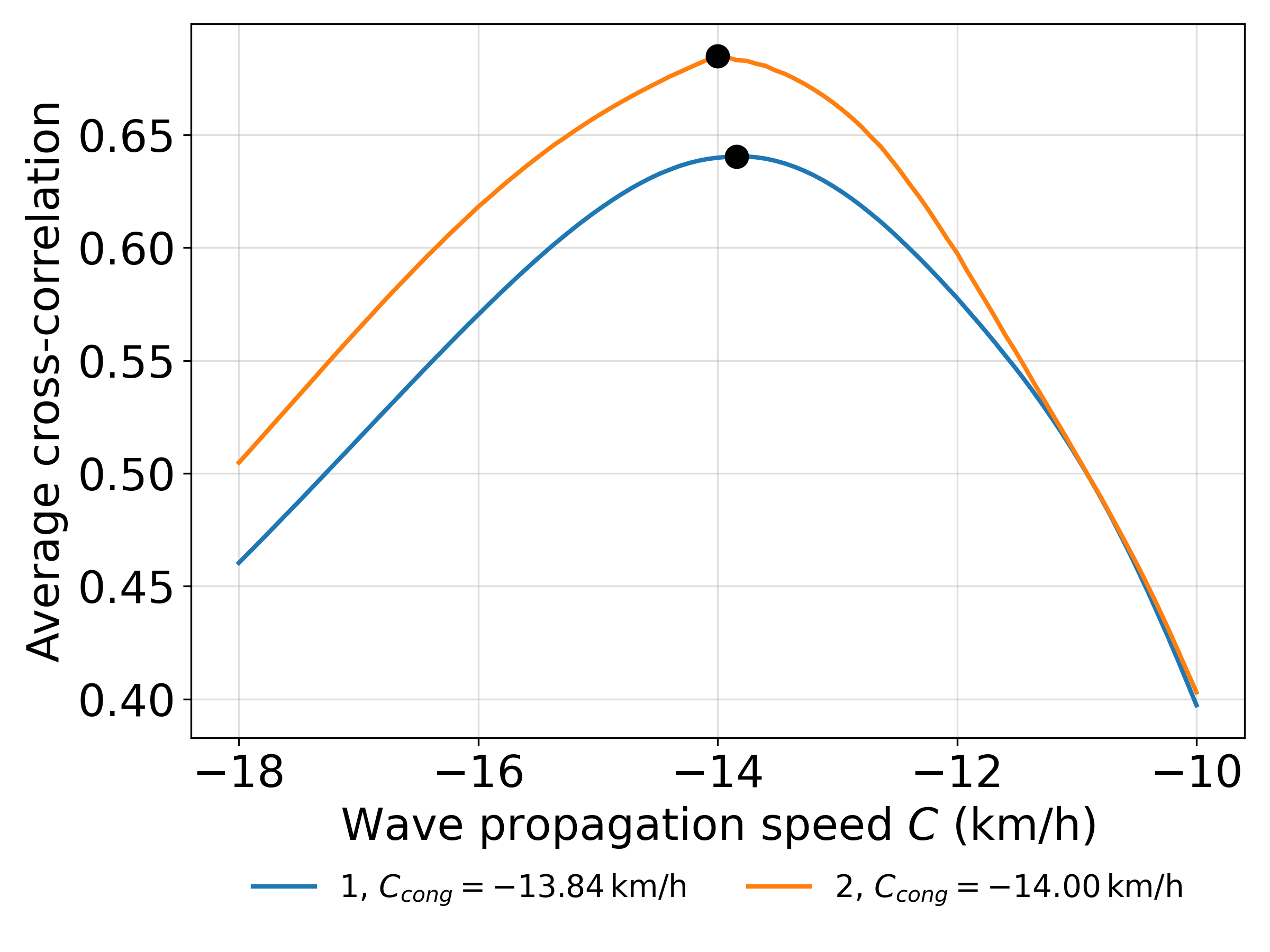}
        \caption{}
        \label{fig:fig_10b}
    \end{subfigure}\hfill
    \begin{subfigure}{0.33\linewidth}
        \centering
        \includegraphics[width=\linewidth]{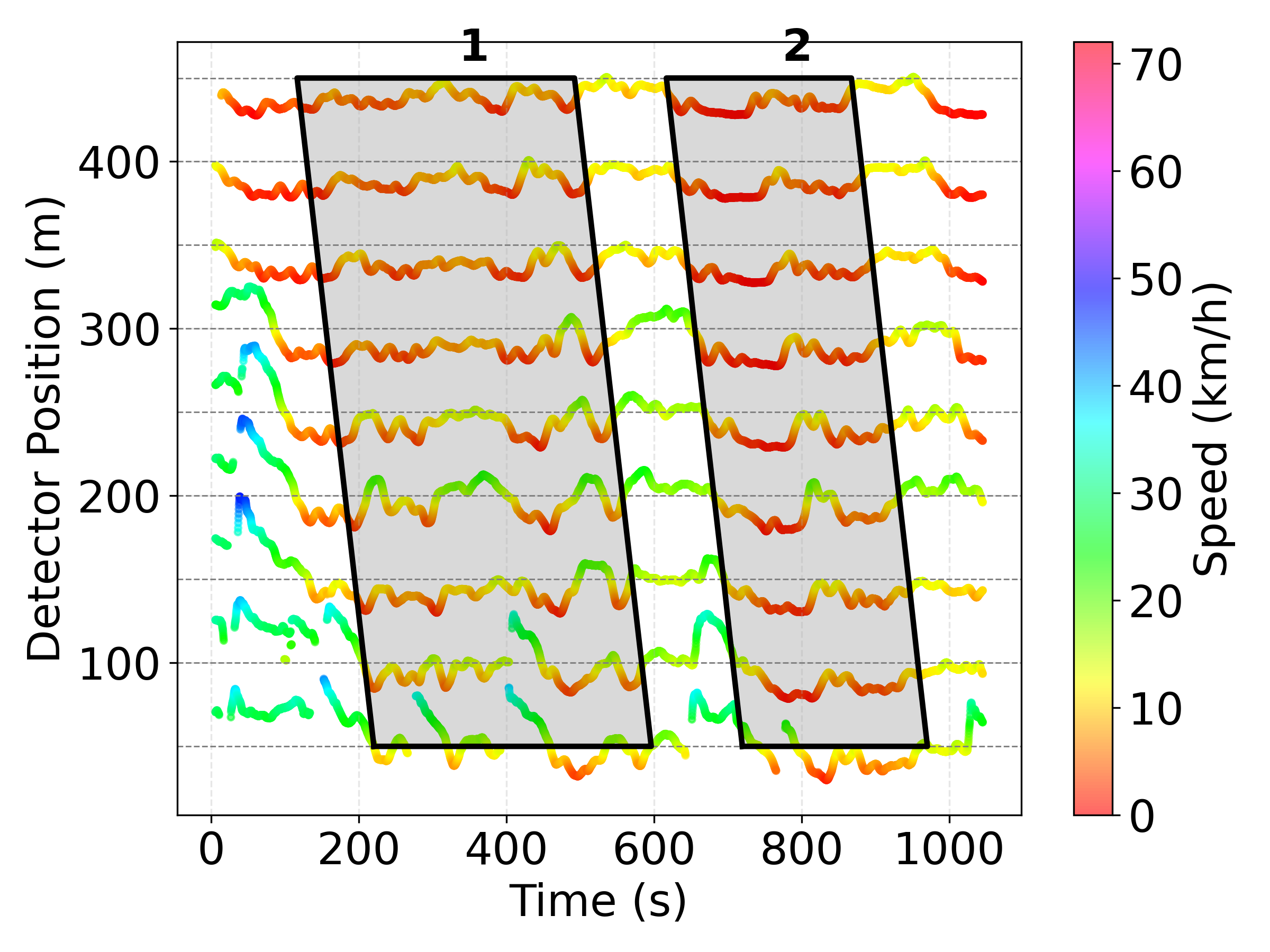}
        \caption{}
        \label{fig:fig_10c}
    \end{subfigure}\hfill

    \vspace{3mm}

    \begin{subfigure}{0.33\linewidth}
        \centering
        \includegraphics[width=\linewidth]{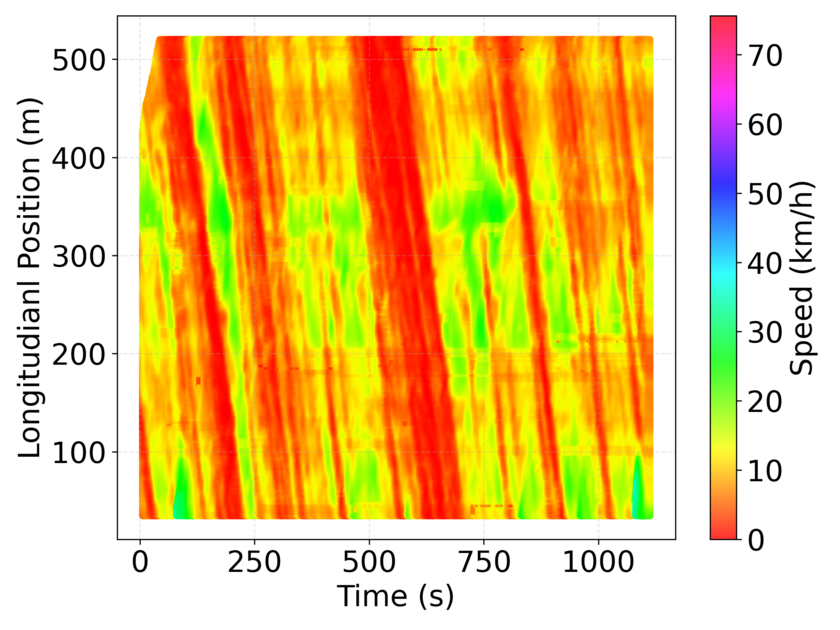}
        \caption{}
        \label{fig:fig_10d}
    \end{subfigure}\hfill
    \begin{subfigure}{0.33\linewidth}
        \centering
        \includegraphics[width=\linewidth]{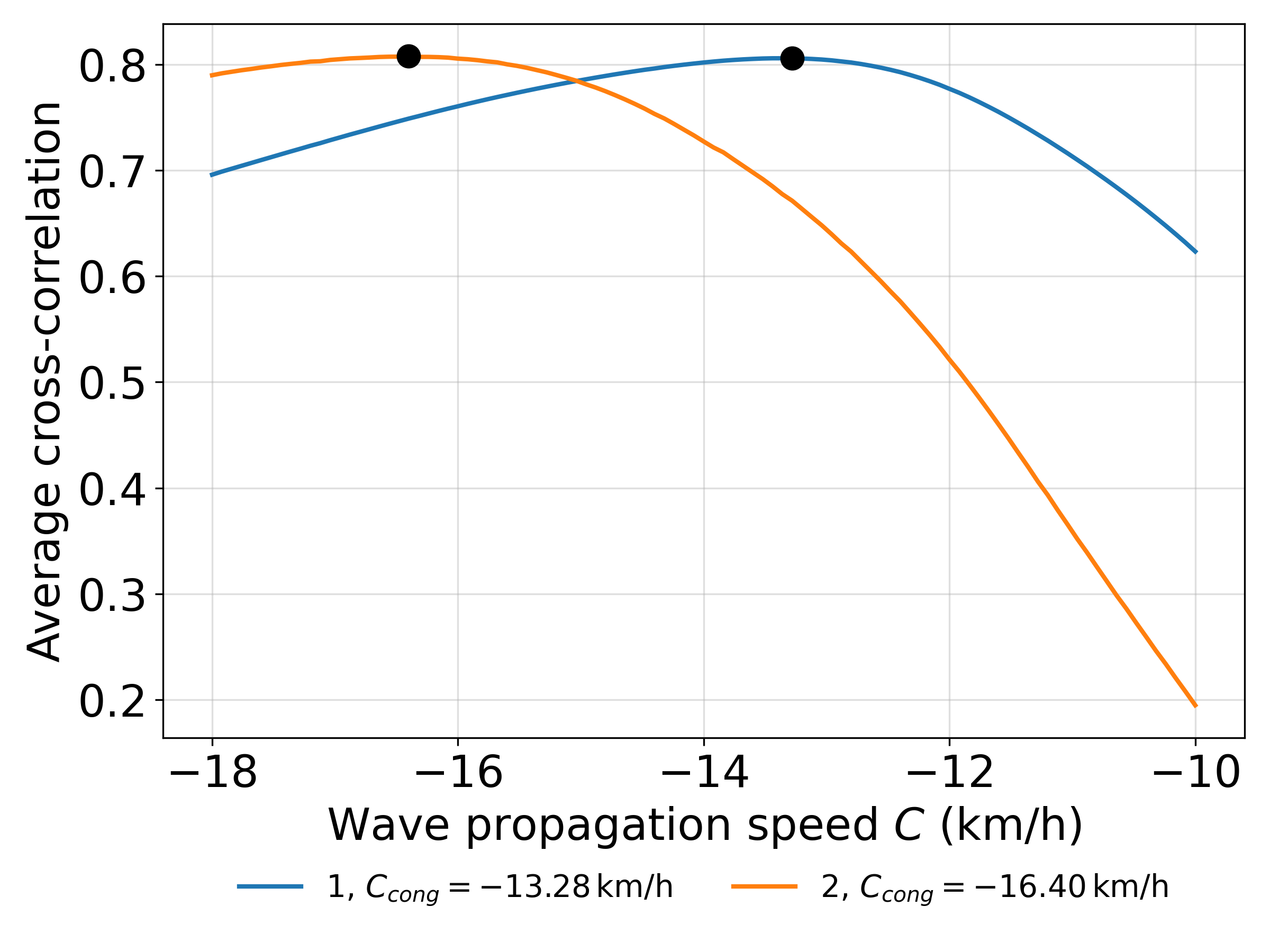}
        \caption{}
        \label{fig:fig_10e}
    \end{subfigure}\hfill
    \begin{subfigure}{0.33\linewidth}
        \centering
        \includegraphics[width=\linewidth]{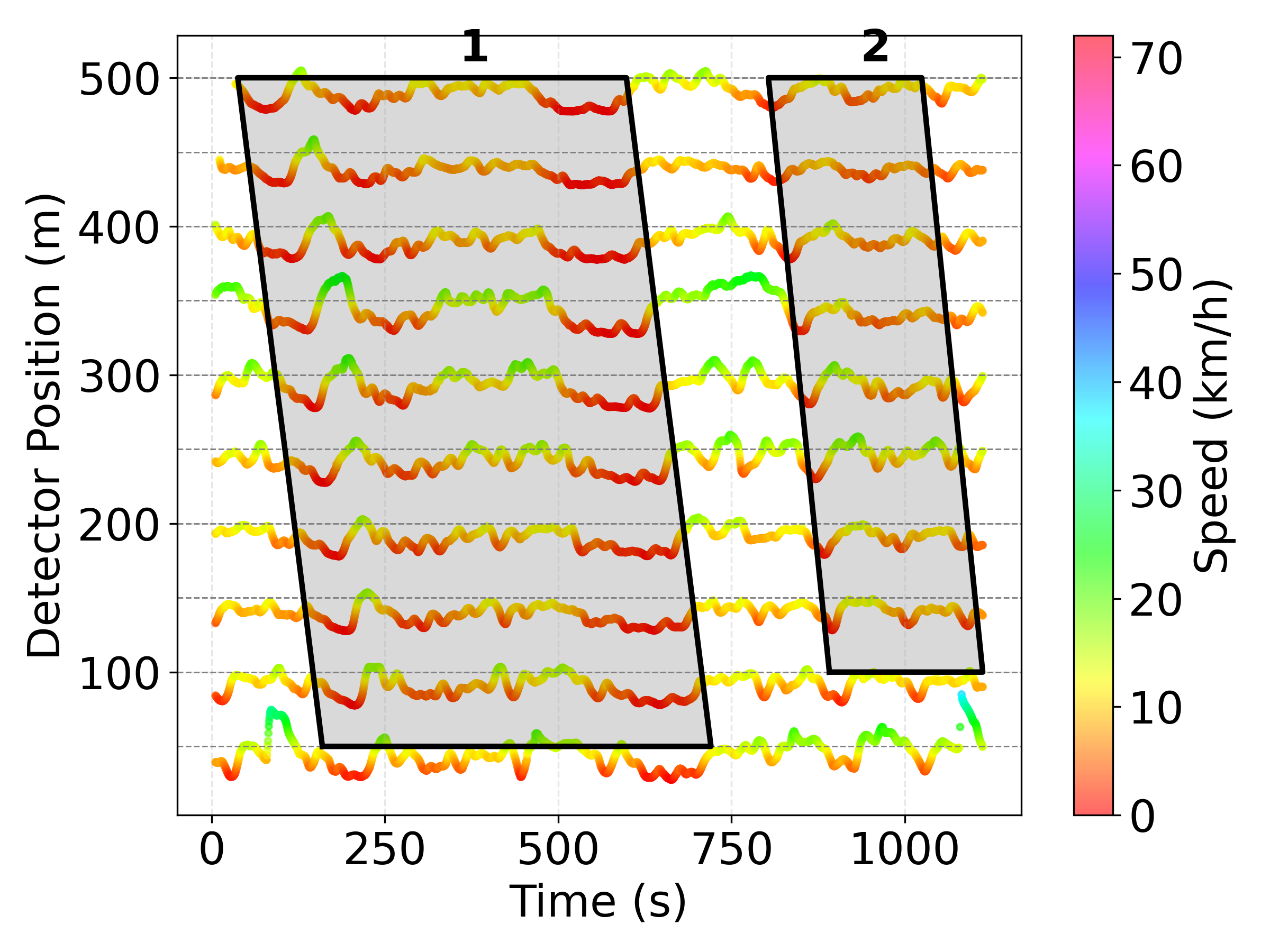}
        \caption{}
        \label{fig:fig_10f}
    \end{subfigure}
    \vspace{3mm}

        \caption{
            (a, d) Spatiotemporal evolution of longitudinal speed along the study section, illustrating the formation and upstream propagation of stop-and-go waves in disordered traffic.
            (b, e) Average cross-correlation values obtained for candidate congestion propagation speeds across identified stop-and-go events. (c, f) Speed time series measured at multiple virtual detector locations; the horizontal dashed line denotes the critical speed threshold ($4.17\,\mathrm{m/s}$) used to identify congestion, and the shaded parallelogram indicates a selected congestion event delineated using the estimated propagation speed $C_{\text{cong}}$. Panels (a)--(c) correspond to Flight~1, and panels (d)--(f) correspond to Flight~2.}
         
    \label{fig:fig_10}
\end{figure}

\noindent Congestion propagation is identified and quantified using longitudinal speed as the primary macroscopic indicator. In mixed, lane-free traffic, density and flow density may vary across the road width due to vehicle heterogeneity and lateral maneuvering. In contrast, longitudinal speed exhibits comparatively smaller lateral variation within a given traffic state, making it a suitable variable for identifying congestion and tracking its propagation. A stop-and-go state is defined when the longitudinal speed at the downstream detector falls below a critical threshold of $4.17~\mathrm{m/s}$. Using this criterion, stop-and-go events are detected from the spatiotemporal speed field estimated using Edie’s space-time formulation. The longitudinal speed is evaluated in a sequence of virtual detectors (space-time Edie volumes with $\dd y$ extended over the whole road width) placed along the study section at uniform longitudinal intervals of $50~\mathrm{m}$, with the upstream detector located at $50~\mathrm{m}$. The resulting spatiotemporal evolution of longitudinal speed is shown in Figure~\ref{fig:fig_10a} \& \ref{fig:fig_10d}, where repeated formation and upstream propagation of low-speed regions corresponding to stop-and-go waves can be observed.\\

\noindent For an identified stop-and-go event, a spatiotemporal domain associated with the propagation of a congestion wave is isolated in the time-space plane. This domain is defined as a parallelogram bounded by lines aligned with the observed wave speed, which is estimated by connecting straight lines to the congestion wave in the speed field (Figures~\ref{fig:fig_10c} and \ref{fig:fig_10f}). An initial estimate of the congestion propagation speed is used solely for delineating this domain and does not constrain the final estimation. Let $t_d^b$ and $t_d^e$ denote the times at which congestion begins and ends at detector $d$, located at position $x_d$. The corresponding start and end times of the congestion event at the upstream reference detector ($d = 1$) are determined as
\begin{equation}
t_1^b = \max_i \left( t_d^b - \frac{x_d - x_1}{C_{\text{cong}}} \right),
\end{equation}
\begin{equation}
t_1^e = \min_i \left( t_d^e - \frac{x_d - x_1}{C_{\text{cong}}} \right),
\end{equation}
and the duration of the congestion event is given by $T = t_1^e - t_1^b$. For speed data sampled at a constant interval $\Delta t$, the time indices corresponding to the congestion interval at detector $d$ are defined as
\begin{equation}
J_d =
\left[
\frac{t_1^b + (x_d - x_1)/C_{\text{cong}}}{\Delta t},
\;
\frac{t_1^e + (x_d - x_1)/C_{\text{cong}}}{\Delta t}
\right].
\end{equation}

\noindent Within each delineated stop-and-go region, longitudinal speed time series from all detectors are analyzed using a cross-correlation approach \citep{Treiber2012Cong}. For a candidate propagation speed $C'$, the speed signals $V_d(t)$ at detector locations $x_d$ and $x_k$ are temporally aligned by shifting one signal by $(x_k - x_d)/C'$. The degree of alignment is quantified using the Pearson cross-correlation coefficient. The congestion propagation speed is estimated as the value of $C'$ that maximizes the cumulative cross-correlation across all detector pairs:
\begin{equation}
C_{\text{cong}} =
\arg\max_{C'}
\sum_d \sum_{k>d}
\mathrm{corr}
\left(
V_d(t),
V_k\!\left(t + \frac{x_k - x_d}{C'}\right)
\right),
\end{equation}

\noindent as illustrated in Figure~\ref{fig:fig_10b} \& \ref{fig:fig_10e}. For the trajectory dataset analyzed in this study, traffic remains predominantly congested over much of the observation period, as indicated by the persistent low-speed regions in Figure~\ref{fig:fig_10}. In the spatiotemporal speed field, two congestion events for flight 1 (Figure~\ref{fig:fig_10c}) and flight 2 (Figure~\ref{fig:fig_10f})) with coherent upstream propagation are identified and delineated using parallelogram-shaped regions. Each parallelogram corresponds to a separate stop-and-go wave exhibiting approximately constant propagation speed. For flight 2, the second congestion wave, the analysis is restricted to detector locations downstream of $100~\mathrm{m}$ as the detector at $50~\mathrm{m}$ does not exhibit a stable propagation structure. Whereas for the other congestion events, all the detectors were considered. Note that within a congestion wave event, the speed may sometimes exceed the critical speed of $4.17~\mathrm{m/s}$. We cannot completely eliminate such speed values inside the congestion wave due to frequent stop-and-go traffic. However, this does not affect the final wave speed, which is obtained from the maximum value of the cross-correlation.\\

\noindent
The congestion events identified across both flights yield consistent estimates of upstream propagation speed, with values ranging from approximately $3.69$ to $4.56~\mathrm{m/s}$, as indicated by the peaks in the cross-correlation curves (Figure~\ref{fig:fig_10b}, \ref{fig:fig_10e}). Specifically, the two events observed in Flight~1 show propagation speeds of $3.84$ and $3.89~\mathrm{m/s}$, while the three events in Flight~2 exhibit speeds of $3.69$ and $4.56~\mathrm{m/s}$. This value is close to propagation speeds reported in previous studies of stop-and-go traffic \citep{Zielke2008StopandGo, Treiber2012Cong}. These results demonstrate that, even under strongly disordered and lane-free traffic conditions characterized by heterogeneous vehicle interactions, congestion waves exhibit a well-defined macroscopic propagation behavior. The estimated propagation speed is comparable to values reported for ordered, lane-based traffic conditions, indicating that the large-scale dynamics of congestion waves are preserved despite fundamental differences in vehicle composition and interaction structure. Moreover, the present analysis provides an empirical estimate of congestion wave propagation speed derived directly from high-resolution trajectory data in mixed, lane-free traffic. Such quantified evidence remains limited in the existing literature and offers a reference value for macroscopic modeling, calibration, and comparative analysis of congestion dynamics in disordered traffic regimes.

\subsection{Microscopic Dynamics}
\noindent The macroscopic traffic states identified in Section~5.1 are generated by vehicle-level interactions that depend on spacing preferences, physical dimensions, and motion regulation under lane-free conditions. In mixed traffic streams, heterogeneity in vehicle size and maneuverability alters both longitudinal spacing and kinematic responses, leading to the dispersion and regime-dependent behavior observed in the two-dimensional macroscopic relationships. To interpret these aggregate patterns, it is therefore necessary to examine microscopic variables that directly govern vehicle following, space occupation, and motion adjustment. This section presents a microscopic analysis based on high-resolution vehicle trajectories, focusing on interaction variables that can be consistently extracted under disordered traffic conditions. The analysis begins with the desired time gap, estimated under steady-state following conditions, to characterize equilibrium longitudinal spacing preferences across vehicle categories. This is followed by an examination of the minimum standstill gap, which captures spatial spacing behavior when vehicles are stopped and provides insight into jam packing and space utilization independent of speed.\\

\noindent 
Subsequently, vehicle length and width distributions are analyzed using heading-corrected bounding-box measurements to quantify the physical heterogeneity of the traffic stream and its implications for longitudinal and lateral space occupancy. Finally, speed and acceleration characteristics are examined to describe how vehicles regulate motion and respond to local interactions under disordered conditions, complementing the spacing-based measures with kinematic evidence. Together, these microscopic analyses provide a quantitative basis for linking vehicle-scale interaction rules to the macroscopic traffic states observed in lane-free mixed traffic, and they supply empirically grounded parameters relevant for the calibration and validation of microscopic traffic flow models that do not rely on lane-based assumptions.

\subsubsection{Desired Time Gap}
\noindent The time gap between successive vehicles is a microscopic variable that describes how spacing is regulated in traffic. It is defined as the time required for the front bumper of a following vehicle to reach the current position of the rear bumper of its immediate leader. Time gap directly influences traffic stability, achievable flow, and the formation of congestion. As a result, the desired time gap is a central parameter in microscopic traffic flow models, where it governs car-following responses and the transition between steady-state and unsteady traffic states. In recent two-dimensional microscopic traffic flow models \cite{Kanagaraj2018SelfDriven,TreiberChaudhari2022IAM}, desired time gap is a key model parameter. To replicate real-world traffic conditions, this parameter must be quantified from empirical trajectory data.\\
\begin{figure}
    \centering
    \includegraphics[width=0.75\linewidth]{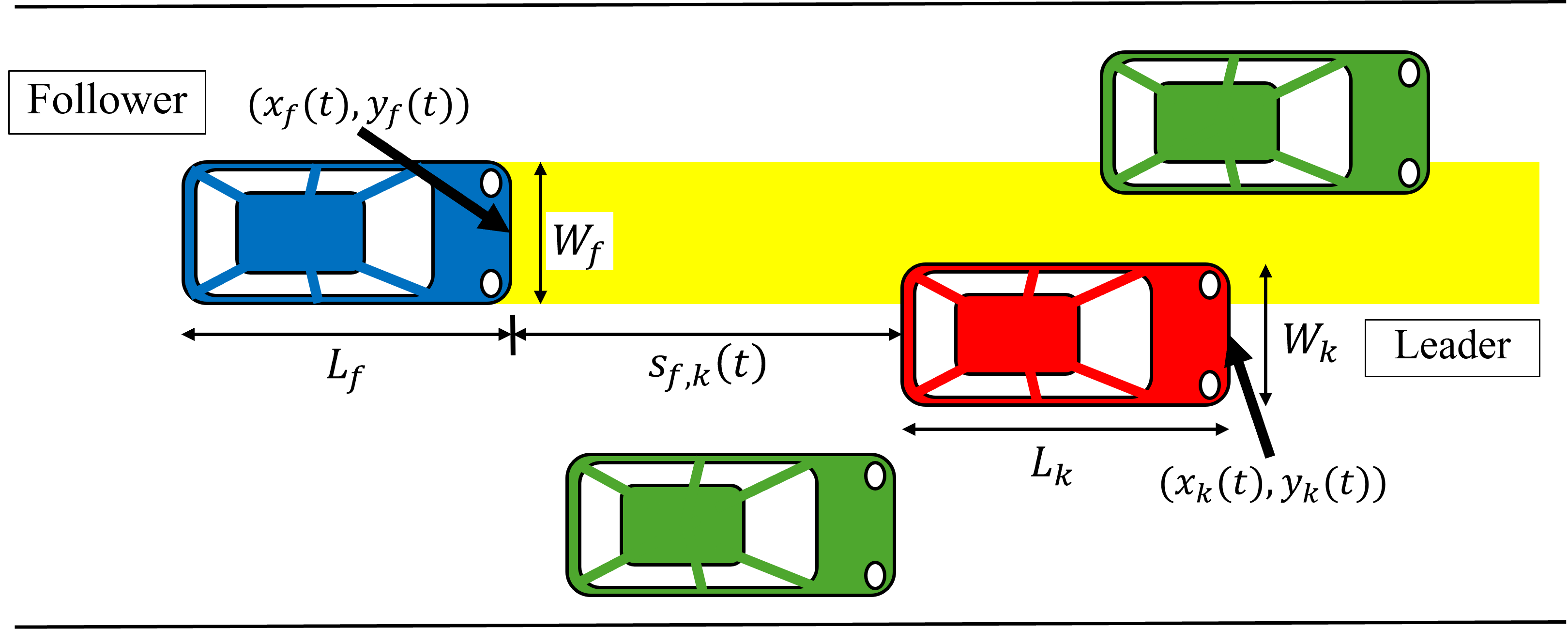}
    \caption{Schematic representation of the follower-leader identification criteria.}
    \label{fig:fig_fl}
\end{figure}

\noindent Hence, in this study, particular emphasis is placed on the desired time gap, which represents the spacing preference of drivers under steady-state conditions. The desired time gap is estimated when a follower-leader pair travels at nearly the same speed over a sustained duration, indicating an equilibrium interaction. To estimate the desired time gap, follower-leader pairs are first identified under steady-state conditions. For a given subject vehicle, indexed as the follower $f$, all vehicles ahead of it within the observation frame are treated as potential leaders and indexed by $k \in \mathcal{L}_f$. A candidate leader must exhibit lateral overlap with the follower to ensure physical interaction (Figure~\ref{fig:fig_fl}). This condition is enforced as
\begin{equation}
|y_k(t) - y_f(t)| \leq \frac{W_f + W_k}{2},
\end{equation}
where $y_f(t)$ and $y_k(t)$ denote the lateral positions of the follower and the candidate leader at time $t$, and $W_f$ and $W_k$ are their respective widths. Among all candidates satisfying the lateral overlap criterion, the effective leader $l$ is selected as the nearest vehicle ahead of the follower in the longitudinal direction. The longitudinal spacing between the follower and each candidate leader is defined as
\begin{equation}
s_{fk}(t) = x_k(t) - x_f(t) - L_k,
\label{eq:distance Gap}
\end{equation}
where $x_f(t)$ and $x_k(t)$ are the longitudinal positions, and $L_k$ is the length of the lead vehicle. The leader is then identified as
\begin{equation}
l = \arg\min_{k \in \mathcal{L}_f} s_{fk}(t), \quad \text{with } s_{fk}(t) > 0,
\end{equation}
ensuring that only vehicles physically ahead of the follower are considered. Once a follower-leader pair $(f,l)$ is identified, steady-state conditions are imposed to isolate equilibrium interactions. Specifically, the relative speed and the longitudinal acceleration of the follower must satisfy
\begin{equation}
|V_l(t) - V_f(t)| \leq \epsilon_v,
\qquad
|a_f(t)| \leq \epsilon_a,
\quad
\forall\, t \in [t_0,\, t_0 + \Delta t_{\min}],
\end{equation}
where $\epsilon_v = 0.5~\mathrm{m/s}$, $\epsilon_a = 0.2~\mathrm{m/s^2}$, and $\Delta t_{\min} = 3~\mathrm{s}$. These conditions are imposed to exclude transient responses associated with braking, lane changes, or short-lived disturbances. A relative speed threshold of $0.5~\mathrm{m/s}$ ($\approx 1.8~\mathrm{km/h}$)
 lies within the perceptual deadband of drivers, while an acceleration threshold of $0.2~\mathrm{m/s^2}$ corresponds to near-coasting motion and excludes active acceleration or braking phases. The steady-state condition is required to persist over a $3~\mathrm{s}$ interval, which exceeds typical driver reaction and vehicle response times while remaining short enough to avoid contamination by lane changes, leader changes, or the onset of stop-and-go oscillations. This ensures that the identified periods correspond to genuine near-equilibrium car-following rather than transient adjustments. For follower-leader pairs that satisfy the above criteria, the desired time gap is computed as
\begin{equation}
T_f(t) = \frac{s_f(t)}{V_f(t)},
\end{equation}
where $s_f(t) = s_{fl}(t)$ denotes the longitudinal spacing to the identified leader. This formulation captures spacing preferences under near-equilibrium conditions and provides a consistent basis for analyzing desired time-gap distributions across vehicle categories in disordered traffic conditions.\\

\renewcommand{\arraystretch}{1.7}
\begin{table}
\centering
\caption{Summary statistics of desired time gap and minimum gap for different vehicle categories under congested, lane-free traffic conditions.}
\label{tab:gap_stats}
\begin{threeparttable}
\begin{tabular*}{\tblwidth}{@{}cllcccccccc@{}}
        \toprule
        Metric & Vehicle Type & Observation Count & Mean & Median & Standard Deviation & P25 & P75 & P90\\
        \midrule
        \multirow{5}{*}{\rotatebox[origin=c]{90}{Desired Time Gap (s)}}
        & TW & 510	& 0.674	& 0.488	& 0.704	& 0.247	& 0.837	& 1.416\\
        & Car	& 553	& 1.339 &	1.213	& 0.772 & 0.813	& 1.695	& 2.288\\
        & AR	& 118	& 0.916	& 0.811	& 0.735 & 0.43	& 1.267	& 1.724\\
        & LCV	& 17	& 1.248	& 1.08	& 0.562	& 0.893	& 1.401	& 2.157\\
        & HCV	& 13	& 1.856	& 1.57	& 1.431	& 0.751	& 2.367	& 3.182\\
        \midrule
        \multirow{5}{*}{\rotatebox[origin=c]{90}{Minimum Gap (m)}}
        & TW & 450 & 0.728 & 0.547 & 0.801 & 0.264 & 0.905 & 1.368\\
        & Car & 552 &1.464 & 1.324 & 1.065 & 0.703 & 1.944 
        & 2.588\\
        & AR & 91 & 1.043 & 0.834 & 1.012 & 0.437 & 1.445 & 1.705\\
        & LCV & 19 & 1.199 & 0.991 & 0.859 & 0.435 & 1.847 & 2.267\\
        & HCV & 11 & 0.833 & 0.713 & 0.51  & 0.356 & 1.244 & 1.351\\
        \bottomrule
    \end{tabular*}
\setlength{\parindent}{0pt}
\footnotesize
Note: P25, P75, and P90 denote the 25th, 75th, and 90th percentile values of the observed distributions, respectively.

\end{threeparttable}
\end{table}

\begin{figure}
    \centering
        \centering
    \begin{subfigure}{0.32\linewidth}
        \centering
        \includegraphics[width=\linewidth]{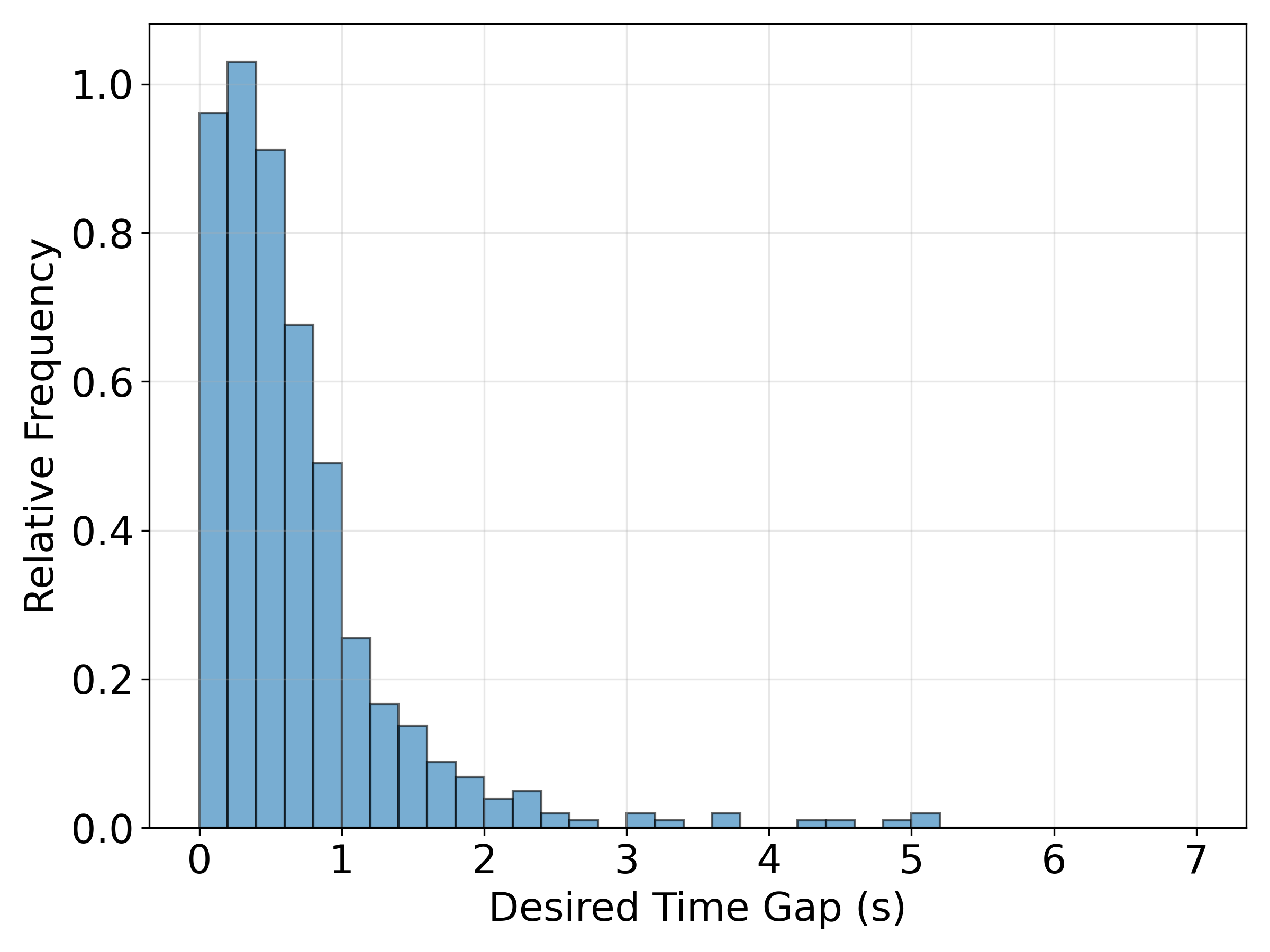}
        \caption{Two-wheelers (TW)}
        \label{fig:fig_DTGa}
    \end{subfigure}
    \hfill
    \begin{subfigure}{0.32\linewidth}
        \centering
        \includegraphics[width=\linewidth]{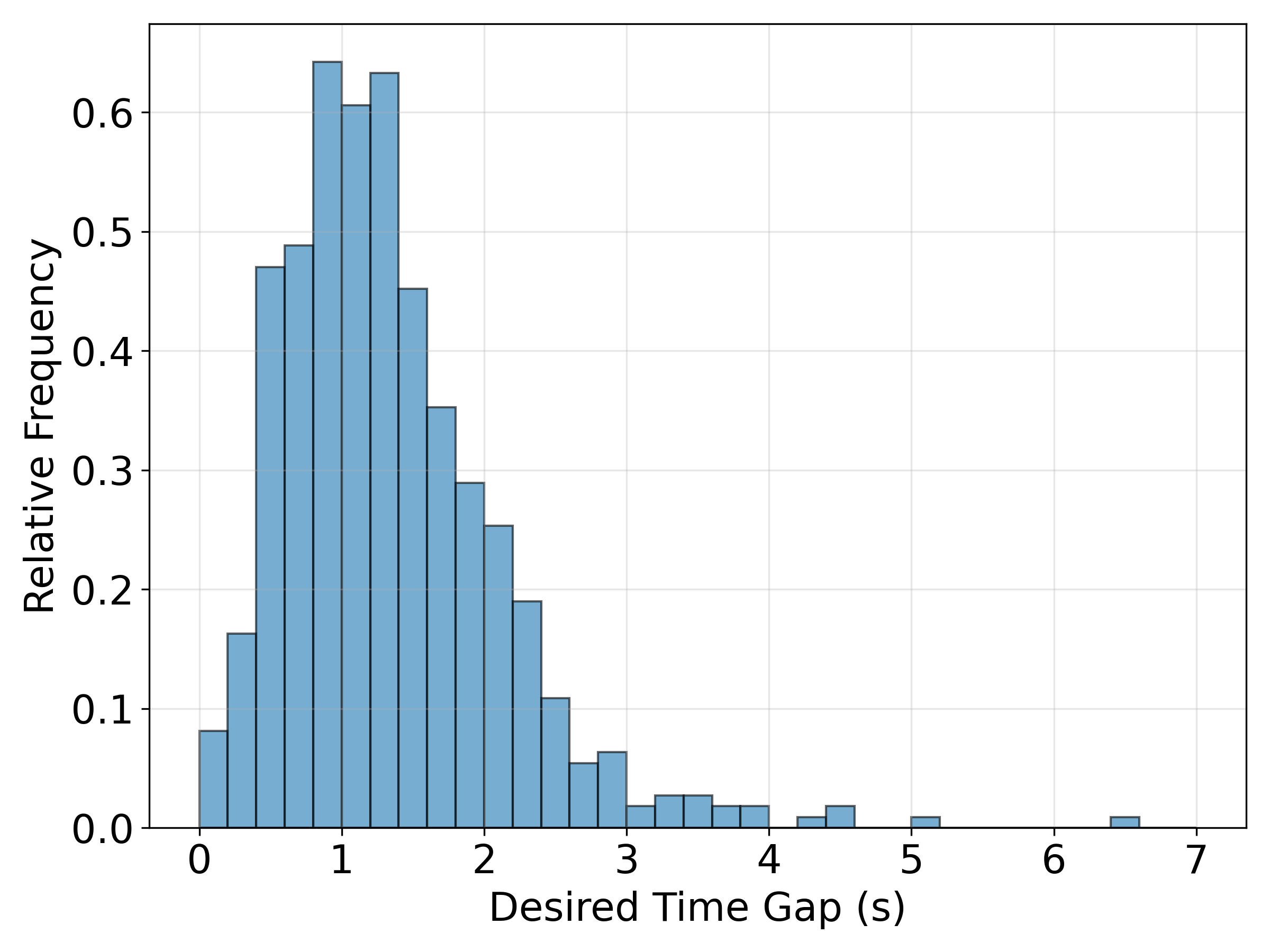}
        \caption{Cars}
        \label{fig:fig_DTGb}
    \end{subfigure}
    \hfill
    \begin{subfigure}{0.32\linewidth}
        \centering
        \includegraphics[width=\linewidth]{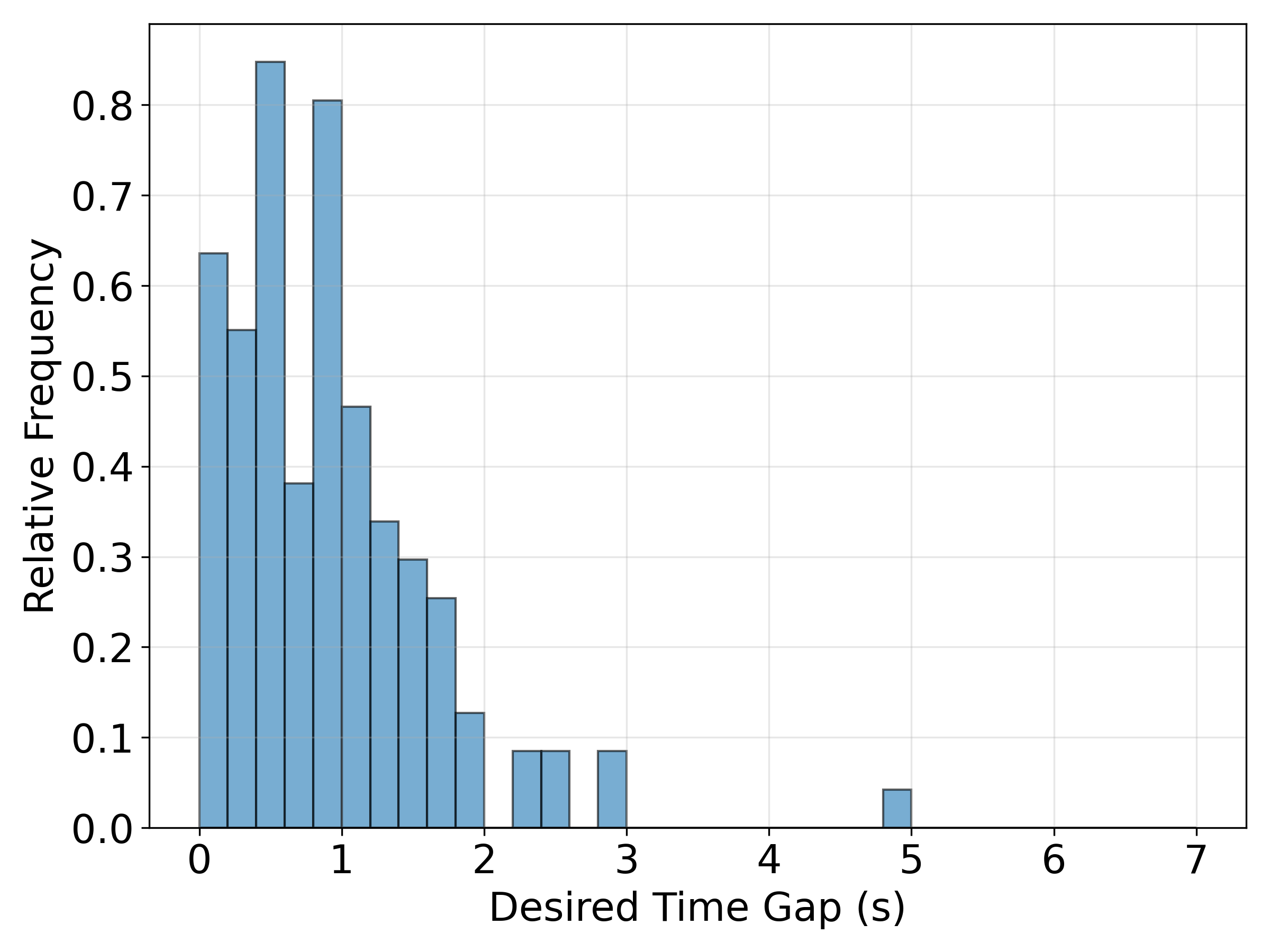}
        \caption{Auto-rickshaws (AR)}
        \label{fig:fig_DTGc}
    \end{subfigure}

    \caption{Distributions of desired time gap for different vehicle categories under congested, lane-free traffic conditions.}
    \label{fig:fig_DTG}
\end{figure}
\noindent
The desired time gap results show clear differences in following behavior across vehicle categories under lane-free conditions. For two-wheelers, the distribution is heavily skewed toward very short gaps, with most observations concentrated below 0.4~s and a steady drop in frequency as the gap increases (Figure~\ref{fig:fig_DTGa}). This trend is reflected in the low median value of $0.49$~s and the narrow interquartile range reported in Table~\ref{tab:gap_stats}. Even the 90th percentile remains close to $1.4$~s, indicating that long gaps are relatively rare for two-wheelers. Such behavior suggests that riders are comfortable maintaining close proximity to the leader in dense traffic. Their small physical size and ability to reposition laterally allow them to sustain movement even when longitudinal spacing is minimal. Cars display a more structured pattern. The distribution gradually rises from very short gaps and reaches its highest concentration around 0.8–1.4~s, after which it tapers off (Figure~\ref{fig:fig_DTGb}). This preference for moderate spacing is also evident in the summary statistics (Table~\ref{tab:gap_stats}), where the median desired time gap is $1.21$~s and the interquartile range spans from $0.81$ to $1.70$~s. These values indicate that cars maintain larger desired time gaps than two-wheelers due to their larger physical dimensions and limited lateral maneuverability. Auto-rickshaws exhibit a more scattered distribution (Figure~\ref{fig:fig_DTGc}), with noticeable concentrations at both short and intermediate gaps. Their median gap of $0.81$~s lies between that of two-wheelers and cars, reflecting their mixed operating behavior. Depending on local space availability and surrounding vehicles, auto-rickshaws may follow closely like two-wheelers or maintain gaps closer to those of cars.\\

\noindent
For light and heavy commercial vehicles, the steady-state following cases are relatively few, which limits detailed interpretation of their desired time gap behavior. Nevertheless, the higher median desired time gaps observed for these vehicle categories (Table~\ref{tab:gap_stats}) are consistent with their larger physical dimensions and restricted maneuverability. Taken together, the results show that desired time gaps in disordered traffic are strongly shaped by vehicle size and maneuverability. Smaller vehicles consistently operate with shorter and more variable gaps, while larger vehicles adopt longer and more stable spacing. To assess the dependence of the desired time gap on follower speed, a correlation analysis was performed. The analysis indicated a weak and statistically non-significant relationship under the analyzed steady-state conditions. At the macroscopic level, this widespread time gap, particularly the dominance of short gaps among smaller vehicles, contributes to the dispersion observed in flow density–density relationships and to the emergence of non-equilibrium traffic states (Figure~\ref{fig:fig_7} and \ref{fig:fig_8}). From a modeling standpoint, the combined distributional patterns and summary statistics provide realistic parameter ranges for calibrating car-following and gap-acceptance models for lane-free traffic, enabling a better representation of following dynamics and lane-changing processes in mixed urban environments.\\

\noindent Although spacing behavior is defined at the level of individual vehicle interactions, its aggregated effect governs the macroscopic state of traffic. This linkage can be formalized through the relationship between longitudinal flow density and average time headway. This relationship holds for both ordered and disordered traffic conditions. For ease of understanding, we derive this relationship here for single-lane, car-only traffic. Consider a reference cross-section of width $\Delta y$ observed over a time interval $\Delta t$. In ordered (single-lane) traffic, $\Delta y$ corresponds to the lane width. In disordered (lane-free) traffic, $\Delta y$ denotes an effective lateral spacing of the order of the vehicle width plus the minimum lateral gap between neighboring vehicles. If $N$ vehicles cross this section, the longitudinal flow density is defined as $Q_x = \frac{N}{\Delta y \, \Delta t}$, with units of veh.$/(\mathrm{m}\cdot\mathrm{s})$. The average time headway $\bar{h}$ represents the mean time interval between successive vehicle passages at this section, such that $N \approx \Delta t / \bar{h}$ per unit width, yielding
\[
Q_x = \frac{1}{\Delta y \bar{h}}.
\]
Further, the average time headway emerges from a sequence of follower-leader interactions and can be expressed as
\[
\bar{h} = \frac{1}{N}\sum_{i=1}^{N} \left( T'_i + \frac{L_{l,i}}{V_{f,i}} \right),
\]
where $i = 1, 2, \ldots, N$ indexes the observed follower-leader pairs, $T'_i$ is the instantaneous time gap, $L_{l,i}$ is the length of the leading vehicle, and $V_{f,i}$ is the corresponding follower speed. These instantaneous gaps reflect local disturbances and contribute to the variability observed in macroscopic flow density-density relationships. Under steady-state conditions, however, the time gap tends toward the desired time gap, which represents the preferred spacing behavior of drivers. In such cases, the average headway, and hence the longitudinal flow density, can be approximated using the desired time gap together with representative vehicle dimensions and speeds. This provides a stable link between microscopic behavior and macroscopic traffic states. In disordered traffic, smaller vehicles with shorter desired gaps and lengths reduce the effective headway and increase longitudinal flow density, whereas larger vehicles increase headway and reduce it, explaining both the spread and structure observed in the fundamental diagram.

\subsubsection{Minimum Gap}

\noindent
This study focuses on the minimum gap ($S_0$), defined as the physical separation between the rear of a leading vehicle and the front of its follower when both vehicles are stopped. Under these conditions, speed-dependent effects are negligible, and the observed spacing primarily reflects the safety margin maintained between vehicles. Analysis of the minimum gap is therefore essential for understanding space utilization in disordered traffic and for estimating jam density, queue formation, and vehicle packing under lane-free urban conditions. The minimum gap is a core parameter in car-following models for ordered traffic \citep{Treiber2000IDM} and in microscopic traffic flow models developed for disordered traffic (\cite{Kanagaraj2018SelfDriven,TreiberChaudhari2022IAM, KASHYAPNR2024HSSFM}). Empirical estimation of this parameter from field data under disordered traffic conditions provides realistic bounds on safety margins across vehicle types and contributes to improved representation of stop-and-go dynamics in microscopic traffic models.\\

\noindent 
\noindent
The minimum gap is evaluated using the same follower-leader identification framework employed in the desired time gap analysis (Figure~\ref{fig:fig_fl}), with the additional requirement that both vehicles are in a stopped condition. For each identified follower-leader pair $(f,l)$, the minimum gap ($S_0$) is computed using Eq.~\eqref{eq:distance Gap}. A stopped state is defined by the condition
\begin{equation}
|V_f(t)| \leq \epsilon_s, \qquad |V_l(t)| \leq \epsilon_s,
\quad
\forall\, t \in [t_0,\, t_0 + \Delta t_{\min}],
\end{equation}
where $\epsilon_s = 0.01~\mathrm{m/s}$ accounts for measurement noise while ensuring near-zero motion, and $\Delta t_{\min} = 3~\mathrm{s}$ enforces persistence of the stopped condition. The minimum duration $\Delta t_{\min} = 3~\mathrm{s}$ is selected to ensure persistence of the stopped state over multiple observation frames, thereby excluding brief halts caused by transient interactions or measurement fluctuations. This duration exceeds typical driver reaction and vehicle response times, while remaining short enough to avoid contamination from leader changes, and therefore reliably isolates genuine queueing conditions.
Within each such interval, the minimum gap ($S_0$) is measured as the longitudinal spacing $s_{fl}(t)$ between the follower and the leader.
\begin{figure}
    \centering
        \centering
    \begin{subfigure}{0.32\linewidth}
        \centering
        \includegraphics[width=\linewidth]{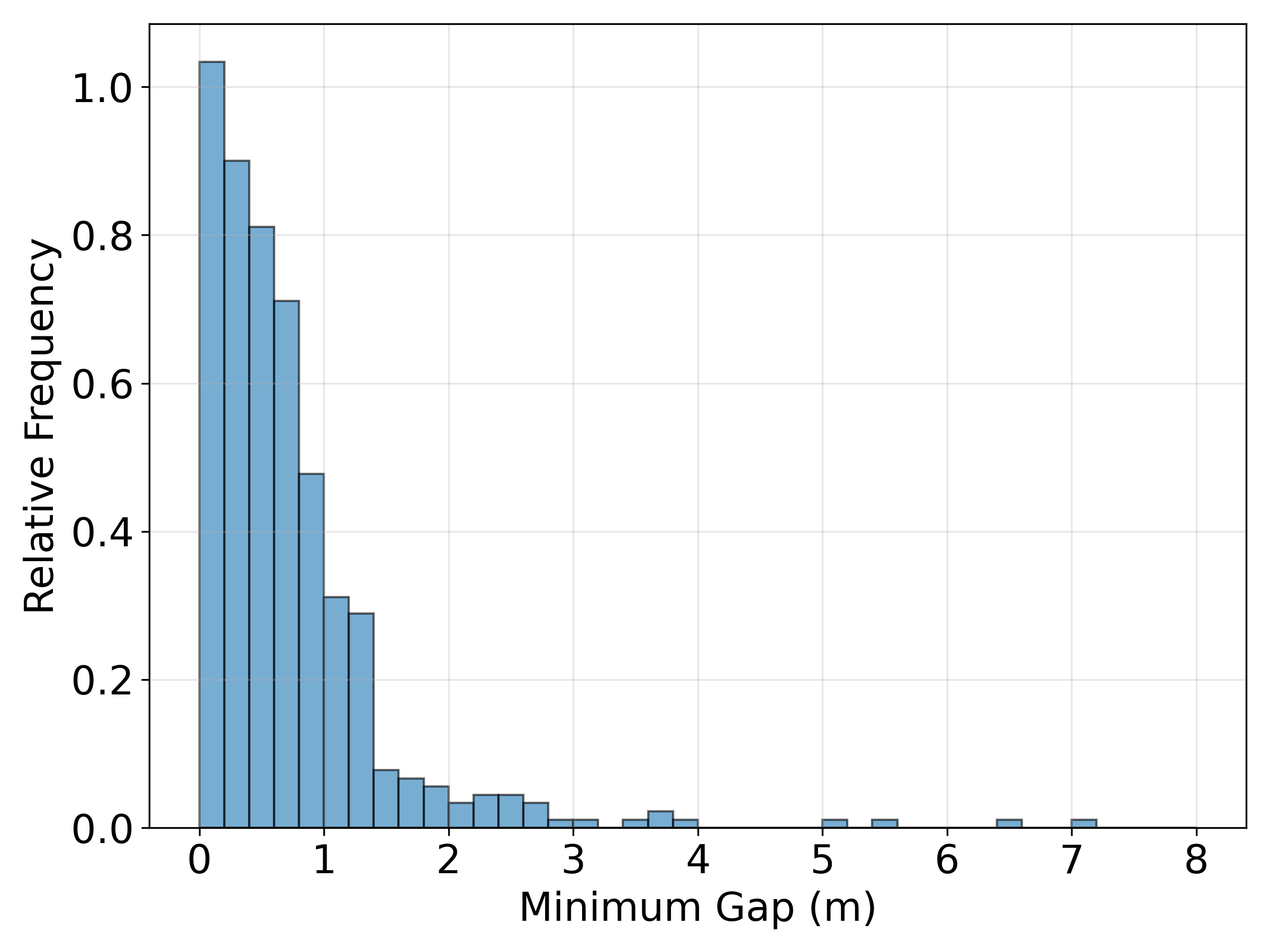}
        \caption{Two-wheelers (TW)}
        \label{fig:fig_MGa}
    \end{subfigure}
    \hfill
    \begin{subfigure}{0.32\linewidth}
        \centering
        \includegraphics[width=\linewidth]{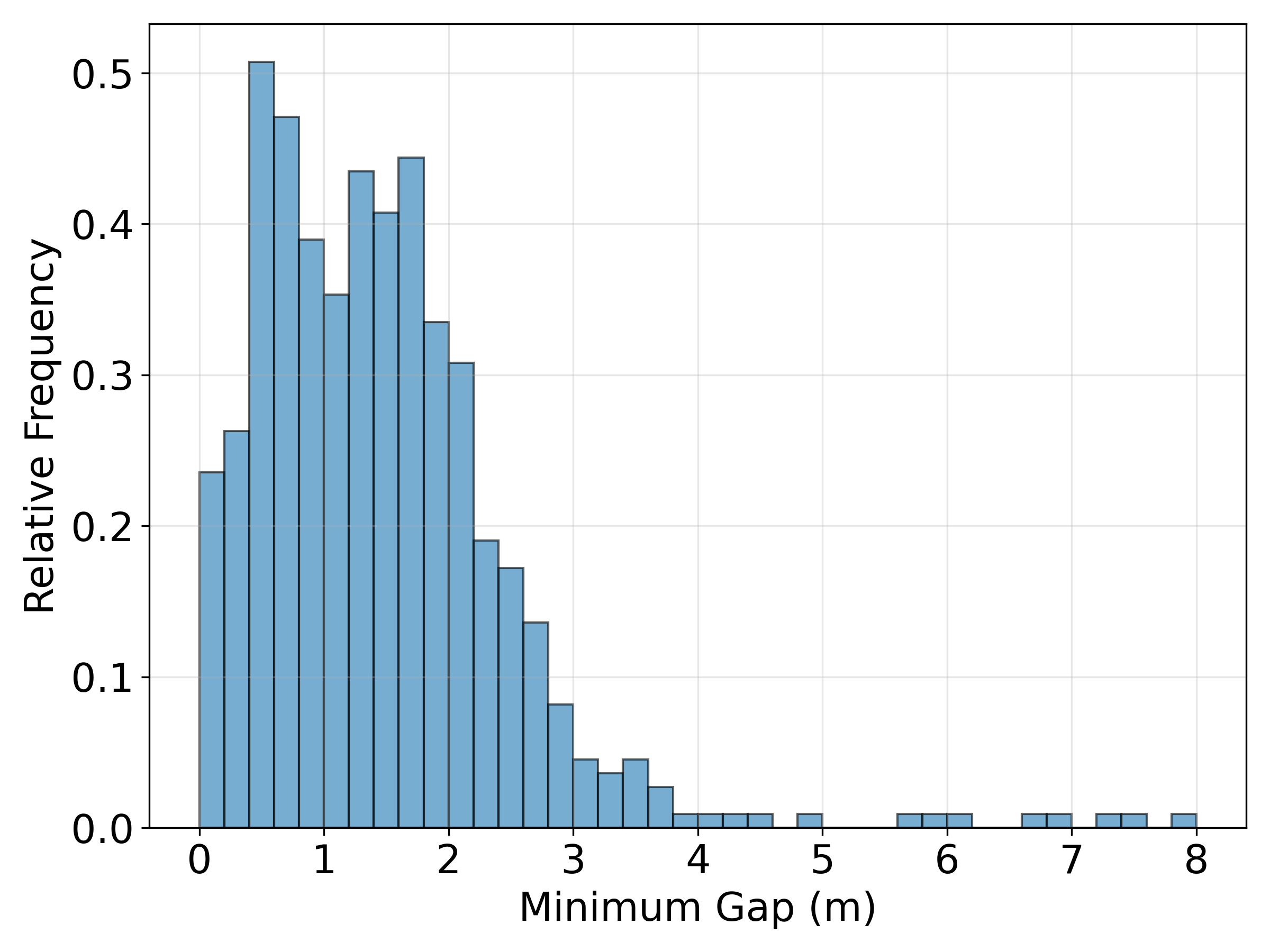}
        \caption{Cars}
        \label{fig:fig_MGb}
    \end{subfigure}
    \hfill
    \begin{subfigure}{0.32\linewidth}
        \centering
        \includegraphics[width=\linewidth]{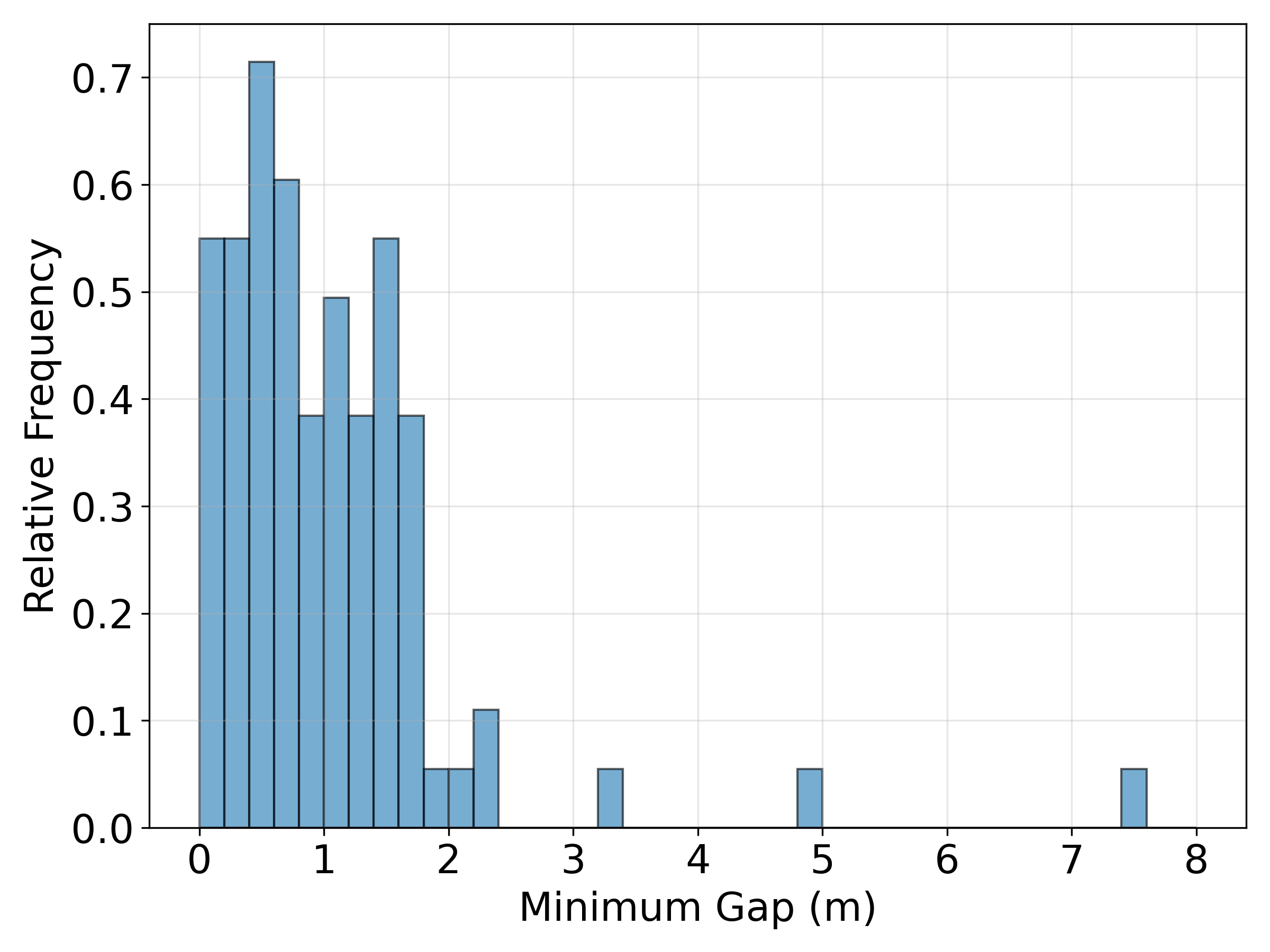}
        \caption{Auto-rickshaws (AR)}
        \label{fig:fig_MGc}
    \end{subfigure}

    \caption{Distributions of minimum gap for different vehicle categories under congested, lane-free traffic conditions.}
    \label{fig:fig_MG}
\end{figure}

\noindent
Both the form and statistical parameters of the minimum gap distributions depend strongly on the vehicle type. For two-wheelers, the distribution declines steadily with increasing gap, with most observations concentrated below $1.0~\mathrm{m}$ (Figure~\ref{fig:fig_MGa}). Both the distribution and the summary statistics (Table~\ref{tab:gap_stats}) indicate a strong tendency toward very small standstill spacing, with a median gap of about $0.55~\mathrm{m}$ and the central half of observations lying below $0.9~\mathrm{m}$. This compact spacing reflects riders’ willingness to stop close to the leader, supported by their small physical dimension and their ability to make fine lateral adjustments even in tightly packed queues. Cars display a more structured standstill spacing pattern. The distribution rises initially and reaches its highest concentration in the $0.4$–$0.8~\mathrm{m}$ range, followed by a gradual decline (Figure~\ref{fig:fig_MGb}). This behavior is consistent with the larger median minimum gap of $1.32~\mathrm{m}$ and a broader interquartile range extending up to nearly $2.0~\mathrm{m}$ (Table~\ref{tab:gap_stats}).These values indicate that cars maintain larger minimum gaps at standstill. Compared to two-wheelers, the larger physical dimensions and limited lateral maneuverability of cars constrain how closely they can position themselves under stopped conditions, resulting in more uniform and larger minimum gaps.\\

\noindent
Auto-rickshaws exhibit the widest spread in minimum gap behavior. The distribution remains relatively even across small gaps up to about $0.8~\mathrm{m}$, with a gradual reduction beyond this range (Figure~\ref{fig:fig_MGc}). This variability is reflected in the intermediate median gap of $0.83~\mathrm{m}$ and a wide dispersion across observations (Table~\ref{tab:gap_stats}). Such behavior is consistent with their intermediate size and operating characteristics, auto-rickshaws can stop very close to the leader when space permits, yet often maintain larger gaps when maneuverability or surrounding traffic conditions limit close positioning. As with desired time gaps, steady standstill observations for light and heavy commercial vehicles are limited, and estimating their statistical distribution does not make sense. Taken together, the minimum gap findings show how spacing preferences translate into physical queue formation once motion ceases. Smaller vehicles not only follow with shorter time gaps but also compress queues more tightly at standstill, increasing packing density. By contrast, larger vehicles maintain greater spacing even when stopped, which can influence jam density, queue length, and discharge characteristics. These minimum gap distributions provide key input for calibrating standstill spacing and queue formation mechanisms in microscopic simulations of mixed disordered traffic.

\subsubsection{Vehicle Dimension Distributions}
\noindent  
\noindent
Vehicle dimensions influence space utilization, maneuverability, and interaction patterns in disordered traffic. Moreover, vehicle width is one of the parameters used to identify potential leaders through lateral overlap with the follower vehicle, and it is also relevant for characterizing lateral shift behavior, indicating how much a vehicle can move laterally to improve speed while maintaining safe lateral spacing. Vehicle length is another important parameter for estimating desired time gap and minimum gap, both of which are crucial variables for simulating traffic systems. In addition, disordered traffic consists of a mix of vehicle types with varying dimensions, and even within a given vehicle class, substantial intra-class variability exists in vehicle dimensions. Hence, vehicle dimensions constitute one of the important parameters for simulating disordered traffic systems. In this section, vehicle length and width are initially obtained from axis-aligned bounding boxes generated using a YOLOv8-based detection and tracking framework. These bounding boxes remain aligned with the image frame and do not rotate with the vehicle. Let $(L, W)$ denote the length and width of the detected bounding box, and $(l, w)$ denote the true vehicle dimensions. When a vehicle travels with a non-zero heading angle, the detected bounding box encloses additional empty space, leading to an overestimation of the true dimensions $(L \neq l,\; W \neq w)$. This effect becomes more pronounced during lateral maneuvers and on curved road segments. To correct this distortion, the heading angle of vehicle $i$ at time $t$, denoted by $\theta_i(t)$, is estimated from successive trajectory positions using the longitudinal and lateral displacements $\Delta X_i(t)$ and $\Delta Y_i(t)$ as
\begin{equation}
\theta_i(t) = \tan^{-1}\!\left(\frac{\Delta Y_i(t)}{\Delta X_i(t)}\right).
\end{equation}
\begin{figure}
    \centering
    \includegraphics[width=0.5\linewidth]{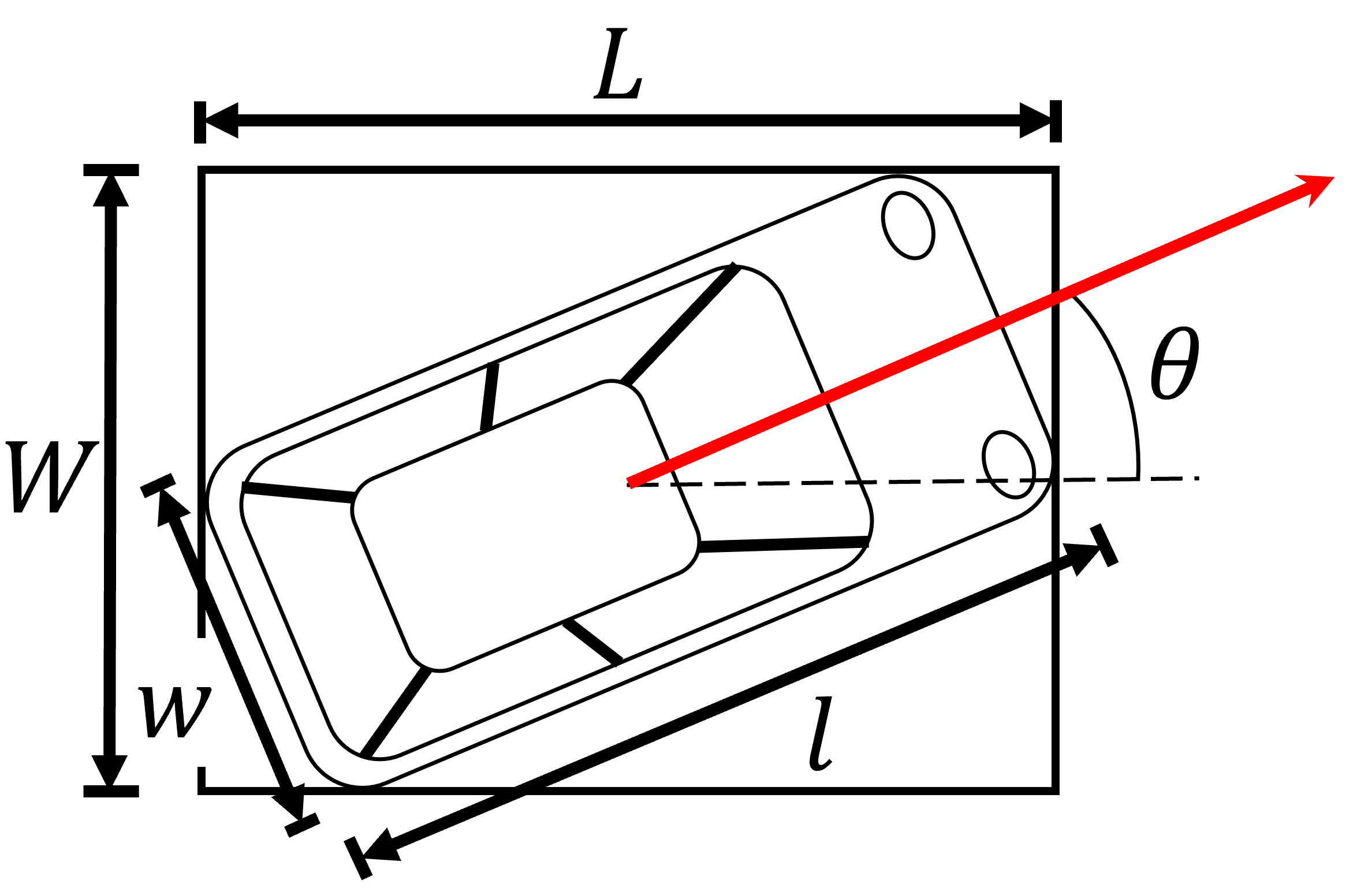}
    \caption{Vehicle Dimension Correction}
    \label{fig:Veh_Dim}
\end{figure}
Since part of the observed heading also arises from road curvature, all angles and corresponding vehicle dimensions are computed in the Cartesian coordinate system (Figure~\ref{fig:fig_3a}). Assuming that the detected bounding box fully circumscribes the vehicle, the true vehicle dimensions are recovered using the geometric correction proposed by \citet{RAJPUT2026Traj} (Figure~\ref{fig:Veh_Dim}). The relationships between the observed bounding box dimensions $(L, W)$, the true vehicle dimensions $(l, w)$, and the heading angle $\theta$ are given by

\begin{equation}
\begin{aligned}
L &= w |\sin\theta| + l |\cos\theta|, \\ \\
W &= w |\cos\theta| + l |\sin\theta|.
\end{aligned}
\end{equation}

Solving for the true width and length yields

\begin{equation}
\begin{aligned}
w = \frac{W|\cos\theta| - L|\sin\theta|}{\cos^2\theta - \sin^2\theta}, \\ \\
l = \frac{L|\cos\theta| - W|\sin\theta|}{\cos^2\theta - \sin^2\theta}.
\end{aligned}
\end{equation}

\noindent  
These corrections are applied to vehicle trajectories for which the estimated heading angle satisfies $|\theta| < \pi/4$. The present analysis focuses on forward-moving vehicles whose headings remain well within this range. As \(|\theta|\) approaches \(\pi/4\), the correction equations become increasingly sensitive to small angular variations because of a near ''zero divided by zero'' situation leading to unstable dimension estimates. Such near-singular cases were excluded in the filtered dataset used for the dimension analysis. Since heading angle estimates are influenced by localization noise and frame-to-frame fluctuations, the corrected dimensions for a given vehicle exhibit small temporal variation. To obtain stable estimates, the median of corrected length and width was taken over the full trajectory of each vehicle. The median is less sensitive to occasional outliers caused by transient detection errors, partial occlusions, or momentary bounding-box distortions, and therefore provides a more robust estimate of the true vehicle size. The resulting length-width distributions are then used to examine the physical heterogeneity of the traffic stream across vehicle classes.\\

\begin{figure}
    \centering
        \centering
    \begin{subfigure}{0.32\linewidth}
        \centering
        \includegraphics[width=\linewidth]{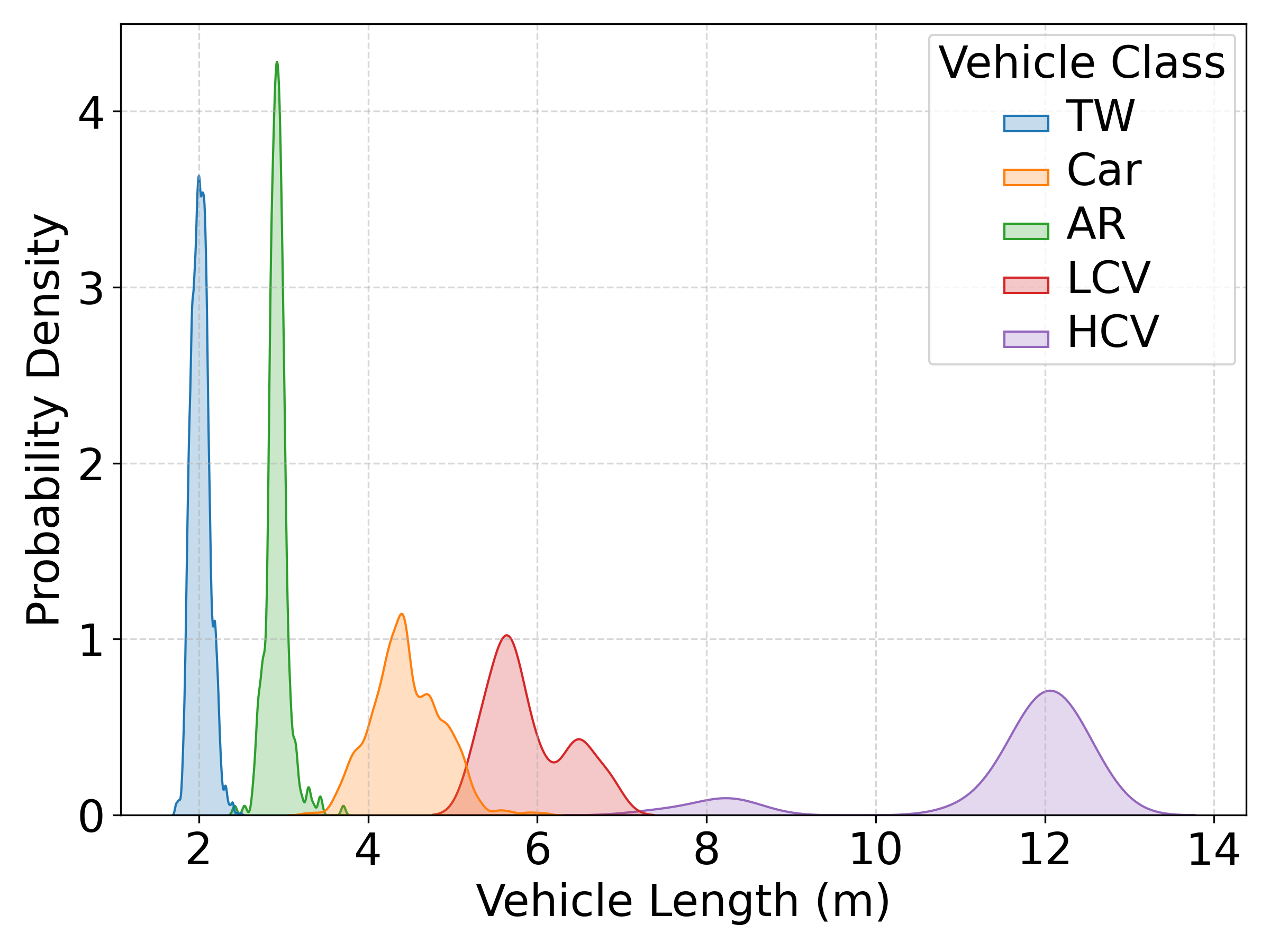}
        \caption{}
        \label{fig:fig_12a}
    \end{subfigure}
    \hfill
    \begin{subfigure}{0.32\linewidth}
        \centering

        \includegraphics[width=\linewidth]{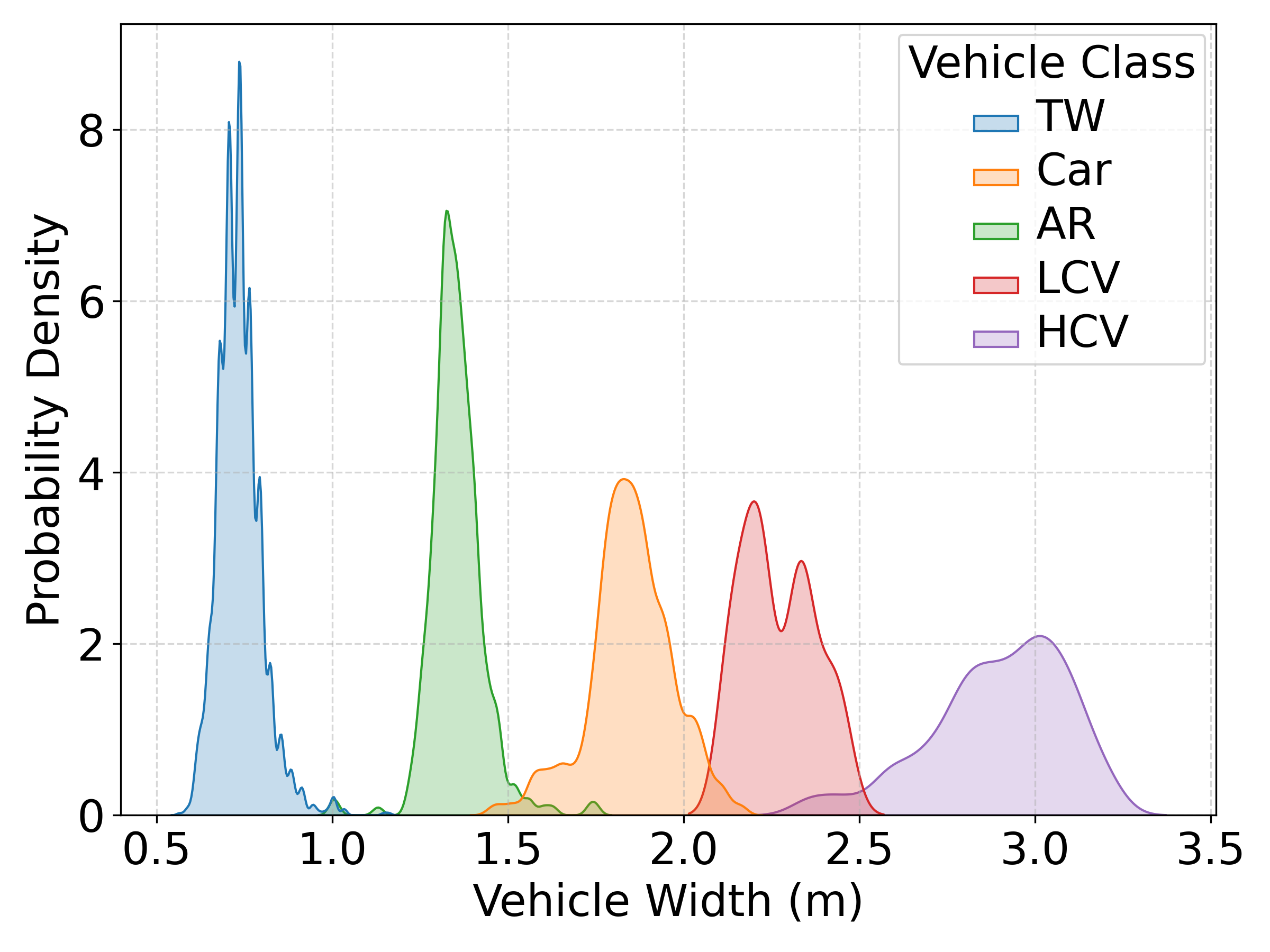}
        \caption{}
        \label{fig:fig_12b}
    \end{subfigure}
    \hfill
    \begin{subfigure}{0.32\linewidth}
        \centering
        \includegraphics[width=\linewidth]{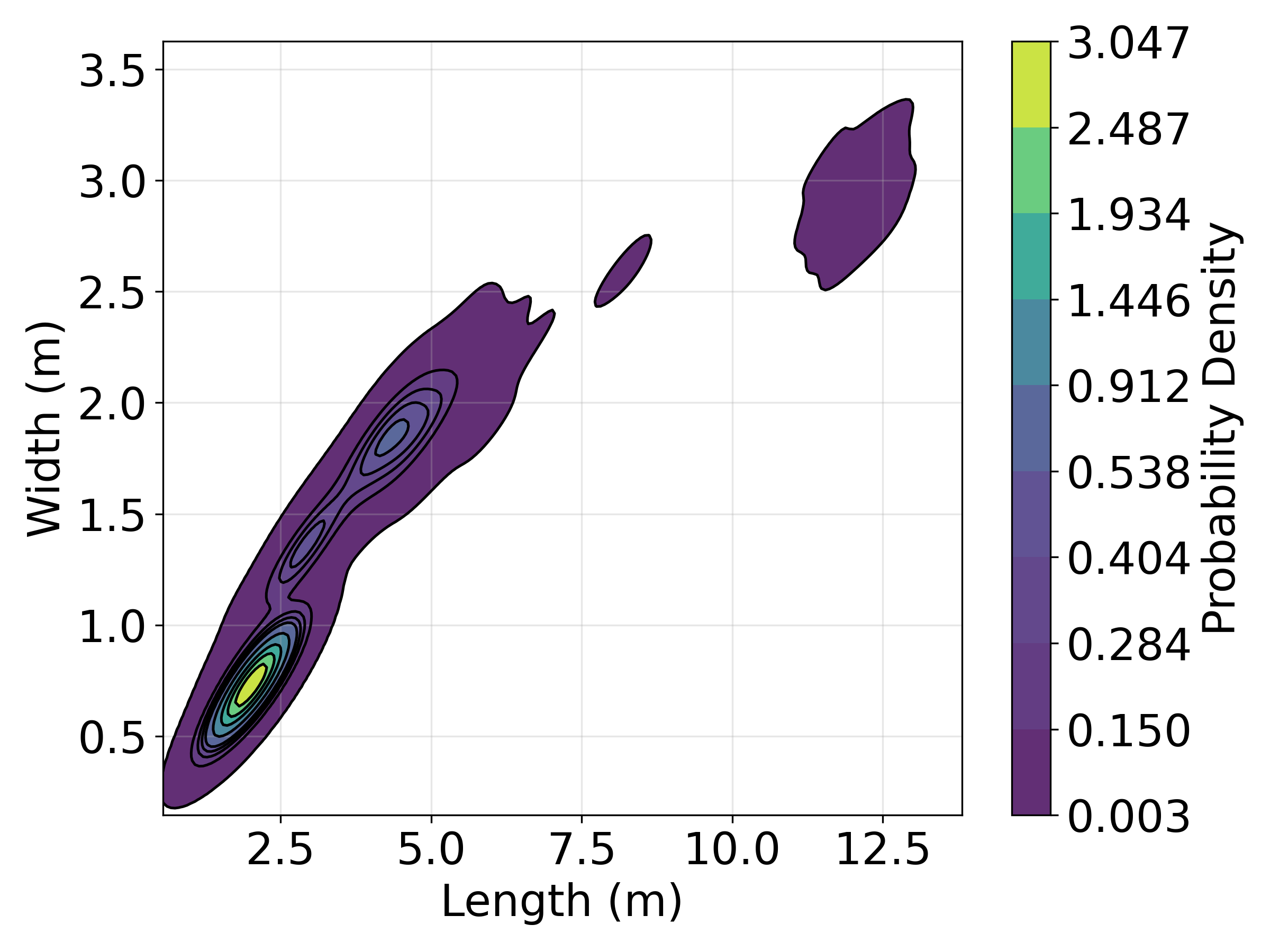}
        \caption{}
        \label{fig:fig_12c}
    \end{subfigure}

    \caption{Vehicle dimension distributions derived from median-based corrected bounding-box measurements: (a–b) marginal distributions of vehicle length and width for different vehicle classes, and (c) joint length–width density.}
    \label{fig:fig_12}
\end{figure}

\renewcommand{\arraystretch}{1.2}
\begin{table}
\centering
\caption{Summary statistics of the dimensions of vehicle categories sharing the same right of way on an urban road.}
\label{tab:dim_stats}
\begin{threeparttable}
\begin{tabular*}{\tblwidth}{@{}cllccccccccc@{}}
        \toprule
        Metric & Vehicle Type & Mean & Median & Standard Deviation & P25 & P75 & P90\\
        \midrule
        \multirow{5}{*}{\rotatebox[origin=c]{90}{Length (m)}}
        & TW  & 2.018 & 2.010 & 0.108 & 1.941 & 2.083 & 2.159\\
        & Car & 4.457 & 4.418 & 0.417 & 4.189 & 4.738 & 5.007\\
        & AR  & 2.922 & 2.915 & 0.131 & 2.854 & 2.978 & 3.043\\
        & LCV & 5.893 & 5.661 & 0.495 & 5.615 & 6.234 & 6.500\\
        & HCV & 11.563 & 11.972 & 1.335 & 11.749 & 12.243 & 12.431\\
        \midrule
        \multirow{5}{*}{\rotatebox[origin=c]{90}{Width (m)}}
        & TW  & 0.736 & 0.733 & 0.063 & 0.699 & 0.766 & 0.808\\
        & Car & 1.848 & 1.850 & 0.120 & 1.784 & 1.918 & 2.000\\
        & AR  & 1.354 & 1.350 & 0.078 & 1.317 & 1.386 & 1.436\\
        & LCV & 2.269 & 2.229 & 0.107 & 2.193 & 2.332 & 2.406\\
        & HCV & 2.896 & 2.923 & 0.196 & 2.785 & 3.038 & 3.111\\
        \bottomrule
    \end{tabular*}
\setlength{\parindent}{0pt}
\footnotesize
Note: P25, P75, and P90 denote the 25th, 75th, and 90th percentile values of the observed distributions, respectively.

\end{threeparttable}
\end{table}

\begin{figure}
    \centering
        \centering
        \begin{subfigure}{0.32\linewidth}
        \centering
        \includegraphics[width=\linewidth, height=0.7\linewidth]{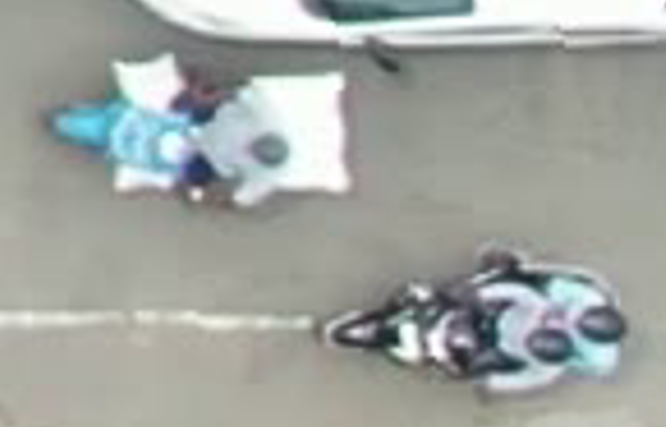}
        \caption{Scooter}
        \label{fig:VC_TW_scooter}
        \end{subfigure}
        \hfill
        \begin{subfigure}{0.32\linewidth}
            \centering
            \includegraphics[width=\linewidth, height=0.7\linewidth]{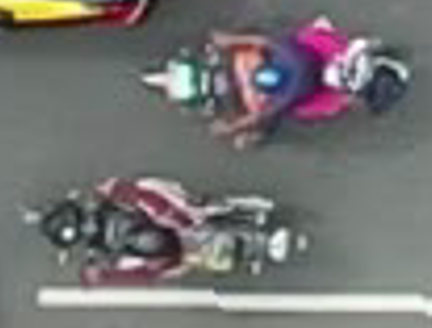}
            \caption{Motorcycle}
            \label{fig:VC_TW_motorcycle}
        \end{subfigure}
        \hfill
        \begin{subfigure}{0.32\linewidth}
            \centering
            \includegraphics[width=\linewidth]{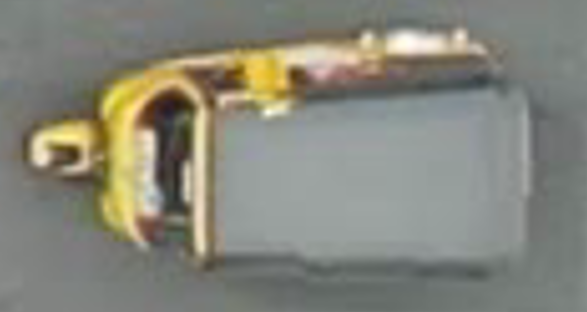}
            \caption{Auto-rickshaw}
            \label{fig:VC_AR}
        \end{subfigure}

    \vspace{3mm}

        \begin{subfigure}{0.32\linewidth}
        \centering
        \reflectbox{\includegraphics[width=\linewidth]{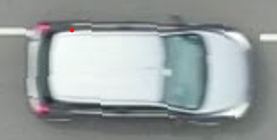}}
        \caption{Hatchback}
        \label{fig:VC_car_hatchback}
        \end{subfigure}\hfill
        \begin{subfigure}{0.32\linewidth}
            \centering
            \includegraphics[width=\linewidth]{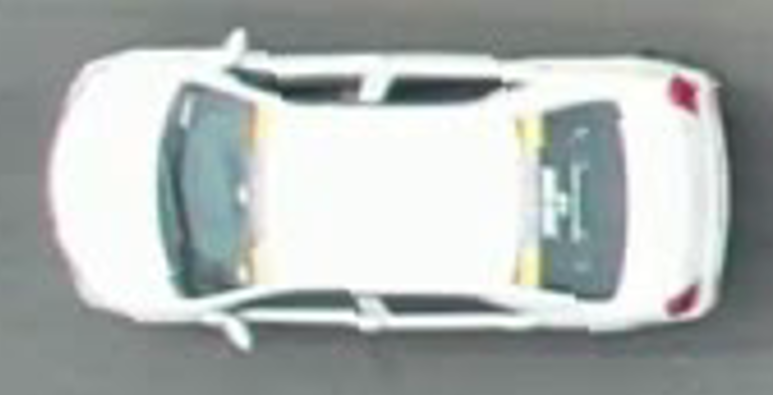}
            \caption{Sedan}
            \label{fig:VC_car_sedan}
        \end{subfigure}\hfill
        \begin{subfigure}{0.32\linewidth}
            \centering
            \includegraphics[width=\linewidth]{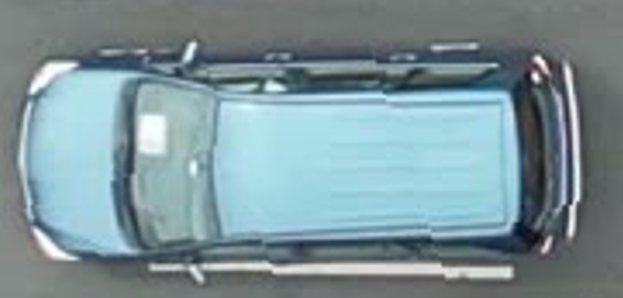}
            \caption{Sport-utility vehicle}
            \label{fig:VC_car_suv}
        \end{subfigure}

    \vspace{3mm}

        \begin{subfigure}{0.32\linewidth}
        \centering
        \includegraphics[width=\linewidth]{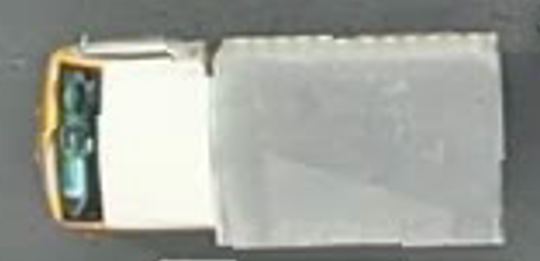}
        \caption{Small goods carrier}
        \label{fig:VC_car_goods}
        \end{subfigure}\hfill
        \begin{subfigure}{0.32\linewidth}
            \centering
            \includegraphics[width=\linewidth]{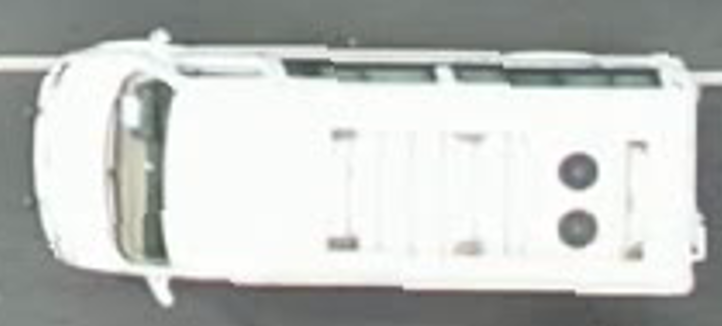}
            \caption{Mini bus}
            \label{fig:VC_LCV_bus}
        \end{subfigure}\hfill
        \begin{subfigure}{0.32\linewidth}
            \centering
            \includegraphics[width=\linewidth]{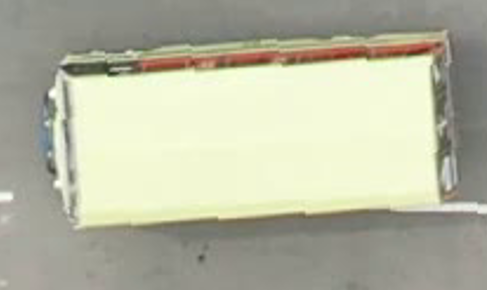}
            \caption{Passenger van}
            \label{fig:VC_LCV_van}
        \end{subfigure}

    \vspace{3mm}

        \begin{subfigure}{0.32\linewidth}
        \centering
        \includegraphics[width=\linewidth]{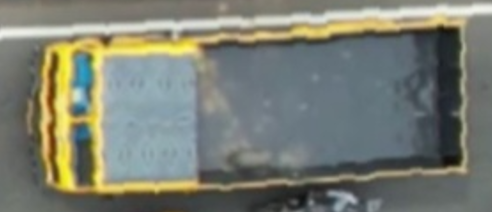}
        \caption{Cargo truck}
        \label{fig:VC_HCV_small_truck}
        \end{subfigure}\hfill
        \begin{subfigure}{0.32\linewidth}
        \centering
        \includegraphics[width=\linewidth]{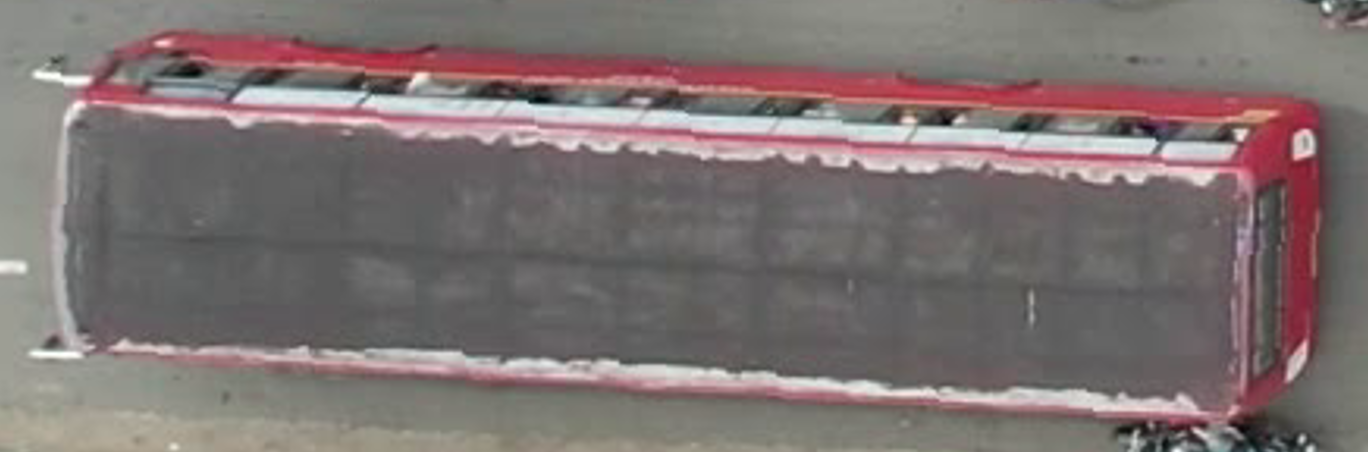}
        \caption{City bus}
        \label{fig:VC_HCV_bus}
        \end{subfigure}\hfill
        \begin{subfigure}{0.32\linewidth}
            \centering
            \reflectbox{\includegraphics[width=\linewidth]{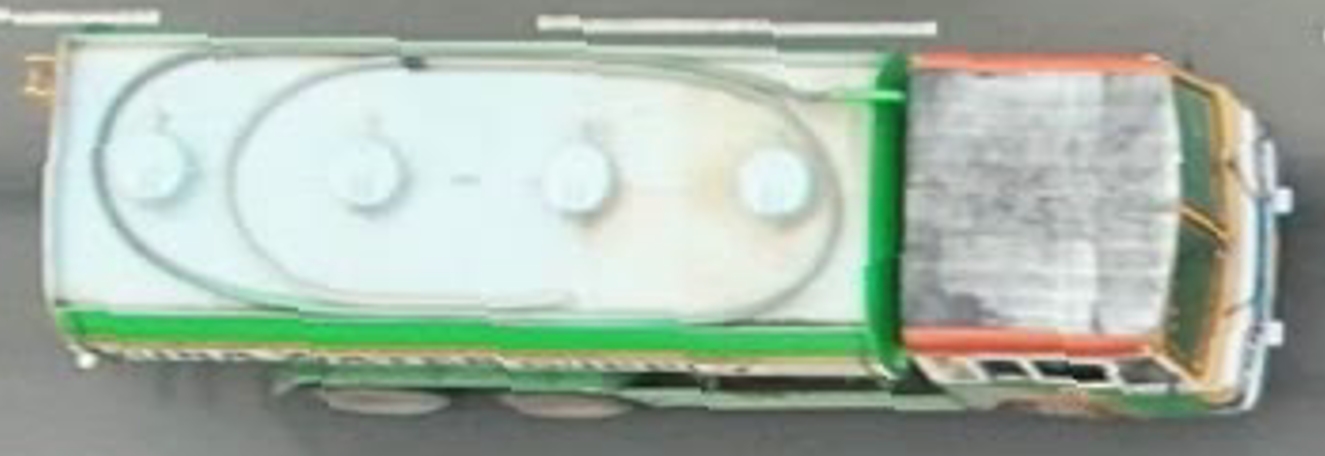}}
            \caption{Tamker truck}
            \label{fig:VC_HCV_truck}
        \end{subfigure}

    \caption{Illustration of geometric heterogeneity across vehicle categories observed in the study area. 
Panels (a)–(b) show representative two-wheelers (scooters and motorcycles); 
(c) depicts the dominant auto-rickshaw type observed in the traffic stream; 
(d)–(g) represent common subcategories within the car class; 
(h)–(i) correspond to light commercial vehicles; 
and (j)–(l) illustrate examples of heavy commercial vehicles.}

    \label{fig:fig_VC}
\end{figure}

\noindent
The marginal distributions of vehicle length and width (Figures~\ref{fig:fig_12a} and~\ref{fig:fig_12b}, Table~\ref{tab:dim_stats}) indicate pronounced geometric heterogeneity both within and across vehicle categories sharing the same roadway. Two-wheelers exhibit relatively limited variation, with lengths concentrated around a median of $2.01$~m and widths near $0.73$~m. The observed spread reflects the presence of both scooters and motorcycles (Figure~\ref{fig:VC_TW_scooter} and \ref{fig:VC_TW_motorcycle}), with scooters generally occupying the shorter end of the length distribution. Auto-rickshaws show a narrow and well-defined size range, with median dimensions of approximately $2.92$~m in length and $1.35$~m in width. The majority of observations correspond to standard passenger auto-rickshaws with seating capacity for three passengers and a driver (Figure~\ref{fig:VC_AR}). A small number of larger three-wheelers appear in the upper tail of the distributions, which are primarily associated with goods-carrying variants. In contrast, cars exhibit substantially higher variability, with a mean length of $4.46$~m and a standard deviation of $0.42$~m, and a mean width of $1.85$~m with a standard deviation of $0.12$~m (Table~\ref{tab:dim_stats}). This spread reflects the coexistence of multiple subcategories, including compact hatchbacks, sedans, sport-utility vehicles, and small goods carriers (Figures~\ref{fig:VC_car_hatchback}, \ref{fig:VC_car_sedan}, \ref{fig:VC_car_suv}, and~\ref{fig:VC_car_goods}), resulting in a broad interquartile range and an extended upper tail.\\

\noindent
Light commercial vehicles occupy an intermediate geometric range between cars and heavy commercial vehicles, with median lengths near $5.66$~m and widths around $2.23$~m. This category includes a mix of passenger vans, small buses, and medium-sized goods carriers (Figure~\ref{fig:VC_LCV_bus} and \ref{fig:VC_LCV_van}), leading to noticeable dispersion in both dimensions. The length distributions of cars and light commercial vehicles also exhibit partial overlap (Figure~\ref{fig:fig_12a}). This overlap primarily arises because the lengths of mini buses and passenger vans are comparable to those of sport utility vehicles \ref{fig:VC_car_suv} and small goods carriers ~\ref{fig:VC_car_goods}. Heavy commercial vehicles form the largest size class, with median lengths around $11.97$~m and widths around $2.92$~m (Table~\ref{tab:dim_stats}). The bimodal length distribution reflects the heterogeneous composition of this class. The dominant peak and upper percentiles are associated with large buses and multi-axle trucks (Figure~\ref{fig:VC_HCV_bus} and \ref{fig:VC_HCV_truck}), whereas the secondary peak around 8~m corresponds primarily to smaller cargo trucks (Figure~\ref{fig:VC_HCV_small_truck}). Overall, the results highlight that geometric heterogeneity in disordered traffic is not only driven by differences between vehicle classes but also by substantial variability within each category. Such variation directly influences lateral space usage, gap availability, and interaction dynamics, underscoring the need to account for realistic vehicle dimension distributions in both microscopic modeling and macroscopic traffic flow analysis.\\

\noindent The joint length–width distribution (Figure~\ref{fig:fig_12c}) represents the joint probability density of observing specific length–width pairs in the traffic stream, without explicit separation by vehicle class. Regions of high probability density correspond to dominant size combinations identified in the marginal distributions, whereas extended low-density regions reflect the presence of larger and more diverse vehicle geometries. Rather than exhibiting sharp class boundaries, the joint distribution emphasizes the continuous nature of geometric heterogeneity in disordered traffic, where vehicles of different sizes overlap across both length and width ranges. This overlap represents one of the factors contributing to the dispersion observed in macroscopic flow density–density and speed–density relationships under lane-free conditions (Figures~\ref{fig:fig_7} and~\ref{fig:fig_8}). From a traffic dynamics perspective, the observed geometric heterogeneity directly influences vehicle interactions across both microscopic and macroscopic scales. Variations in vehicle size affect how space is utilized locally, with smaller vehicles able to occupy narrower lateral gaps and adjust position more flexibly in dense traffic, while larger vehicles require greater spatial clearance and impose stronger constraints on the surrounding traffic stream. In traffic simulation and modeling, the empirically derived two-dimensional size distributions provide realistic geometric inputs for identifying effective leaders, estimating desired time gaps and minimum standstill distances, and representing lateral shift behavior in mixed urban traffic streams where longitudinal and lateral interactions are strongly coupled.

\subsubsection{Speed and Acceleration Trends}

\begin{figure}
    \centering
    \begin{subfigure}{0.48\linewidth}
        \centering
        \includegraphics[width=\linewidth]{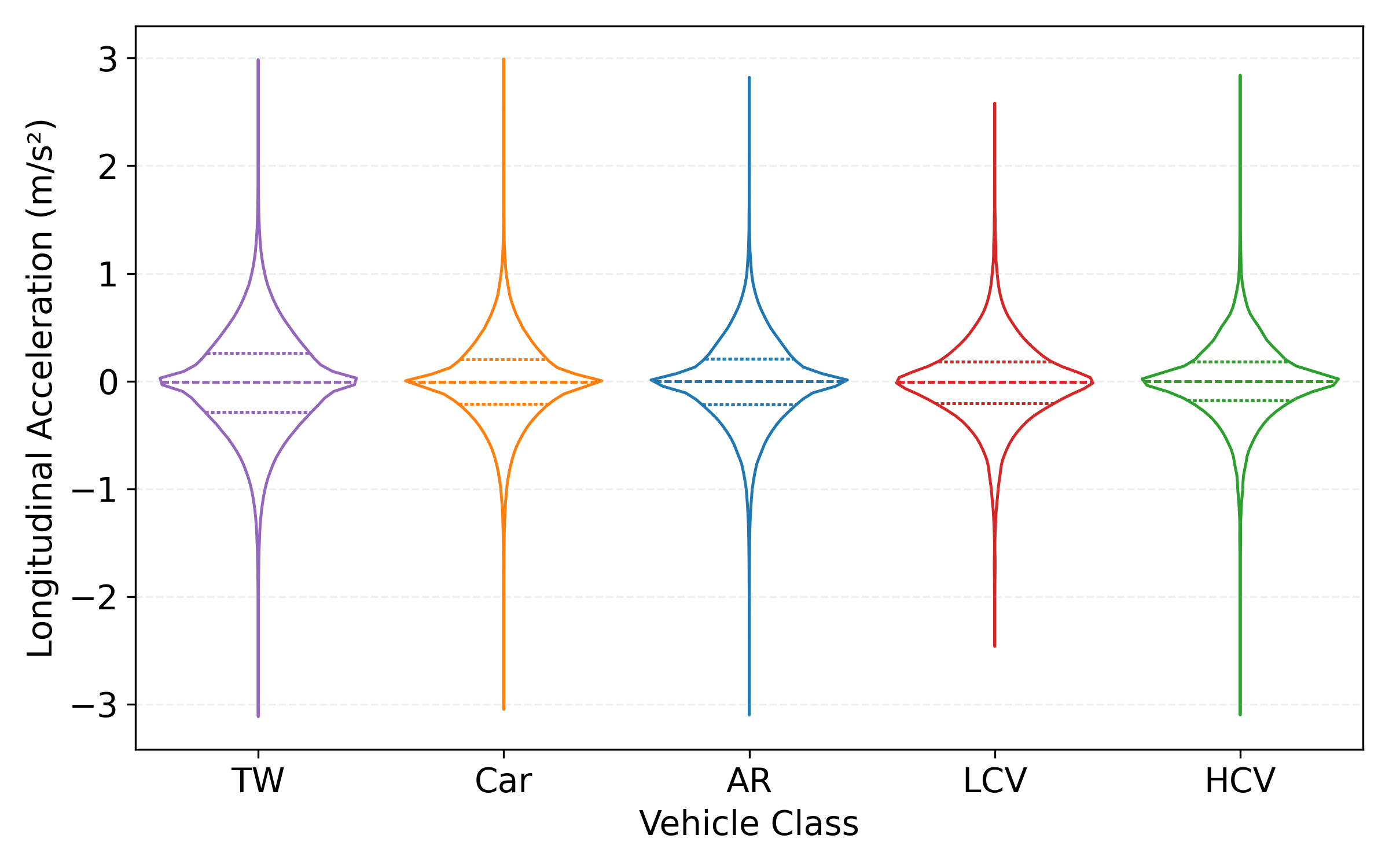}
        \caption{Longitudinal acceleration distributions}
        \label{fig:fig_13a}
    \end{subfigure}\hfill
    \begin{subfigure}{0.48\linewidth}
        \centering
        \includegraphics[width=\linewidth]{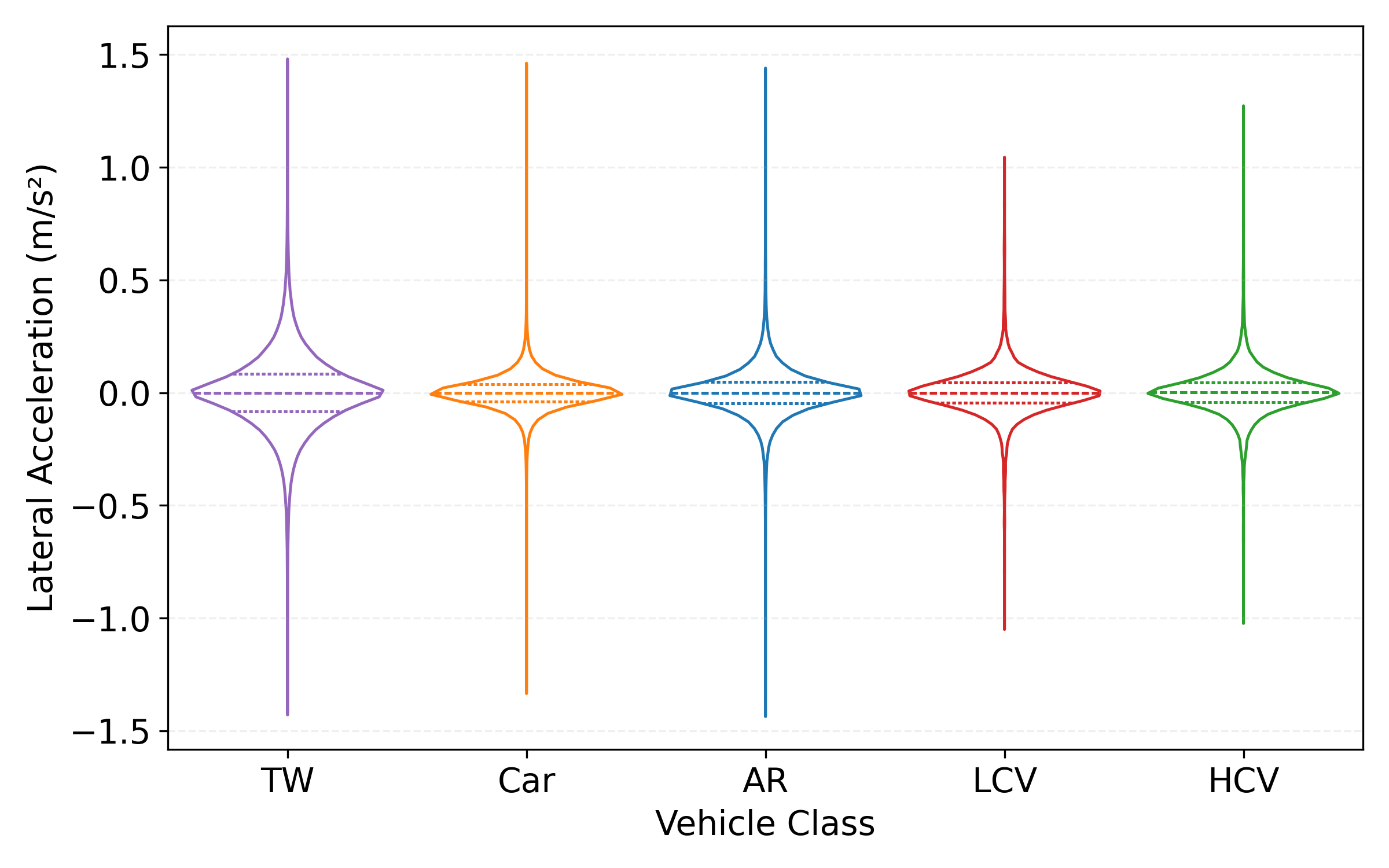}
        \caption{Lateral acceleration distributions}
        \label{fig:fig_13b}
    \end{subfigure}

    \vspace{3mm}

    \begin{subfigure}{0.48\linewidth}
        \centering
        \includegraphics[width=\linewidth]{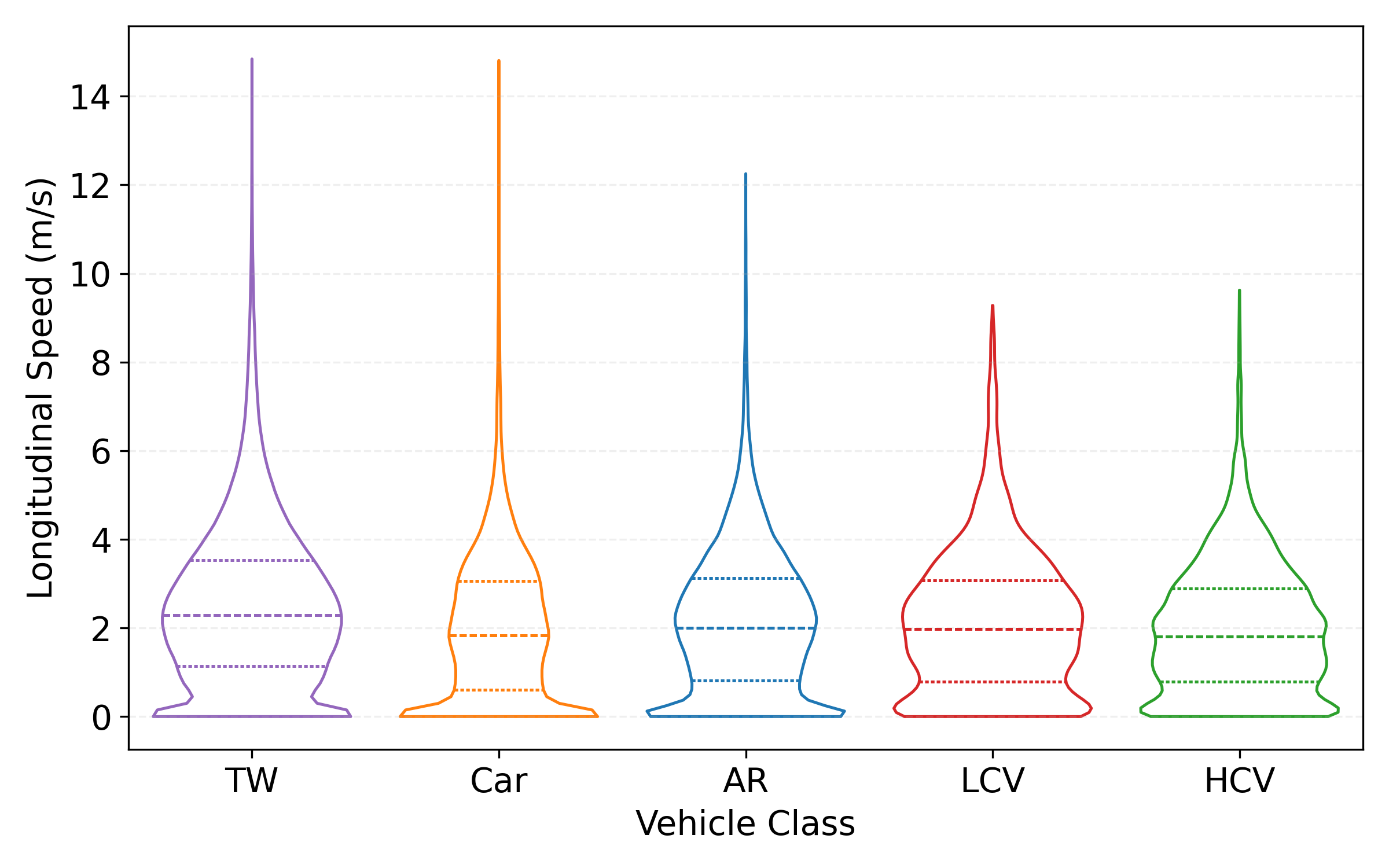}
        \caption{Longitudinal speed distributions}
        \label{fig:fig_13c}
    \end{subfigure}\hfill
    \begin{subfigure}{0.48\linewidth}
        \centering
        \includegraphics[width=\linewidth]{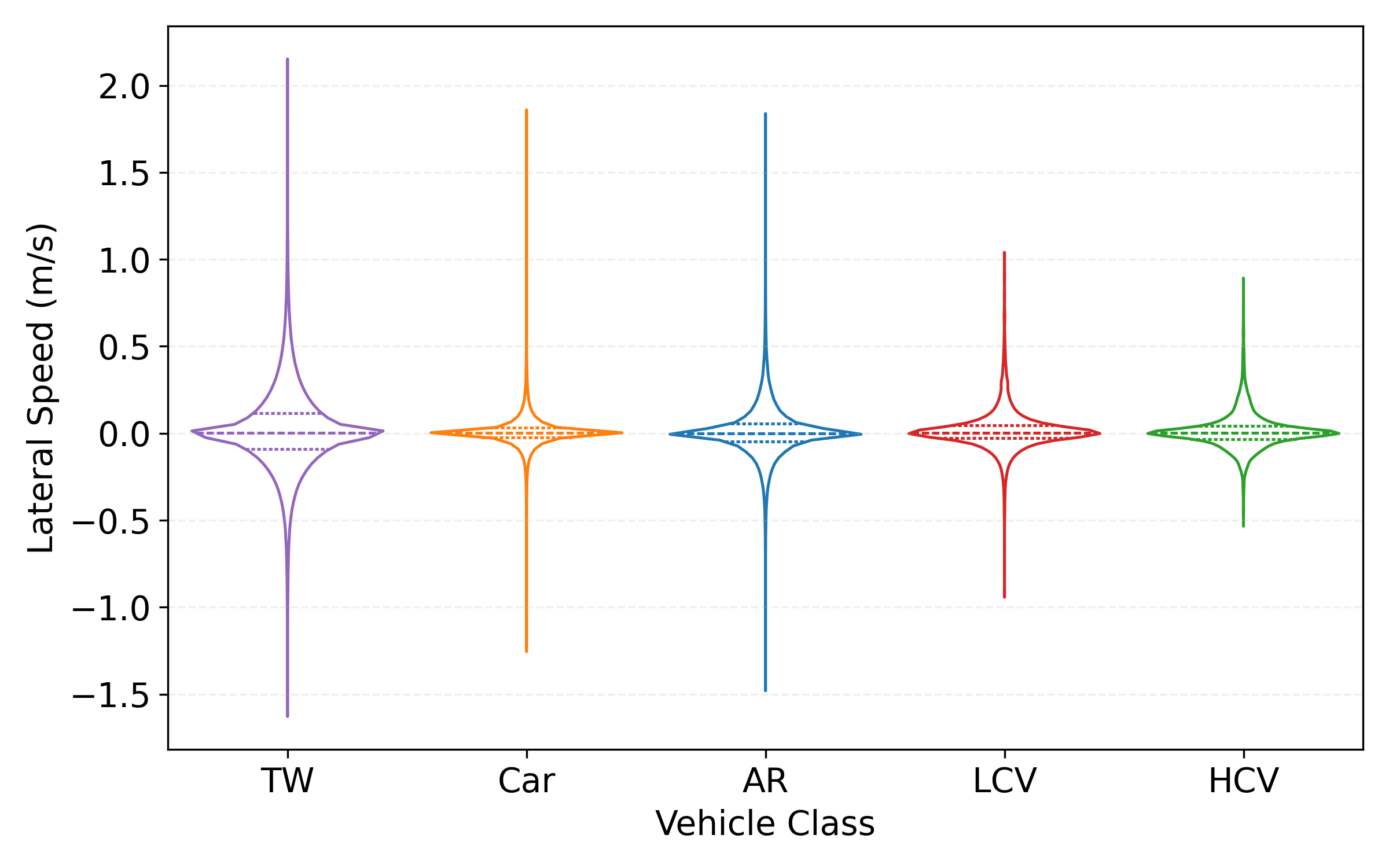}
        \caption{Lateral speed distributions}
        \label{fig:fig_13d}
    \end{subfigure}
        \caption{
        Distributions of kinematic behavior for different vehicle classes under congested, disordered traffic conditions. The violin plots illustrate the spread and variability of, highlighting differences in motion patterns across vehicle types. Further, the horizontal line in each violin plot represent the respective first, second and third quartiles.
        }

\label{fig:fig_13}
\end{figure}
\noindent In disordered traffic, drivers constantly adjust their speed and acceleration, not only in the direction of travel but also laterally, in response to surrounding vehicles and available gaps. These adjustments reflect how traffic evolves under different traffic conditions, and vehicles interact when lane discipline is weak. Studying speed and acceleration trends, therefore, helps capture the dynamics of traffic behavior that cannot be inferred solely from vehicle geometry. In the following analysis, longitudinal and lateral speed and acceleration patterns are examined for congested traffic states to understand how vehicles move, slow down, and maneuver in lane-free conditions. Under congested conditions, vehicle motion is characterized by persistently low longitudinal speeds and small-amplitude kinematic adjustments. As summarized in Table~\ref{tab:kinematic_stats}, mean longitudinal speeds remain below $2.6~\mathrm{m\,s^{-1}}$ for all vehicle categories, with median values clustered between $1.8$ and $2.3~\mathrm{m\,s^{-1}}$, confirming sustained congestion across the traffic stream. Two-wheelers exhibit the highest mean longitudinal speed ($2.53~\mathrm{m\,s^{-1}}$), followed by auto-rickshaws ($2.14~\mathrm{m\,s^{-1}}$), whereas cars and heavy commercial vehicles travel slowly, with mean speeds of $2.03~\mathrm{m\,s^{-1}}$ and $1.96~\mathrm{m\,s^{-1}}$, respectively. This systematic reduction in speed for larger vehicles is consistent with their greater physical dimensions and reduced maneuverability.\\

\begin{table}
\centering
\caption{Statistical summary of kinematic variables for different vehicle categories.}
\label{tab:kinematic_stats}
\begin{tabular*}{\tblwidth}{@{}llcccccccc@{}}
\toprule
Metric & Vehicle Category & Mean & SD & Median & P25 & P75 & P90 & Minimum & Maximum \\
\midrule

\multirow{5}{*}{\makecell[l]{Longitudinal\\Acceleration ($\mathrm{m}/\mathrm{s}^2$)}} 
& TW  & -0.014 & 0.475 & -0.009 & -0.286 & 0.262 & 0.569 & -3.115 & 2.986 \\
& CAR & -0.010 & 0.389 & -0.005 & -0.212 & 0.202 & 0.463 & -3.047 & 2.992 \\
& AR  & -0.009 & 0.395 & -0.005 & -0.219 & 0.209 & 0.472 & -3.101 & 2.823 \\
& LCV & -0.014 & 0.383 & -0.007 & -0.205 & 0.181 & 0.427 & -2.463 & 2.581 \\
& HCV & -0.007 & 0.367 & -0.001 & -0.177 & 0.179 & 0.419 & -3.100 & 2.841 \\
\midrule

\multirow{5}{*}{\makecell[l]{Lateral\\Acceleration ($\mathrm{m}/\mathrm{s}^2$)}} 
& TW  & 0.002 & 0.175 &  0.000 & -0.083 & 0.085 & 0.195 & -1.429 & 1.481 \\
& Car & 0.000 & 0.078 & -0.000 & -0.038 & 0.038 & 0.084 & -1.335 & 1.462 \\
& AR  & 0.001 & 0.112 &  0.000 & -0.047 & 0.048 & 0.114 & -1.437 & 1.440 \\
& LCV & 0.001 & 0.104 & -0.000 & -0.045 & 0.046 & 0.100 & -1.050 & 1.044 \\
& HCV & 0.002 & 0.102 &  0.001 & -0.042 & 0.046 & 0.103 & -1.023 & 1.273 \\
\midrule

\multirow{5}{*}{\makecell[l]{Longitudinal\\Speed ($\mathrm{m}/\mathrm{s}$)}} 
& TW  & 2.531 & 1.873 & 2.291 & 1.146 & 3.526 & 4.907 & 0.0 & 14.84 \\
& CAR & 2.031 & 1.682 & 1.839 & 0.604 & 3.062 & 4.126 & 0.0 & 14.807 \\
& AR  & 2.141 & 1.640 & 2.009 & 0.816 & 3.119 & 4.256 & 0.0 & 12.249 \\
& LCV & 2.117 & 1.635 & 1.972 & 0.789 & 3.073 & 4.141 & 0.0 & 9.276 \\
& HCV & 1.962 & 1.486 & 1.803 & 0.785 & 2.885 & 3.923 & 0.0 & 9.622 \\
\midrule

\multirow{5}{*}{\makecell[l]{Lateral\\Speed ($\mathrm{m}/\mathrm{s}$)}} 
& TW  & 0.018 & 0.248 & 0.002 & -0.092 & 0.116 & 0.298 & -1.629 & 2.153 \\
& CAR & 0.007 & 0.090 & 0.001 & -0.023 & 0.031 & 0.090 & -1.256 & 1.860 \\
& AR  & 0.009 & 0.152 & 0.000 & -0.048 & 0.056 & 0.167 & -1.481 & 1.838 \\
& LCV & 0.016 & 0.124 & 0.003 & -0.029 & 0.045 & 0.128 & -0.943 & 1.040 \\
& HCV & 0.012 & 0.103 & 0.002 & -0.033 & 0.042 & 0.126 & -0.534 & 0.892 \\

\bottomrule
\end{tabular*}
\end{table}

\noindent Despite these low average speeds, the upper tails of the longitudinal speed distributions reveal marked differences in residual mobility across vehicle classes. Two-wheelers and cars attain substantially higher maximum longitudinal speeds, reaching $14.84~\mathrm{m\,s^{-1}}$ and $14.81~\mathrm{m\,s^{-1}}$, respectively, whereas auto-rickshaws, light commercial vehicles, and heavy commercial vehicles exhibit markedly lower maxima of $12.25~\mathrm{m\,s^{-1}}$, $9.28~\mathrm{m\,s^{-1}}$, and $9.62~\mathrm{m\,s^{-1}}$. These differences indicate that, even under congested conditions, two-wheelers and cars are occasionally able to exploit transient gaps to accelerate over short distances. In contrast, the lower maximum speeds of larger and heavier vehicles reflect stronger physical and operational constraints, and limited lateral repositioning, which restrict their ability to capitalize on brief space availability. The dispersion in longitudinal speed is substantial, particularly for smaller vehicles. Two-wheelers show a standard deviation of $1.87~\mathrm{m\,s^{-1}}$, compared to $1.68~\mathrm{m\,s^{-1}}$ for cars, indicating intermittent short forward movements even under dense conditions. These fluctuations are also reflected in the upper percentiles, with the $90^{\mathrm{th}}$ percentile longitudinal speed reaching $4.91~\mathrm{m\,s^{-1}}$ for two-wheelers, compared to $4.13~\mathrm{m\,s^{-1}}$ for cars, highlighting the persistence of stop-and-go dynamics in lane-free congestion. Longitudinal accelerations across all vehicle classes are tightly centered around zero, with mean values close to $0~\mathrm{m\,s^{-2}}$ and medians within $\pm0.01~\mathrm{m\,s^{-2}}$, indicating gradual speed regulation rather than abrupt braking or acceleration. Two-wheelers again exhibit the highest acceleration variability ($0.48~\mathrm{m\,s^{-2}}$), compared to cars ($0.39~\mathrm{m\,s^{-2}}$) and heavy commercial vehicles ($0.37~\mathrm{m\,s^{-2}}$), reflecting their tendency to make frequent micro-adjustments to navigate small gaps efficiently, as also evident from the broader distributions in Figure~\ref{fig:fig_13a} and Figure~\ref{fig:fig_13c}.\\

\noindent Lateral motion, although smaller in magnitude than longitudinal movement, remains a persistent feature of disordered traffic. The median lateral speeds for all vehicle categories are close to zero (below $0.01~\mathrm{m\,s^{-1}}$), indicating the absence of sustained lateral drift. However, the variability around these medians reveals continuous side-to-side repositioning. Two-wheelers exhibit the highest lateral speed variability, with a standard deviation of $0.25~\mathrm{m\,s^{-1}}$ and a $90^{\mathrm{th}}$ percentile value approaching $0.30~\mathrm{m\,s^{-1}}$. Auto-rickshaws also show notable lateral motion, with a standard deviation of approximately $0.15~\mathrm{m\,s^{-1}}$. In contrast, cars display much narrower lateral speed distributions, with a standard deviation of about $0.09~\mathrm{m\,s^{-1}}$, while light and heavy commercial vehicles exhibit similarly constrained ranges. A similar pattern is observed for lateral accelerations. Although mean and median lateral accelerations remain close to zero for all vehicle types, two-wheelers show the largest variability ($\sigma = 0.18~\mathrm{m\,s^{-2}}$), compared with $0.08~\mathrm{m\,s^{-2}}$ for cars and approximately $0.10~\mathrm{m\,s^{-2}}$ for heavy commercial vehicles. These fluctuations indicate frequent low-amplitude steering corrections rather than abrupt lateral maneuvers.\\

\noindent The pronounced lateral variability among smaller vehicles reflects their ability to exploit narrow gaps and continuously adjust position to access locally available space in the absence of strict lane discipline. Larger vehicles, such as buses and trucks, require greater lateral clearance and are constrained by vehicle width and turning radius, which limits lateral maneuverability and results in narrower distributions of lateral speed and acceleration. Consequently, traffic streams with comparable densities may exhibit distinct macroscopic speed and flow characteristics, driven by differences in lateral space utilization and vehicle composition. The observed longitudinal and lateral speed and acceleration distributions serve as error measures in the objective function for calibrating model parameters and validating microscopic traffic models to replicate field conditions, while also providing a quantitative basis for developing congestion management strategies tailored to mixed, lane-free urban traffic.

\section{Conclusions}
\noindent This study presented a unified macroscopic--microscopic investigation of traffic flow dynamics under disordered, lane-free conditions using high-resolution UAV-based trajectory data collected over an extended urban arterial segment. By combining geometric trajectory refinement, validation against manually annotated ground truth, and systematic space-time aggregation, the analysis established a robust empirical foundation for examining both aggregate and vehicle-level behavior under sustained congestion (90th percentile of speed below \unit[5]{m/s} in all vehicle classes). At the macroscopic level, a two-dimensional extension of Edie’s framework was implemented to explicitly account for longitudinal and lateral vehicle motion. The resulting fundamental relationships demonstrate that traffic states under weak lane discipline cannot be adequately represented by one-dimensional density-flow formulations. The two-dimensional fundamental relationships were observed over densities up to 0.32 veh/m$^{2}$. Longitudinal speeds reached approximately 14--16 m/s in low-density conditions (<0.03 veh/m$^{2}$) and decreased rapidly with increasing density, with speeds typically below 2 m/s for densities exceeding 0.20 veh/m$^{2}$. Longitudinal flow density reached peak values of approximately 0.7 veh/(m$\cdot$s) at intermediate densities ($\approx$0.05--0.12 veh/m$^{2}$). Lateral motion remained non-negligible over a wide density range, with lateral velocities between $-0.3$ and $0.6$ m/s at low densities and lateral flow density values up to $\pm 0.05$ veh/(m$\cdot$s), highlighting the persistent role of lateral redistribution in lane-free traffic. The identified traffic states show that macroscopic behavior is governed not solely by density but by the coupled effects of vehicle composition, spatial occupancy, and maneuverability, particularly the dominance of two-wheelers in high-density regimes.\\

\noindent Congestion propagation was quantified directly from trajectory-derived speed fields using a cross-correlation approach applied to virtual detectors at 50 m intervals. Coherent upstream stop-and-go waves were observed even under strongly heterogeneous and lane-free conditions. The estimated upstream propagation speeds, ranging from approximately $3.69$ to $4.56~\mathrm{m/s}$, are consistent with values reported for ordered, lane-based traffic, indicating that large-scale kinematic wave dynamics are preserved despite fundamentally different interaction structures at the microscopic level. This provides empirical evidence that congestion wave propagation is a robust macroscopic property, largely independent of lane discipline.
At the microscopic level, steady-state follower-leader identification enabled the estimation of desired time gaps and minimum standstill gaps across vehicle classes. The mean desired time gap was 0.674 s for two-wheelers, 1.339 s for cars, 0.916 s for auto-rickshaws, 1.248 s for light commercial vehicles, and 1.856 s for heavy commercial vehicles. Corresponding mean minimum standstill gaps were 0.728 m (two-wheelers), 1.464 m (cars), 1.043 m (auto-rickshaws), 1.199 m (LCV), and 0.833 m (HCV). These results quantify strong inter-class heterogeneity in spacing behavior and demonstrate the role of vehicle size and maneuverability in determining packing density and sustained flow under congestion. Vehicle dimension distributions further quantified geometric heterogeneity in the traffic stream. Heading-corrected measurements yielded mean vehicle lengths of 2.02 m (two-wheelers), 4.46 m (cars), 2.92 m (auto-rickshaws), 5.89 m (LCV), and 11.56 m (HCV), with corresponding mean widths of 0.74 m, 1.85 m, 1.35 m, 2.27 m, and 2.90 m, respectively. These large inter-class differences, together with measurable intra-class variability, explain the observed dispersion in macroscopic flow density-density relationships. Smaller vehicles are able to exploit narrow lateral gaps and sustain mobility under dense conditions, whereas larger vehicles impose stronger geometric constraints on surrounding traffic.\\

\noindent Under congested conditions, mean longitudinal speeds remained below 2.6 m/s for all vehicle classes, with two-wheelers exhibiting the highest mean speed (2.53 m/s) and heavy commercial vehicles the lowest (1.96 m/s). Maximum observed speeds reached 14.84 m/s for two-wheelers and 14.81 m/s for cars but were substantially lower (9-12 m/s) for larger vehicles, indicating limited residual mobility in the short episodes without congestion. Longitudinal accelerations were centered near zero (medians within $\pm 0.01$ m/s$^{2}$), confirming gradual speed regulation, while acceleration variability was highest for two-wheelers (0.48 m/s$^{2}$) compared to cars (0.39 m/s$^{2}$) and heavy vehicles (0.37 m/s$^{2}$). Lateral motion persisted as continuous micro-adjustments, with lateral speed standard deviation of 0.25 m/s for two-wheelers versus 0.09 m/s for cars, and lateral acceleration variability of 0.18 m/s$^{2}$ for two-wheelers compared to 0.08 m/s$^{2}$ for cars. Overall, the study demonstrates that traffic flow under weak lane discipline emerges from the coupled evolution of longitudinal progression, lateral redistribution, and heterogeneous spacing behavior. The two-dimensional macroscopic formulation, combined with empirically derived microscopic interaction metrics, provides a consistent framework for modeling, calibration, and analysis of disordered traffic systems. The results contribute rare trajectory-based evidence on congestion propagation and equilibrium spacing in lane-free mixed traffic and establish reference values that can inform both macroscopic continuum models and interaction-based microscopic formulations.\\

\textbf{Declaration of Generative AI and AI-assisted technologies in the writing process}\\
The authors declare that a chatbot was used to assist with grammar and language editing of this manuscript. All content generated by the tool was carefully reviewed, revised, and approved by the authors, who take full responsibility for the content of this publication.\\

\textbf{CRediT authorship contribution statement}\\
Shrey Agrawal: Writing – original draft, Software, Formal analysis, Data curation, Methodology.  Gowri Asaithambi: Writing – review \& editing, Methodology, Funding acquisition, Conceptualization. Venkatesan Kanagaraj : Writing – review \& editing, Methodology, Funding acquisition, Conceptualization. Martin Treiber: Writing – review \& editing, Methodology, Conceptualization. Ostap Okhrin: Writing – review \& editing, Methodology, Conceptualization. Harish Babu Kumara: Formal analysis, Data curation.\\ 

\textbf{Acknowledgments}\\
The authors gratefully acknowledge the support received through the Scheme for Promotion of Academic and Research Collaboration (SPARC) project, funded by the Ministry of Education, Government of India (Project No. SPARC/2019–2020/P2384/SL). The third author acknowledges the fellowship support provided by the Alexander von Humboldt Foundation, Germany, during 2025, hosted at the Technical University of Dresden, Germany.











\bibliographystyle{apalike}

\bibliography{bibliography}



\end{document}